\documentclass[superscriptaddress,twocolumn,aps,prb,longbibliography]{revtex4-1}
\usepackage{float}
\usepackage{comment}
\usepackage{booktabs}
\usepackage[percent]{overpic}
\usepackage{microtype}
\usepackage{array}
\usepackage{amsmath}
\usepackage{amsthm}
\usepackage{amssymb}
\usepackage{physics}
\usepackage{graphicx}
\usepackage{bbold}
\usepackage{hyperref}
\usepackage[usenames,dvipsnames]{xcolor}
\usepackage{dsfont}

\bibpunct{[}{]}{,}{n}{}{}
\usepackage{hyperref} 
\hypersetup{colorlinks=true,citecolor=Red,linkcolor=Blue,urlcolor=Blue}

\newcommand{\bcen}{\begin{center}}
\newcommand{\ecen}{\end{center}}
\newcommand{\btab}{\begin{tabular}}
\newcommand{\etab}{\end{tabular}}
\newcommand{\bdes}{\begin{description}}
\newcommand{\edes}{\end{description}}

\newcommand{\beq}{\begin{equation}}
\newcommand{\eeq}{\end{equation}}
\newcommand{\bea}{\begin{eqnarray}}
\newcommand{\eea}{\end{eqnarray}}

\newcommand{\bary}{\begin{array}}
\newcommand{\eary}{\end{array}}
\newcommand{\benum}{\begin{enumerate}}
\newcommand{\eenum}{\end{enumerate}}
\newcommand{\bitem}{\begin{itemize}}
\newcommand{\eitem}{\end{itemize}}

\newcommand{\bK} {{\textbf{K}}}

\newcommand{\Fig}[1]{Fig.~\ref{#1}}

\makeatletter

\newcommand{\Rmnum}[1]{\expandafter\@slowromancap\romannumeral #1@}
\makeatother

\definecolor{ForestGreen}{HTML}{668000}
\definecolor{red1}{HTML}{FF4136}
\definecolor{green1}{HTML}{00802b}

\begin{document}

\title{Anyon dynamics in field-driven phases of the anisotropic Kitaev model}
\author{Shi Feng}
\email[E-mail:]{feng.934@osu.edu}
\affiliation{Department of Physics, The Ohio State University, Columbus, Ohio 43210, USA}
\author{Adhip Agarwala}
\affiliation{Department of Physics, Indian Institute of Technology, Kanpur 208016, India}
 \author{Subhro Bhattacharjee}
\affiliation{International Centre for Theoretical Sciences, Tata Institute of Fundamental Research, Bengaluru 560089, India}
\author{Nandini Trivedi}
\email[E-mail:]{trivedi.15@osu.edu}
\affiliation{Department of Physics, The Ohio State University, Columbus, Ohio 43210, USA}

\begin{abstract}
The Kitaev model on a  honeycomb lattice with bond-dependent Ising interactions offers an exactly solvable model of a quantum spin liquid (QSL) with fractionalized excitations: gapped $Z_2$ fluxes and gapless linearly dispersing majorana fermions in the isotropic limit ($K_x=K_y=K_z$). We explore the phase diagram along two axes, an external magnetic field, $h$, applied out-of-plane of the honeycomb, and anisotropic interactions, $K_z$ larger than the other two. For $K_z/K\gg 2$ and $h=0$, the matter majorana fermions have the largest gap, and the system is described by a gapped $Z_2$ Toric code. One of the central questions we address is whether the fractionalized excitations in the different phases have sharp signatures that can be detected in experiments. 
We show that while the response to single spin excitations is broad, the spectral function corresponding to two-spin excitations across a bond 
has sharp signatures that can be attributed to specific anyons.
In the toric code regime, the $\epsilon=e\times m$ fermion, formed from the bosonic Ising electric (e) and magnetic (m) charge, disperses along a specific one-dimensional direction that provides a fingerprint of fractionalization. 
At lower $K_z$ in the center of the abelian phase, in a regime we dub the primordial fractionalized (PF) regime, the field generates a hybridization between the $\epsilon$ fermion and the majorana matter fermion, resulting in a $\psi$ fermion which too has a distinct quasi-one-dimensional dispersion. All the other phases in the field-anisotropy plane are naturally obtained from this primordial soup. 
These highly constrained fracton-like dispersions can be observable by inelastic light and neutron scattering, thereby providing ``smoking gun" signatures of fractionalization in the QSL phase.  
Our analysis is based on calculations of susceptibilities, topological entanglement entropy, and excitation dynamics, obtained using exact diagonalization and density matrix renormalization group, and supported by perturbation theory.

\end{abstract}
\date{\today}
\maketitle

\section{Introduction}
A Mott insulator is formed because of strong local repulsive interactions in a system with an odd number of electrons in a unit cell~\cite{Mott_1949,Vishwanath}. The fate of the resulting local magnetic moments, interacting with its neighbours  on a specified lattice, can progress along two paths as the temperature is lowered: the moments can undergo long range ordering, spontaneously breaking the spin rotation and/or lattice symmetries, leading to a conventional ordered phase; or the moments can remain disordered but get quantum mechanically entangled with long range patterns of many-body entanglement and form a quantum spin liquid (QSL) \cite{Wen1990,wen2002quantum,Wen2010,savary2016quantum}. The possibility of obtaining QSL phases is enhanced by having a low spin, which leads to greater quantum fluctuations, and frustration arising from the lattice geometry and/or competing exchange interactions~\cite{wen2002quantum,ZhouRMP,KnolleReview}.

The Kitaev model is a paradigmatic model for QSLs \cite{kitaev2006anyons}, consisting of $S=\frac{1}{2}$ local moments or qubits on a two-dimensional honeycomb lattice with very specific bond-dependent compass interactions given by the Hamiltonian 
\begin{align}
    \mathcal{H}_K = \sum_{\langle i,j\rangle_\alpha}{K_\alpha \sigma^\alpha_i \sigma^\alpha_j}
    \label{eq_kitham}
\end{align}
where $\alpha=x,y,z$ refer to the three bonds of the honeycomb lattice. The ground state of the Kitaev model is known to be a topologically non-trivial QSL with fractionalized excitations that have anyon statistics \cite{Vala2008,Bolukbasi_2012}. Topological order in a QSL is reflected in multiple ground state degeneracy characterizing the genus of the lattice manifold. The promise of utilizing these anyons for robust quantum computing has led to considerable excitement and activity in the field, all the way from fundamental theories to applications~\cite{Sankar2008,Semeghini2021,Kato2021,Jason2022}. 

In this model each spin fractionalizes into itinerant majoranas and static $Z_2$ fluxes of the emergent Ising gauge field that are minimally coupled with the itinerant majoranas. The conservation of fluxes at each plaquette allows for an exact solution of the interacting spin model. In particular the ground state lies in the flux free sector \cite{Lieb1994} where, upon fixing a translation invariant gauge, the problem reduces to a nearest neighbor tight-binding model for the itinerant majoranas. At the isotropic point $K_x=K_y=K_z \equiv K$, the matter majorana excitations are gapless, whereas the flux excitations have a small but finite energy gap $\Delta_f \sim 0.26 K$. Upon increasing one of the three bond exchange interactions relative to the others (e.g. $K_z > K_{x/y}$), the matter majoranas remain gapless whereas the gap to flux excitations starts decreasing. At $K_z=2 K$ there is a phase transition to a gapped abelian phase where the matter majoranas also get gapped. For large $K_z$, the gap for matter majoranas increases as $\sim K_z$ and greatly exceeds the flux gap which decreases as $\sim K^4/K_z^{3}$. In this regime the system maps to a Toric code (TC) gauge theory~\cite{kitaev2006anyons}. 

In the TC regime the excitations can conveniently be understood by dividing the Ising fluxes into two classes of bosonic quasi-particles-- the Ising electric charges, $e$, and magnetic charges (also referred to as fluxes), $m$, with mutual semionic statistics that see each other as sources of $\pi$-flux \cite{kitaev2003fault, kitaev2006anyons,nanda2020phases}.

Another experimentally relevant handle that can affect the energetics of the phases is an external magnetic field. In particular, we consider a Zeeman field, $h$, along the $[111]$ direction given by 
\begin{align}
    \mathcal{H}_z =  - h \sum_{i}(\sigma_i^x+\sigma_i^y+\sigma_i^z)
    \label{eq_zham}
\end{align}
added to the Kitaev Hamiltonian (Eq. \ref{eq_kitham}). It is known that a perturbatively small [111] magnetic field in the isotropic limit stabilises the chiral spin liquid (CSL) by gapping out the majoranas, while the fluxes still remain gapped \cite{kitaev2006anyons}. In the anisotropic limit, the primary role of a small magnetic field, however, is to provide dispersion to the $Z_2$ fluxes while the majoranas remain gapped. 

Despite the proliferating investigations on the itinerant majoranas in candidates of Kitaev QSL, the dynamics of $Z_2$ fluxes is less explored. Recently there have been both phenomenological and microscopic theories which predict thermal Hall effects of flux/vison bands of perturbed $Z_2$ QSLs near the isotropic limit of the Kitaev model \cite{Joy2022,Song2022}.
This sets up an interesting question whether one can use the anisotropy Kitaev interaction and the magnetic field to manipulate the nature of the low energy fractionalized quasi-particles and understand the nature of the resultant phases that they favor along with associated phase transitions. 
\begin{figure*}[t]
    \centering
    \includegraphics[width=0.9\textwidth]{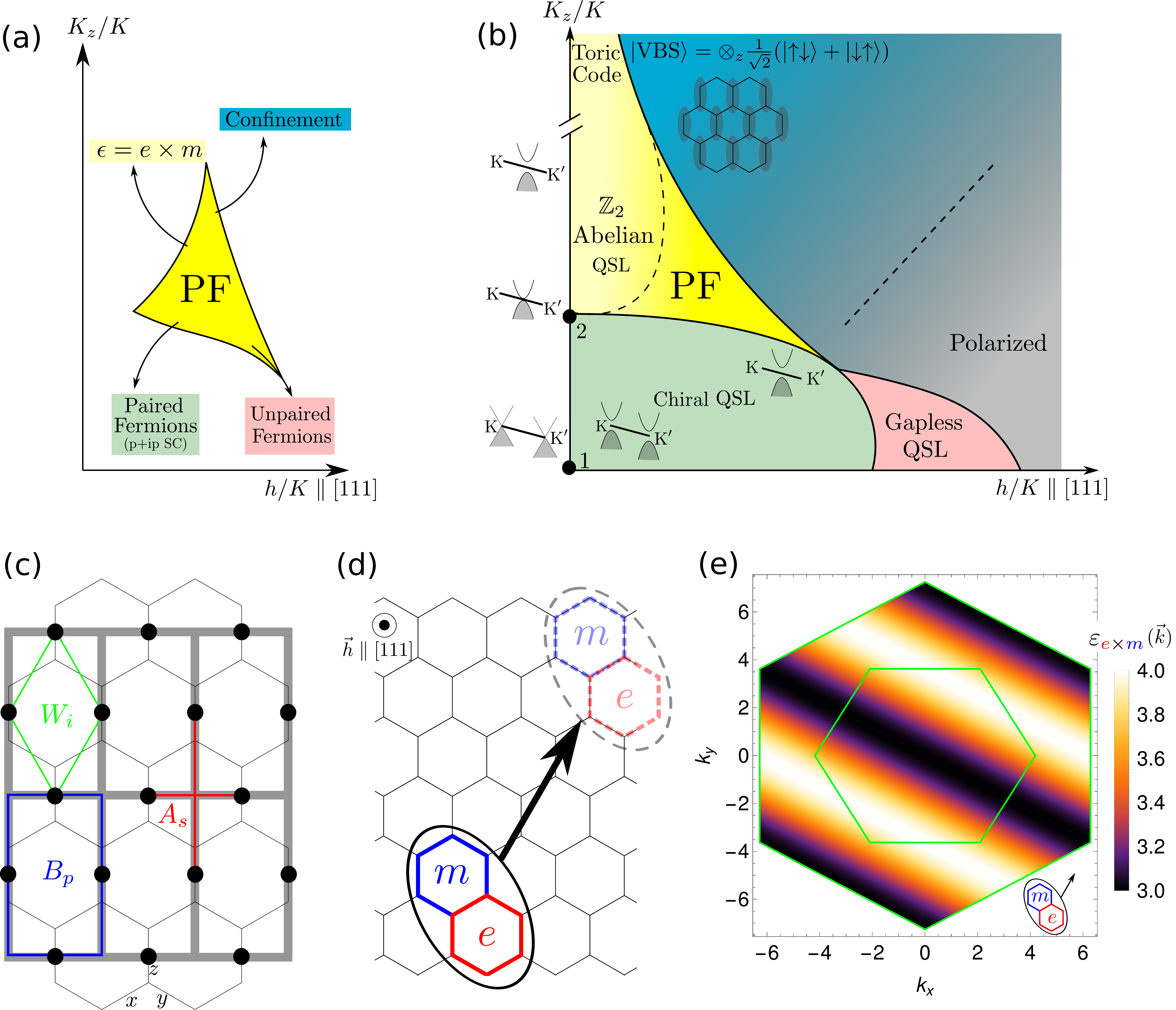}
    \caption{(a,b) Schematic phase diagram in the bond strength anisotropy $K_z/K$ vs magnetic field $h$ along the [111] direction. We identify the general gapped $Z_2$ phase dubbed the primordial fractionalized phase (PF) in yellow color in the center, from which all the other phases around originate. The key low energy excitation of the PF regime is a gapped fermion, $\psi$, that carries $Z_2$ gauge charge and is obtained via magnetic field mediated hybridization of the majorana fermion, $c$, of the Kitaev QSL and the $\epsilon=e\times m$ fermion obtained in the TC limit where $e$ and $m$ are the Ising electric and magnetic charges respectively of gapped $Z_2$ QSL obtained in the TC limit at large $K_z$.  
    Starting from the PF phase with dynamic $Z_2$ matter and gauge fields, for large $K_z$ and small $h$, there is a cross over to the $Z_2$ abelian regime as the $c$ fermion gap increases beyond the flux gap with a concomitant reduction in the magnetic field induced hybridization, approaching the pure TC gauge theory as $K_z/K\rightarrow \infty$.  For large $K_z$ and large $h$ there is a first order phase transition to a z-bond dimer or VBS phase (in blue) formed due to the confinement of the gauge theory~\cite{Nanda_PRB_2021}. The VBS is smoothly connected to the polarized state at large magnetic field at lower anisotropy. In the lower left region for small $K_z$ and small $h$, the flux sector has a larger gap while the majorana sector is gapless at $h=0$ and a finite smaller gap for small $h$. In this phase, the majorana fermions form a gapped non-abelian CSL (shown in green) equivalent to a gapped $p+ip$ superconductor. With increasing field the pairs break forming a gapless QSL with spinon fermi surfaces~\cite{Patel12199} (in pink) and ultimately a polarized phase for large $h$ (in gray).
    (c) Mapping of the anisotropic Kitaev model of $\sigma$ qubits on a honeycomb lattice to an effective square lattice with $\tau$ qubits on links made out of the two $\sigma^\alpha$ spins on the $z$-bond. (d) The 1D soft mode of the composite $\epsilon = e\times m$ fermion induced by a weak out-of-plane magnetic field shown in the honeycomb lattice, whose dispersion within the second Brillouin zone is shown in (e) in arbitrary unit.}
    \label{fig:fig1}
\end{figure*}
Specifically, we would like to ask the following questions:
\begin{enumerate}
    \item Aside from the prediction of quantized edge modes in the chiral spin liquid, are there sharp and direct signatures of fractionalized anyons in the bulk? And are these experimentally detectable? 

    \item What is the interplay between the matter majorana and flux excitations induced by an external magnetic field?  

    \item What is the nature of different phases, the low energy excitations in these phases, and the dynamical mechanisms that drive the phase transitions?
\end{enumerate}
We address the above questions with a set of complementary tools starting from strong coupling perturbative expansions to exact diagonalization (ED) and density matrix renormalization group (DMRG) calculations, to elementary ideas of low energy effective quantum field theory appropriate for topologically ordered phases. For the purposes of this paper, $K_x= K_y = K$ are set to unity, while we vary $K_z>0$ and $h$. 

The most important finding in our results shows that it is possible to find sharp dynamical signatures of the gapped abelian anyons and majoranas through spectral functions of appropriate local spin flips. For low fields, the fractionalized fluxes propagate as composite gauge fermions $\epsilon$ $(e\times m)$ that show distinct one-dimensional dispersion, in contrast to the dispersion of majoranas and bosonic fluxes previously investigated in \cite{Joy2022,Song2022}, illustrated schematically in Fig. \ref{fig:fig1}(d,e). Such highly constrained fracton-like mobility of the gauge excitation $\epsilon$ provides a direct and sharp signature of fractionalized excitations within linear response potentially detectable by inelastic light and neutron scattering experiments, without involving higher-order non-linear correlations~\cite{Armitage2019,Choi2020}.

For higher fields, we find that there is significant matter-gauge flux hybridization in the gapped abelian QSL creating a regime that we dub the primordial fractionalized (PF) regime. In this regime too the signature of anyons is dominated by the one-dimensional fracton-like dispersion, but different from the behavior at low fields where matter and gauge degrees of freedom were uncoupled.           
The importance of the PF phase can be seen from the fact that all the other phases surrounding the PF region in the phase diagram can be obtained from the instabilities of this region along different axes (see Fig. \ref{fig:fig1}(a) and (b)).
The surrounding phases include the CSL and TC phases with increasing bond anisotropy at low fields, the intermediate gapless phase at low bond anisotropy and intermediate fields, and the valence bond solid and polarized phases at high fields and high anisotropy.

The paper is organized as follows. We first discuss the phase diagram in Sec. \ref{sec:phasediag} as a function of $(h/K, K_z/K)$: the [111] magnetic field and the anisotropy in the exchange couplings.  We briefly review the existing knowledge of the phase diagram and discuss the ground state properties using various diagnostics with 24-site ED under periodic boundary condition, revealing two new regions that have not been scrutinized previously: the valence bond solid (VBS) state consisting of isolated dimers on z bonds, which is smoothly connected to the partially polarized state; and the primordial fractionalized (PF) region that features the interplay between majoranas and gauge fluxes. In Sec. \ref{sec:dynAny} we discuss the anyon dynamics induced by [111] magnetic field inside the abelian QSL phase, whose signature is sharply dispersing bound excitations that should be observable within linear response theory. We show numerical evidence of hybridized anyonic excitations inside the PF regime by ED, which is supplemented by a perturbation analysis and by dynamical DMRG on a cylindrical system of $12\times 4$ unit cells (96 sites) with truncation error $\sim 10^{-8}$. We also discuss how excitations in the PF regime are connected to those in adjacent phases, the chiral spin liquid (CSL), the partially polarized magnetic (PPM) phase, and the field-induced intermediate gapless QSL phase. Finally we conclude our investigation in Sec. \ref{sec:discussion} with outlook for future directions; and explain the low-field integrable limit and computational details in the appendix.

\section{Phases and Excitations} \label{sec:phasediag}
In absence of time reversal (TR) symmetry breaking perturbations, {\it i.e.}, for ${h}=0$, the Kitaev model is integrable due to the extensive number of conserved $Z_2$ gauge fluxes resulting in an effective quadratic hopping problem for the majorana fermions in each flux sector  with the ground state belonging to the zero flux sector in accordance with Lieb's theorem \cite{Lieb1994}. For $1< K_z/K \le 2 $ the majorana fermions are gapless, while in the highly anisotropic regime $K_z/K>2 $ they are 
gapped. This exact solution is easily obtained following Kitaev's original prescription~\cite{kitaev2006anyons} of representing the spin degrees of freedom in terms of majorana fermions. 
The nature of the lowest energy excitations in the gapless and gapped $Z_2$ liquid are very different: For $K_z\approx K$, the majorana fermions form linearly dispersing gapless excitations, similar to graphene. However,  
deep inside the anisotropic phase, $K_z/K\gg 2$ the model approaches the TC limit since the effective Hamiltonian can be written in terms of mutually commuting Ising stabilizers~\cite{kitaev2003fault}. In this regime the lowest energy excitations are gapped $Z_2$ fluxes of the honeycomb model which now form abelian anyons: bosonic Ising electric and magnetic charges with mutual semionic statistics \cite{kitaev2003fault, kitaev2006anyons,nanda2020phases}, while the majorana fermions have a gap much larger than the fluxes.  

In this background, it is rather interesting to understand the response of both the gapped and gapless $Z_2$ QSLs to the simplest experimental probe of spin systems -- an external magnetic field -- represented by  Eq. \ref{eq_zham}. Given the fractionalized nature of the low energy excitations and the fact that they couple to the magnetic field differently, a rich set of novel phases can emerge beyond the integrable limit.

For large anisotropy of the bond strengths, {\it i.e.,} $K_z/K\gg 2$, the gapped Kitaev QSL can be mapped to the TC defined on an underlying square lattice as shown in Fig. \ref{fig:fig1}(c)~\cite{nanda2020phases}, where the ground state manifold for large $K_z/K$ is given by $\{\ket{\uparrow\downarrow},~\ket{\downarrow\uparrow}\}$; and the excited states by $\{\ket{\uparrow\uparrow},~\ket{\downarrow\downarrow}\}$. The fourth order perturbation in this 
ground state manifold gives the four-point interaction between these new degrees of freedom $\tau^z = (\sigma_A^z-\sigma^z_B)/2$ which is equivalent to the TC model~\cite{nanda2020phases,nanda2020phases}. 

The quantum phase transition between the gapless and the gapped QSLs arises due to a modification of the majorana band structure at $K_z/K = 2$ in the zero flux  sector. At the isotropic point $K_z/K = 1$ the majorana band has Dirac-like gapless modes located at K and K$'$ points of the Brillouin zone (BZ) similar to electrons in graphene. As is shown in Fig. \ref{fig:fig1}(b), upon increasing the anisotropy, the two Dirac points move towards each other and ultimately combine into a single \emph{semi}-Dirac gapless mode at the M$'$ point when $K_z/K = 2$ is reached \cite{Burnell2011}, which, interestingly, would become linear Dirac modes under weak magnetic field (see Appendix. I); and furthermore the bands get gapped out immediately after the transition point with a gap that continues to increase monotonically with $K_z$ in the zero flux sector by $\mathcal{O}(K_z)$ for $K_z/K\gg 2$, while the energy to create a $\pi$ flux scales as $\sim K^4/K_z^3$. 
These low energy fluxes in the TC limit are canonically best described in terms of a deconfined Ising gauge theory in which these fluxes form electric and magnetic charges~\cite{nanda2020phases}.

\begin{figure*}[t]
    \centering
    \includegraphics[width=\textwidth]{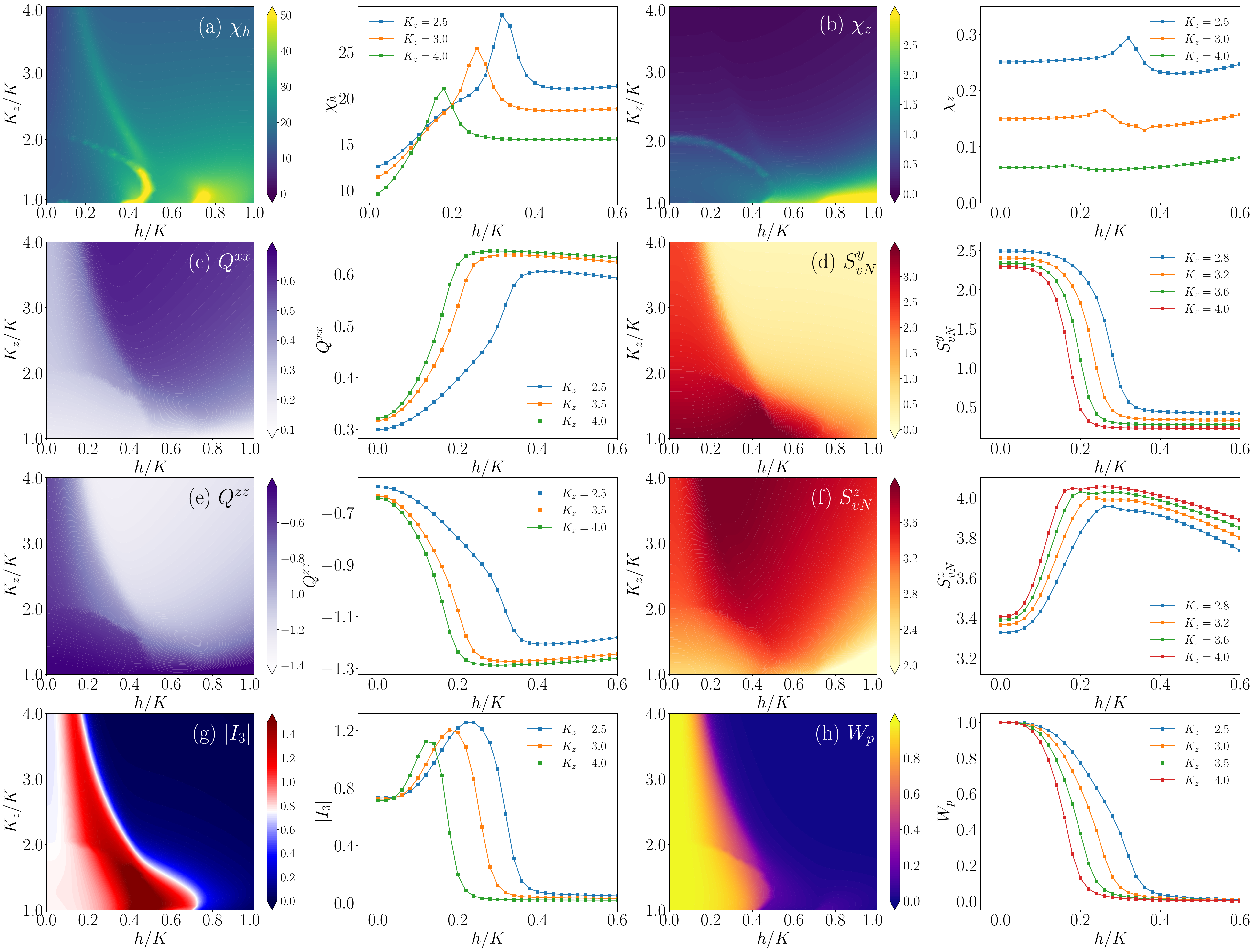}
    \caption{Various diagnostics of the phase diagram as a function of $(K_z, h)$; and their cuts along $h$ at different $K_z$ correspondingly shown on the right side of each density plot. (a) The magnetic susceptibility $\chi_h = \partial_h^2 E_{\text{gs}}/\partial h^2$; (b) Fidelity susceptibility $\chi_z = \partial_{K_z}^2 E_{\text{gs}}/\partial K_z^2$; (c) The $xx$ component of VBS order parameter $Q^{xx}$; (d) Bipartite von-Neumann entropy $S_{vN}^y$ with subsystem boundary cutting only through y bonds; (e) The $zz$ component of VBS (of $z$ dimers) order parameter $Q^{zz}$; (f) Bipartite von-Neumann entropy $S_{vN}^z$ with subsystem boundary cutting only through z bonds; (g) Tripartite entanglement entropy $I_3$, equivalent to the topological entanglement entropy $\gamma$ at small $h$ where the correlation length scale $\xi$ is smaller than the size of the subsystem used in the Kitaev-Preskill construction; (h) Ground state average of the flux operator $W_p$. Data are obtained by diagonalizing a twenty-four-site cluster with torus geometry.}
    \label{fig:static_all}
\end{figure*}

Thus the primary effect of a small [111] external magnetic field is to couple the majorana fermions in the zero flux sector to the low energy Ising gauge charges in the TC limit. This opens up the window for interplay and selective tunability of the different fractionalized excitations which, as we shall show is best understood in terms of a generic primordial $Z_2$ QSL of which the exactly solvable model is a special limit. This primordial fractionalized liquid allows for 
instabilities along different channels 
as a function of bond anisotropy and magnetic field well beyond the currently known small field perturbative regimes -- the CSL at low magnetic field in the isotropic regime obtained by the gapping out the linearly dispersing majoranas via a Chern mass while the fluxes still remains gapped~\cite{kitaev2006anyons}. In addition, we expect a dimerised short range entangled phase formed out of TC topological order~\cite{nanda2020phases,Nanda_PRB_2021} in the highly anisotropic limit via a first order transition owing to the confinement of the $Z_2$ gauge charges. 
These perturbative limits miss the physics of phases at intermediate anisotropy and magnetic field -- the central topic of this work. 

Central to our observation is, while the TC is traditionally described in terms of the bosonic Ising electric, $e$, and magnetic charges, $m$, an equally valid description is, according to the fusion rule
\begin{equation}
    e \times e = 1, ~m \times m = 1, ~e \times m = \epsilon 
\end{equation}
in terms of the fermion $\epsilon=e\times m$ and either $e$ or $m$~\cite{PhysRevResearch.3.023120}.  These are also quasiparticle excitations of the $Z_2$ emergent gauge field that are created when the topological order of the QSL is disturbed, and are able to carry topological charges with emergent braiding statistics. Notably the $\epsilon$ fermion lies in the same super-selection sector as the matter majorana~\cite{kitaev2006anyons} and hence can hybridize with it via local spin operators. As we show, the two complementary effects of the [111] magnetic field are: (1) dispersion of $\epsilon$ fermions, and, (2) hybridization of the $\epsilon$ gauge fermions and matter majoranas leading to the emergence of a low energy hybridized fermion, the $\psi$ fermion, carrying a $Z_2$ gauge charge whose dynamics determines the fate of the system in the intermediate regime far from the perturbative limits.
It is useful here to consider the analogy of the Anderson model in context of heavy fermion system where there are two species of fermions (electrons) -- one almost localized and the other itinerant -- that hybridizes via the Anderson coupling. The situation is somewhat similar except for the fact that both the bandwidth of the localized fermions and the magnitude of the hybridization are fostered by the magnetic field. The PF liquid is therefore the analog of the heavy fermion phase, albeit in this case, it is gapped. From such point of view, the Kitaev model is the exactly solvable limit where the hybridization goes to zero concomitantly with the gapping of the $\epsilon$ fermion, while the majorana fermions form a gapped or gapless spectrum depending on the anisotropy.

\subsection{Diagnostics of the Phases and Phase Transitions}
As the $[111]$ magnetic field is turned on, there emerges a far richer phase diagram as a function of both $K_z$ and $h$ than known previously. Fig.~\ref{fig:static_all} summarises various diagnostic calculations, which together determine the phase boundaries and crossovers shown in the schematic diagram Fig. \ref{fig:fig1}. 

Our discovery lies in the behavior of the gapped $Z_2$ abelian phase in a finite magnetic field obtained by making the coupling $K_z$ larger compared to the other two. For $K_z\gg K$ in the TC limit, the $Z_2$ abelian QSL undergoes a first order transition to a confined short range entangled dimerized phase discussed in greater detail in next section, which, in turn gives way to a polarized phase at even larger fields $h\sim K_z$. However, we find that there exists a remarkably richer physics between the integrable $Z_2$ abelian phase at zero field and the polarized phase at large field as already indicated schematically in Fig. \ref{fig:fig1}(b).  This includes -- (1) a CSL phase at small magnetic field, (2) A U(1) gapless quantum spin liquid at intermediate magnetic field and almost isotropic limit, (3) centrally, for weak fields a gapped abelian QSL in which the $\epsilon$ anyon excitations are effectively one-dimensional, and (4) a gapped abelian $Z_2$ QSL at intermediate magnetic field and anisotropy that we dub the PF regime also showing characteristically dispersing one-dimensional hybridized $\psi$ fermions. Here we would like to note that there is still ongoing discussion about the nature of the intermediate field phase near the isotropic limit. In addition to the gapless U(1) QSL revealed by various approaches including DMRG \cite{Jiang_arXiv_2018,Patel12199,hickey2019emergence,zhang2023machine}, a gapped abelian $Z_2$ QSL with non-zero Chern bands within mean field theory has been suggested in \cite{Jiang2020Phys.Rev.Lett.,Zhang2022NatCommun}. The DMRG result can be affected by finite size effects and the mean field result is susceptible to fluctuations; further investigations are needed to understand the dichotomy arising from the different gauge groups: the presence of low lying pockets of fermionic excitations in DMRG points to a U(1) gauge group whereas mean field theory finds a $Z_2$ gauge group arising from fermion pairing. In the following text we discuss different diagnostics of them. 

\medskip

\noindent {\it Susceptibilities:} We first calculate the susceptibilities as the conventional measure of quantum phases and phase transitions:
\begin{equation}
    \chi_h = \frac{\partial^2 E_{\text{gs}}(K_z,h)}{\partial h^2},~~~ 
    \chi_z = \frac{\partial^2 E_{\text{gs}}(K_z,h)}{\partial K_z^2}
\end{equation}
The results are shown in Fig. \ref{fig:static_all}(a,b) and their cuts on the right side of the contours. 
Figure. \ref{fig:static_all}(a) shows the zero-temperature magnetic susceptibility
$\chi_h$, which marks out several boundaries highlighted in the schematic phase diagram. At finite field in about the range $0.4 \lesssim  h/K \lesssim 0.7$ and for near-isotropic interaction, there is a gapless QSL sandwiched between two singularities of $\chi_h$, reported in our previous works \cite{David2019,Patel12199,Pradhan_PRB_2020} as well as others
~\cite{Jiang_arXiv_2018,Nasu_PRB_2018,Liang_PRB_2018,Gohlke_PRB_2018,hickey2019emergence,Berke_PRB_2020,Zhu_PRB_2018}.
For even larger $h$ the system becomes partially polarized, whose phase boundary is also clearly visible in Fig. \ref{fig:static_all} as singularities in different measures (see below). $\chi_h$ marks out the four phases: the CSL in the lower left region, the abelian $Z_2$ QSL at high anisotropy and finite field, the gapless QSL and the VBS/polarized phase; while in $\chi_z$ (Fig. \ref{fig:static_all}(b)) the boundary between abelian $Z_2$ QSL and the VBS/polarized phase are not as clear, since the transition is being driven by the field. 

\medskip

\noindent {\it von-Neumann Entanglement Entropy:} The distinction between the VBS and the polarized state is not revealed by the susceptibility measurements. To characterize the emergence of the VBS phase we calculate the von-Neumann entropies $S_{vN}^y$ and $S_{vN}^z$ for subsystems obtained by cutting along $y$ bonds or $z$ bonds respectively. These are shown in Fig. \ref{fig:static_all}(d) and (f).  While $S_{vN}^y$ drops off once the QSL is destroyed, $S_{vN}^z$, persists in the VBS phase indicating the presence of dimers on them, which finally gives way to the polarized phase at even larger magnetic fields where $S_{vN}^z$ falls off to zero. The entanglement entropies are also generically sensitive to all gap-closing transitions such as the one out of the CSL.

\medskip

\noindent {\it Quadrupolar order:}
We define the following operator to probe dimerization:
\begin{align} \label{eq:nem_para}
\hat{Q}^{\alpha \beta}_{pp^{\prime}} = \left(\frac{\sigma^{\alpha}_p\sigma^{\beta}_{p^{\prime}} + \sigma^{\beta}_p\sigma^{\alpha}_{p^{\prime}}}{2} - \frac{\delta_{\alpha\beta}}{3}\boldsymbol{\sigma}_p.\boldsymbol{\sigma}_{p^{\prime}}\right)
\end{align}
where $pp'$ stands for the z-bond and $\alpha,\beta \in \{x,y,z\}$. This is formally equivalent to a quadrupolar order or spin nematic order, though the symmetry is explicitly broken away from the isotropic limit. The behavior of the order parameter is discussed in more detail in the next section.  A finite value of the quadrupolar order parameter as shown in Fig. \ref{fig:static_all}(c) and (e), marks out the distinction between VBS and polarized phase as schematically shown in Fig. \ref{fig:fig1}(b).

\medskip

\noindent {\it Mutual Information $I_3$:}
The PF regime is not visible in standard measurements based on two-point correlations, e.g. $\chi_z$ and $\chi_h$ shown in Fig. \ref{fig:static_all}(a,b), however, its nature is explicitly revealed by the third order mutual information $I_3$ :
\begin{equation}
    I_3(\mathcal{A}_1:\mathcal{A}_2:\mathcal{A}_3) = I(\mathcal{A}_1:\mathcal{A}_2) + I(\mathcal{A}_1:\mathcal{A}_3) - I(\mathcal{A}_1:\mathcal{A}_2 \mathcal{A}_3)
\end{equation}
where $I(\mathcal{A}_1:\mathcal{A}_2 \mathcal{A}_3)$ is the quantum mutual information between the region $\mathcal{A}_1$ and $\mathcal{A}_2\cup \mathcal{A}_3$, as shown in Fig. \ref{fig:static_all}(g). The physical meaning of $I_3$ converges to the topological entanglement entropy $\gamma$ \cite{kitaev2006topological,levin2006detecting,feng2023statistical}:
\begin{equation}
\begin{split}
        I_3 \rightarrow -\gamma = &S(\mathcal{A}_1) + S(\mathcal{A}_2) + S(\mathcal{A}_3) - S(\mathcal{A}_1\mathcal{A}_2)\\ &-S(\mathcal{A}_1\mathcal{A}_3) - S(\mathcal{A}_2\mathcal{A}_3) + S(\mathcal{A}_1\mathcal{A}_2\mathcal{A}_3)
\end{split}
\end{equation}
when the correlation length satisfies $\xi/\abs{\partial \mathcal{A}_i} \rightarrow 0$ and $\mathcal{A}_i$ are chosen such that they share boundaries with each other. This has been used in characterizing various of $Z_2$ topological orders \cite{Furukawa2007,David2019} where $\gamma = \ln 2$.  We will discuss details in the forthcoming sections. 

\medskip

\noindent {\it Plaquette Fluxes:} 
We calculate the flux expectation for ground states at different $(K_z,h)$ (Fig. \ref{fig:static_all}(h)) which marks out the deconfined phases -- both the gapped and gapless QSLs -- with the ground states belonging to the zero flux sector.
In the PF regime, there are strong fluctuations of flux excitations that give rise to a definitive dispersion of fractionalized anyons, discussed in the next section. 
Based on the dynamics of the excitations and their dispersion, we show below how all the other phases surrounding the PF region emerge by confinement or by Fermi surface construction of composite fermions. 

\subsection{The strong $K_z$ limit: Toric Code}

The low energy degrees of freedom in the limit of $K_z/K\gg 2$ are obtained by considering only the $z$-bonds and the interaction $K_z\sigma_A\sigma_B$ which results in the ground state doublet $\{\ket{\uparrow\downarrow},~\ket{\downarrow\uparrow}\}$ and excited states $\{\ket{\uparrow\uparrow},~\ket{\downarrow\downarrow}\}$. The ground state manifold is spanned by eigenstates of $\tau^z = (\sigma_A^z-\sigma^z_B)/2$~\cite{nanda2020phases}.

The effective Hamiltonian for the $\tau^\alpha$ operators is systematically obtained via degenerate perturbation theory on the Kitaev Hamiltonian in the presence of a [111] magnetic field $h$. To leading order in $h$ it is given by,
\begin{align}
    \mathcal{H} = -J_{\rm TC}\sum_{i} W_i- \frac{2h^2}{K_z}\sum_i   \tau^x_i
    \label{eq_tcham}
\end{align}
where $W_i$ and $J_{\rm TC}$ are given by
\begin{equation}
    W_i \equiv \tau^z_{i+d_1}\tau^z_{i-d_2}\tau^y_i\tau^y_{i+d_1-d_2}, ~J_{\rm TC}\equiv\frac{K^4}{16|K_z|^3}
\end{equation}
While the first term is the TC Hamiltonian (in Wen's representation~\cite{wen2002quantum}), the second term arises due to the $[111]$ magnetic field. Note that this term is in fact time-reversal symmetric since it is quadratic in $h$ and $\tau^x$ is even under time reversal given that  $\tau^\alpha$ is a non-Kramers doublet~\cite{Nanda_PRB_2021,nanda2020phases}. Therefore Eq.~\ref{eq_tcham} retains the time reversal (and some other lattice symmetries) that are lifted at higher order in perturbation theory. Indeed, the next order in perturbation $h$ contributes the term
\begin{equation}
     \sim \frac{h^5}{K^4_z} \sum_{i} \Big( \tau^x_{i+d_2} \tau^z_{i} \tau^x_{i+d_1} - \tau^x_{i-d_1} \tau^z_{i} \tau^x_{i-d_2} \Big)
\end{equation}
which is odd under ${\cal T}, \sigma_v, C_{2z}$ and even under $R_\pi$. This operator is significantly smaller than the leading order contribution in Eq. \ref{eq_tcham} which dominates the essential dynamics of the abelian QSL phase.

It is useful to understand the physics described above in terms of $\tau$ operators also in terms of the $\sigma$ spins.  For instance the TC ground state is given by
\begin{equation}  \label{eq:tcgs}
 |\text{GS}\rangle = \prod_i \Big( \frac{1 + W_{i}}{2} \Big) \otimes_z | \uparrow \downarrow \rangle 
\end{equation}
where $W_i$ (see Eq. \ref{eq_tcham}) flips both the spins on two consecutive $z$ bonds albeit with a decorated sign structure that depends on the other $z$ bonds. This highly entangled state leads to a finite topological entanglement entropy with low energy excitations above it being comprised of bosonic $e$ and $m$ charges corresponding to $\langle W_i \rangle = -1$ on vertices and plaquettes of the square lattice (in Fig. \ref{fig:fig1}(c)) respectively. These excitations are static in the absence of a field but gain field-dependent dispersions. The instability of these excitations lead to various phase transitions. 

We now perform unitary transformations on the horizontal and vertical bonds according to
\begin{align}
    {\rm horizontal ~bonds:~~} &\{\tau^x, \tau^y,\tau^z \} \rightarrow     \{\tau^y, \tau^x,-\tau^z \} \\
    {\rm vertical ~bonds:~~} &\{\tau^x, \tau^y,\tau^z \} \rightarrow     \{\tau^y, \tau^z,\tau^x \}
\end{align}
The transformed Hamiltonian is given by
\begin{align} 
    \tilde{\mathcal{H}}=&-J_{\rm TC}\left[\sum_{s} A_s+ \sum_{p} B_p\right] -\sum_i  \frac{2h^2}{K_z} \tau^y_i
    \label{TChpert}
\end{align}
where $J_{\rm TC}=\frac{K^4}{16\abs{K_z^3}}$, and $A_s$, $B_p$ are the star and the plaquette operators
\begin{equation}
	A_s = \prod_{i\in +_s} \tau_i^x,~ B_p = \prod_{i\in \square_p} \tau_i^z
\end{equation}
of the standard TC lattice as shown in Fig. \ref{fig:fig1}(c). Therefore, to leading order in perturbation theory, the honeycomb model is equivalent to the TC Hamiltonian perturbed by a Zeeman field in the $y$ direction shown in Eq. \ref{TChpert}. Next, we will discuss both the static and dynamical effects of this effective Hamiltonian in the honeycomb model.

\subsection{The VBS and the polarized phase}

The polarized state in the $\tau^y$ direction, in terms of $\sigma$-spins, is given by 
\beq \label{eq:nematic}
|\Psi_{\rm VBS}\rangle = \otimes_z \frac{1}{\sqrt{2}} \Big( \ket{ \uparrow \downarrow }  +  \ket{\downarrow \uparrow } \Big)
\eeq
which is a short-range entangled dimer state with zero total magnetization, {\it i.e.,} $\langle\Psi_{\rm VBS}|\sigma^\alpha_i|\Psi_{\rm VBS}\rangle=0$. However, the expectation value of the bond operator $\langle\Psi_{\rm VBS}|Q^{\alpha\beta}|\Psi_{\rm VBS}\rangle\equiv\langle Q^{\alpha\beta}\rangle_{\rm VBS}\neq 0$ 
as shown in Fig. \ref{fig:static_all}(c) for $\langle Q^{xx}\rangle_{\rm VBS}=2/3$. 
This state, even while being a short range entangled (product over the $z$-bonds), is however distinct from a completely polarized state
found as $h \gg K_z$, where all the spins are polarized in the $[111]$ direction. In this state the magnetization is finite but $\langle Q^{xx}\rangle_{\rm PM}=0$.
For the dimerized VBS and polarized states, the expectation of the order parameter in
Eq. \ref{eq:nem_para} is found to be
\begin{align}
\expval*{\hat{Q}}_{\text{VBS}}=
\begin{pmatrix}
\frac{2}{3} & 0 & 0\\
0 & \frac{2}{3} & 0\\
0 & 0 & -\frac{4}{3}\\
\end{pmatrix}
,~
\expval*{\hat{Q}}_{\text{PM}} =
\begin{pmatrix}
0 & \frac{1}{3} & \frac{1}{3}\\
\frac{1}{3} & 0 & \frac{1}{3}\\
\frac{1}{3} & \frac{1}{3} & 0\\
 \end{pmatrix}
\label{eq_nematicop}
\end{align}
Note however that the VBS and polarized phases are smoothly connected and unlike a symmetry broken state -- here time-reversal, spin-rotation, and lattice symmetries are explicitly broken due to a $K_z$ anisotropy and a field. 
This gives rise to excitations distinct from the trivial PP. Above the VBS ground state, the gapped excitations are singlet state ($\ket{\uparrow\downarrow} - \ket{\downarrow\uparrow}$) at energy scale $\sim h^2/K_z$, and ($\ket{\uparrow\uparrow}, \ket{\downarrow\downarrow}$) at $\sim K_z/K$.

Notably the fully polarized state is a direct product state in terms of the individual $\sigma$-spins such that the bipartite entanglement between any subsystem bipartition is zero. On the other hand, the dimer state, $|\Psi_{\rm VBS}\rangle$, is a product state over the $z$-bonds and hence any bipartition cutting a $z$ bond would contribute $\ln 2$ per $z$ bond to the von-Neumann entanglement entropy. Such Bell-pair contribution is absent when the bipartition is made across the $y$ bonds. This is shown in Fig. \ref{fig:static_all}(d) and (f).
An alternative insight to the physics of Eq. \ref{TChpert} including the nature of the transition between the gapped QSL and the VBS in the $K_Z/K\gg 2$ limit, can be obtained via a series of mappings~\cite{Vidal_PRB_2009} starting with Eq. \ref{TChpert} via Xu-Moore models~\cite{PhysRevLett.93.047003} leading to the compass model~\cite{PhysRevB.71.195120}.

\subsection{The PF region and $Z_2$ phases}

The PF region can be targeted by $I_3$. 
For small $h$ the fluxes are approximately conserved, and we retrieve the exact mutual information $I_3 = -\gamma = -\ln 2$ in both CSL and abelian $Z_2$ QSL as is shown in the white area of Fig. \ref{fig:static_all}(g). 
Note that for the Kitaev QSL phase at $h=0$, the spin-spin correlation  length $\xi$ is short ranged because of conserved $Z_2$ charges despite the presence of gapless majorana modes.  For small $h$, $\xi$ continues to be extremely small due to the (approximate) orthogonality between different gauge configurations so it is possible to retrieve the topological entanglement entropy locally \cite{Feng2022}. 

For $1<K_z/K<2$ with intermediate magnetic field, the mutual information exceeds $\abs{I_3} =  \ln 2$  and marks out a gapless QSL phase with a large correlation length $\xi$.  Interestingly, for relatively large $h$ within the abelian $Z_2$ QSL, $\abs {I_3} \simeq 2\ln 2$, indicating strong scrambling of the Hilbert space and delocalization of information \cite{Hosur2016,Ryu2020}, arising from a strong mixing between gauge and majorana sectors. This is the region we dub the primordial fractionalized (PF) regime, which is connected to the abelian $Z_2$ QSL by a cross-over illustrated in Fig. \ref{fig:fig1}(b). This region becomes a thin sliver which ultimately disappears as the anisotropy increases $K_z/K \gg 2$, because in this region the large majorana gap prevents the matter-gauge hybridization, as is also visible in Fig. \ref{fig:static_all}(g).

\section{Anyon dynamics} \label{sec:dynAny}
In this section we discuss the spectral function of spin excitations under a [111] magnetic field and with varying anisotropy obtained by ED and DMRG, followed by their interpretation based on perturbation theory.

It is well known that the spectrum of single-spin excitations is broad in a QSL~\cite{balents2010spin,Knolle2014,KnolleReview,Sato2021} in contrast to sharp well-defined dispersing modes in energy and momentum in an ordered magnet. The main question we address below is whether it is possible to have a sharp diagnostic of a QSL that could be measured in a linear response experiment.
We show that the linear response spectrum of two-spin excitations across a bond indeed shows very particular 1D dispersion at low energies. The reason for considering two-spin excitations can be seen from Eq.~\ref{eq:tcgs}. 
While the magnetic field that couples to a single spin operator $\sigma^x$ or $\sigma^y$ projects the $z$-dimer into high energy configurations $\{\ket{\uparrow\uparrow}, \ket{\downarrow\downarrow}\}$ at the order of $O(K_z/K)$; the two-spin operator can project it back into the low energy manifold $\{\ket{\uparrow\downarrow}, \ket{\downarrow\uparrow}\}$, hence is potentially capable of probing excitations within the low energy sector of the high-anisotropy QSL.
Also, when projected into the TC model, the magnetic field to the lowest order couples to a $\tau^y$ as shown in Eq. \ref{TChpert}, leading to low energy excitations above the TC ground state with tractable dynamics. This $\tau^y$ perturbation in TC defined in a square lattice is in fact a two-spin operator in terms of the underlying honeycomb spin system which therefore provides a concrete reason to  probe low energy excitations of Kitaev spin liquids.
\begin{figure*}[t]
    \centering
    \includegraphics[width=\textwidth]{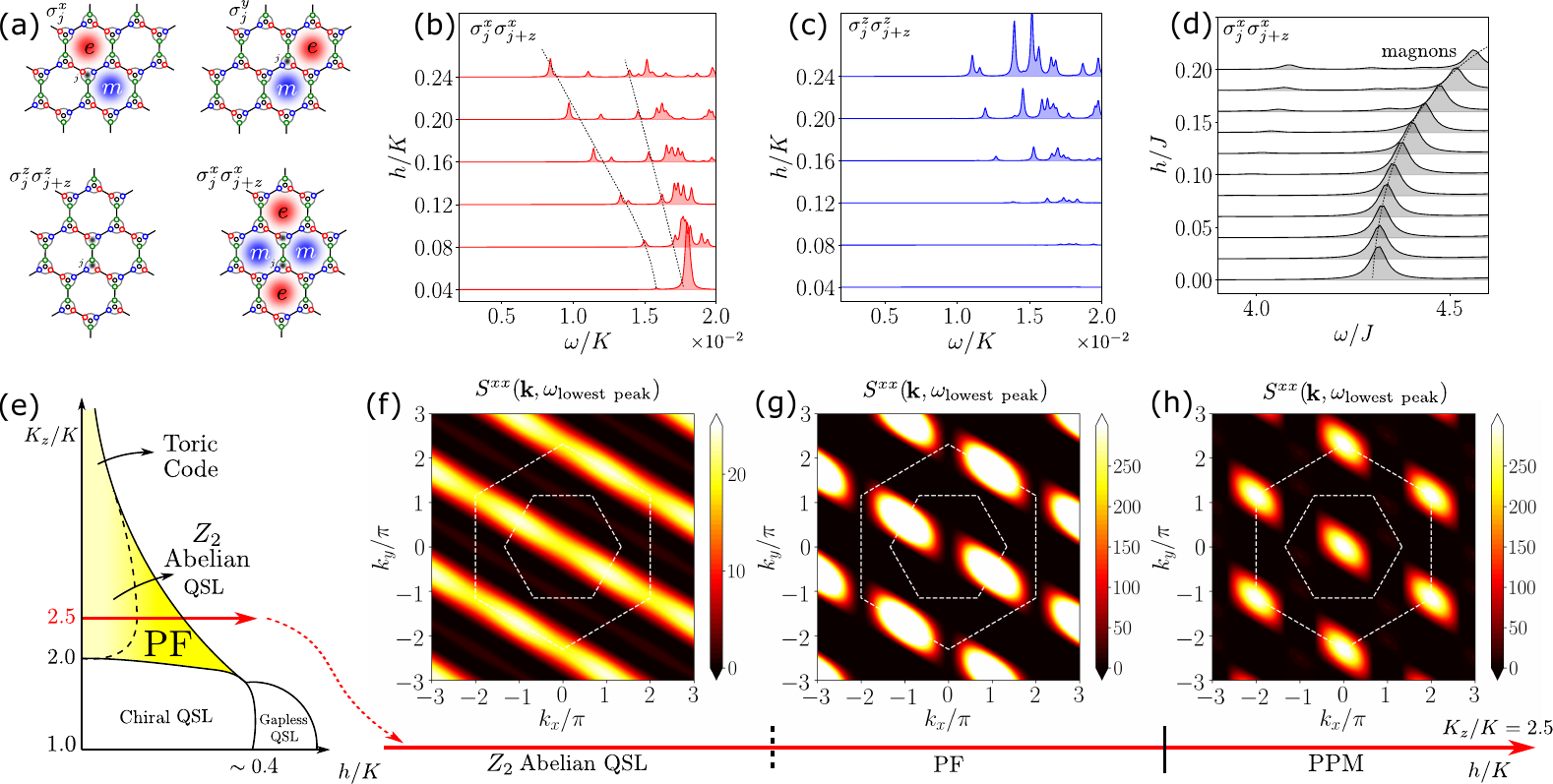}
    \caption{(a) Identifying spin operators that create specific fractionalized excitations. (b) Excitation spectrum $S^{xx}(\omega)$ for the operator $\sigma_j^x \sigma_{j+z}^x$ at $K_z/K=2.5$ for different values of $h$. The dashed black lines track the evolution of the different peaks as a function of $h$. 
    (c) Excitation spectrum $S^{zz}(\omega)$ for the operator $\sigma_j^z \sigma_{j+z}^z$ with the same cut along $K_z/K = 2.5$. (b,c) are rescaled in magnitude so as to make peaks clear; the intensity of the peak in (b) is the order of $10^{3}$ times that of the peak in (c). The origin of multiple peaks  at finite field is a distinct signature of the fractionalization of spins. This is in sharp contrast to the sharp magnon mode of a Heisenberg magnet with anisotropic Heisenberg exchange observed in this field range, as shown in panel (d). Panel (e) shows the schematic phase diagram focusing on the crossover from the region with the lower flux gap to the PF phase where the matter fermions and flux degrees of freedom hybridize. We calculate the evolution of dynamics along the cut marked by the red arrow. (f-h) show the momentum-resolved dynamical structure factor $S^{xx}({\bf k},\omega)$ of the first peak only of panel (b) under different magnetic fields. The dotted hexagons mark the first and second Brillouin zones. $S^{xx}({\bf k}, \omega)$ for the operator $\sigma_j^x \sigma_{j+z}^x$ for three different field: (f) $h/K = 0.08$ in the abelian phase with fluxes; (g) $h/K = 0.20$ in the PF phase with hybridized majoranas and fluxes; (h) $h/K = 0.40$ in the partially polarized phase.
    The excitation in the first peak is dominated by the $\epsilon=e\times m$ particle that only disperses along the fixed $d_1$ direction, giving the 1D dispersion defined by Eq. \ref{eq:dispersion1} as seen in (g). Inside the 
    PF region the hybridization between the majorana and the $\epsilon$ fermions leads to missing intensity near the $\Gamma$ points as seen in (h). The first order phase transition to the polarized phase shows peaks at the $\Gamma$ points as expected for a ferromagnet.}
    \label{fig:dyn}
\end{figure*}

We show from perturbation theory that the aforementioned 1D dispersion arises from the fractionalized $\epsilon$ fermions induced by the [111] magnetic field in the low energy gauge sector of the anisotropic abelian QSL. However, as the anisotropy $K_z/K$ is lowered towards the gapped to gapless  transition point, the majoranas and fluxes hybridize as the majorana gap decreases and becomes on the order of the flux gap. We show that the dynamical structure factor and the dispersion of $\epsilon$ anyon changes qualitatively in this regime -- to the leading order the dispersion is dominated by the one dimensional dynamics of $\epsilon$ anyon, however there are significant changes to the spectral weights in momentum space. These changes in the spectral properties can be understood as arising from the hybridization between the $\epsilon$ anyons and matter majoranas. 

We next present our results on the effects of two-spin excitations on the dynamical signatures in energy and momentum within linear response and propose their identification by  probes such as inelastic light and neutron scattering as “smoking-gun” signatures of fractionalization and anyon excitations in the abelian QSL phase.

\begin{figure*}[t]
    \centering
    \includegraphics[width=\textwidth]{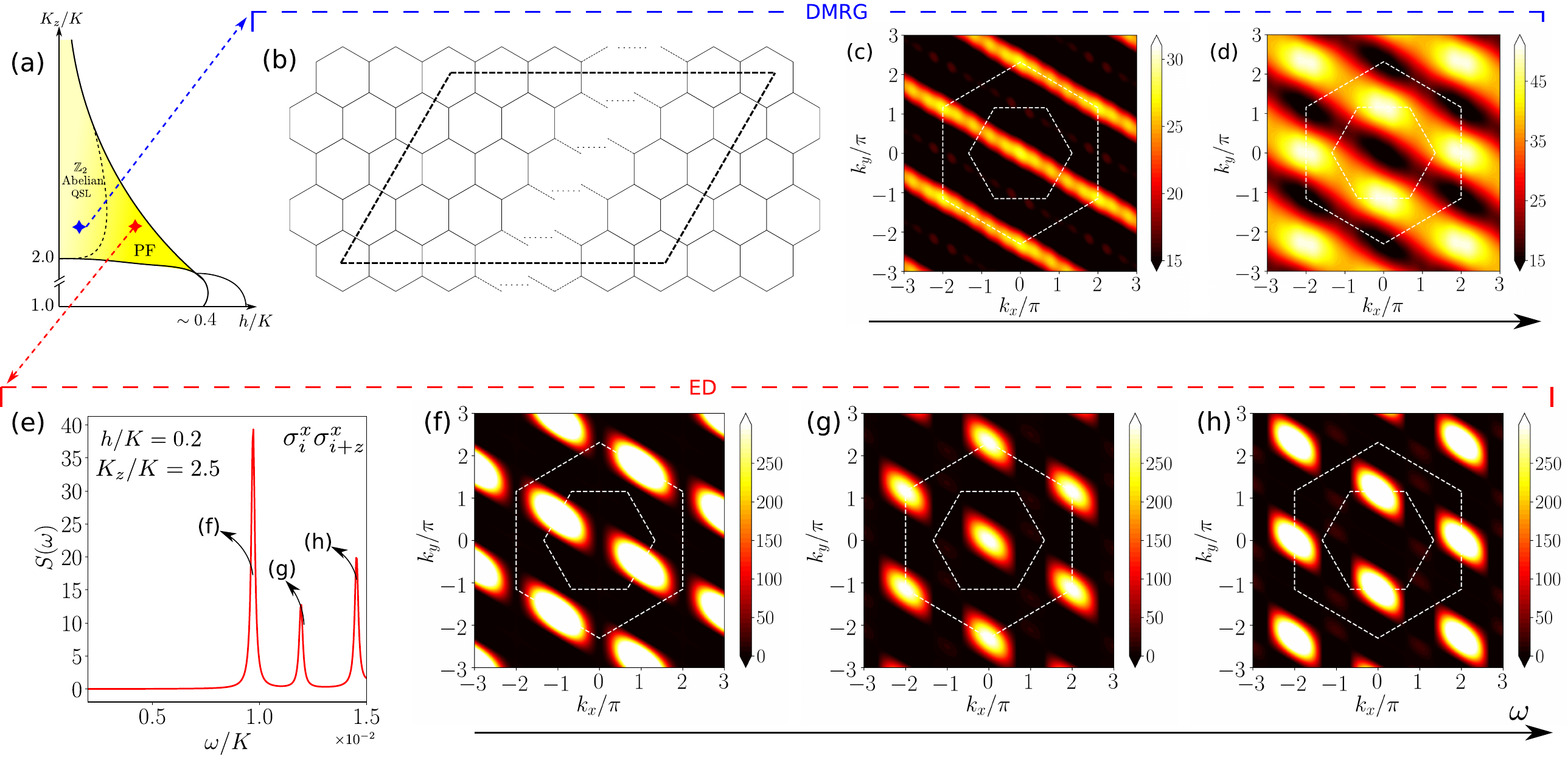}
    \caption{Dynamics of higher-energy quasiparticle excitations, in the abelian phase with only gauge degrees of freedom and in the PF phase with coupled matter and gauge degrees, using ED and DMRG. (a) Schematic diagram with the blue point near the PF regime used to obtain the dynamics with 96-site DMRG. The red point marked inside the PF regime is used to obtain the dynamics with 24-site ED. (b) the $12\times 4$ lattice geometry (96 sites) used in dynamical DMRG, with PBC in $y$ and OBC in $x$ direction. (c,d) Dynamical structure factor $S^{xx}(k, \omega)$ of $\sigma_j^x \sigma_{j+z}^x$ near PF regime. (c) shows the lowest excited mode of operator $\sigma_j^x \sigma_{j+z}^x$ that corresponds to the 1D $\epsilon$ dispersion; and (d) shows the second lowest excited mode i.e. the hybridized $\psi$ particle. (e) The spectral function of operator $\sigma_j^x \sigma_{j+z}^x$ at $K_z/K = 2.5, h/K = 0.2$ in arbitrary units. The first three modes are further revealed in the momentum-resolved $S^{xx}(\mathbf{k}, \omega)$ in (f-h).  }
    \label{fig:dyndmrg}
\end{figure*}


\subsection{Dynamics in the abelian QSL and PF regimes: ED and DMRG results}

To understand the nature of the low energy excitations,
we focus on two excitations: (i) created by $\hat{\mathcal{O}}_i^x = \sigma_i^x \sigma_{i+z}^x$, and (ii) created by $\hat{\mathcal{O}}_i^z = \sigma_i^z \sigma_{i+z}^z$ across a $z-$ bond. The response of these composite particles to an external field is revealed by the dynamical momentum-resolved structure factor:
\begin{align}
    S^{\alpha\beta}(\mathbf{k}, \omega) = - \frac{1}{\pi} \mathfrak{Im} \left[ \mel{\Psi} {\hat{\mathcal{O}}_{\mathbf{k}}^\alpha \frac{1}{\omega - \mathcal{H} + i \eta} \hat{\mathcal{O}}_{-\mathbf{k}}^\beta} {\Psi}\right]
\end{align}
and the spectral density of states for the two-spin operator:
\begin{equation}
        S^{\alpha\beta}(\omega) = -\frac{1}{\pi} \mathfrak{Im} \sum_i \left[ \mel{\Psi} {\hat{\mathcal{O}}_{i}^\alpha \frac{1}{\omega - \mathcal{H} + i \eta}
    \hat{\mathcal{O}}_{i}^\beta} {\Psi}\right]
\end{equation}
excluding the ground state contribution, 
where the operator $\hat{\mathcal{O}}^{\alpha(\beta)}$ is the two-spin operator, $\mathcal{H}$ is the total Hamiltonian with Zeeman field and $\eta$ is a small spectral broadening. Figure.~\ref{fig:dyn}(a) illustrates fractionalized excitations created by single and double Pauli matrices.

The spectral density of states $S^{xx}(\omega)$ and $S^{zz}(\omega)$ for the two operators are shown in Fig. \ref{fig:dyn}(b,c). For a small $h/K \approx 0.04$, $\sigma_{j}^x \sigma_{j+z}^x$ has a large spectral peak at an energy of $O(J_{\text{TC}})$. Remarkably, with increasing $h$, the peak fractionalizes into many sub-peaks. The the lowest energy peak develops around $\omega = 1.6\times 10^{-2}$, as shown in Fig. \ref{fig:dyn}(b), whose intensity grows as the field increases. Indeed, as we discuss below, the $k$-resolved dynamics at the energy values of these spectral peaks reveals the 1D dispersion of fractionalized anyons.  
On the other hand, $S^{zz}(\omega)$ for $\sigma_j^z \sigma_{j+z}^z$, has non-zero weight only at $\omega \sim O(K_z/K)$ for small $h$ and therefore gives no signal for $\omega \sim J_\text{TC}$, as shown in Fig. \ref{fig:dyn}(c). A non-zero signal begins to develop at $h \gtrsim 0.16$, whose intensity, however, is much smaller than that of $S^{xx}(\omega)$.



We would like to emphasize that such spectral weight fractionalization is precisely a signature of spin fractionalization. The external magnetic field acts like a ``dispersive prism'' that reveals the identity of dynamical modes of each constituent fractionalized degrees of freedom. This is in sharp contrast to the signature of a quasiparticle in an ordered Heisenberg magnet with a magnon peak that disperses in a field but remains intact throughout, as shown in Fig. \ref{fig:dyn}(d). In addition the field dependence of the peaks in the ordered magnet and the QSL shows opposite behavior, as marked by dashed lines in Fig. \ref{fig:dyn}(b).


The momentum-resolved $S^{xx}(\mathbf{k}, \omega)$ further reveals the sharp dispersion of the fractionalized quasiparticles. Figure~\ref{fig:dyn}(f) shows the dynamical structure factor of the lowest peak of the corresponding $S^{xx}(\omega)$ at $h=0.08$, $K_z/K = 2.5$ shown in Fig. \ref{fig:dyn}(b). These results are consistent with the large scale 96-site dynamical DMRG calculations at $h = 0.1$ \cite{re:white92,re:white93,alvarez09,alvarez0311,Alvarez2016} as shown in Fig. \ref{fig:dyndmrg}(c), whose details are discussed in Appendix. II. 
Remarkably, $S^{xx}(\mathbf{k}, \omega)$ under small magnetic field exhibits a readily discernible one-dimensional pattern. 
We would like to emphasize that this sharp 1D signature of fractionalization of anyon excitations appears within linear response. In previous work, the linear response of single spin operators yielded broad continua \cite{Knolle2014,KnolleReview,Sato2021} due to fractionalized quasi-particles and sharp signatures were reported only in the non-linear response regime \cite{Armitage2019,Choi2020}. 

In a larger field, the system is driven into the PF regime where we expect strong majorana-flux hybridization, as is shown in Fig. \ref{fig:dyn}(g) and Fig. \ref{fig:dyndmrg}(f). 
Spectral weight is found to be missing near the $\Gamma$ points as shown in Fig. \ref{fig:dyndmrg}(f) and is pushed to higher energy reflected in the second peak of $S^{xx}(\omega)$ at $h=0.2$ cut (Fig. \ref{fig:dyndmrg}(e,g)), whose momentum-resolved intensity pattern resembles that of the PPM shown in Fig. \ref{fig:dyn}(h). Indeed, the hybridization between fluxes and gapped majoranas inside the PF regime at this energy scale leads to a confined magnonic mode immediately above the lowest-lying anyon excitation as the precursor of the forthcoming transition into PPM, which is made clear by the comparison between Fig. \ref{fig:dyn}(h) and Fig. \ref{fig:dyndmrg}(g). These lowest-lying dynamical signatures together establish the interesting connection between the PF regime, the TC and the PPM phase. The hybridization within the PF regime gives rise to trivial bosonic modes at low energy as the field is increased into the PPM phase. On the other side with lowering of the field, the PF regime smoothly crosses over to the TC regime as the gauge excitations ($\epsilon$) and matter majoranas separate out.

Higher $\omega$ cuts of $S^{xx}(\mathbf{k}, \omega)$ further reveal the relation between PF and the CSL phase, as is indicated by the intense peaks in the spectral function around the M$'$ points shown in Fig. \ref{fig:dyndmrg}(d) and Fig. \ref{fig:dyndmrg}(h).
The third lowest mode in the PF regime, as shown in Fig. \ref{fig:dyndmrg}(e), whose dynamical signal centers around the M$'$ points of the first Brillouin zone, is a signature of itinerant majorana fermions residing near the non-abelian to abelian transition at $K_z/K = 2$. 

We therefore predict that the momentum distribution at specific $\omega$ cuts in the low energy region of the dynamical structure factor reveals sharp signatures of different fractionalized excitations or partons -- deconfined majoranas, gauge anyons and emergent bosonic modes -- in the PF regime that should be observable in inelastic scattering experiments. 
We will discuss the details regarding the nature of excitations and perturbation theory in the forthcoming sections.

\subsection{Perturbation analysis: The dispersion of $\epsilon$ fermions in TC}
In this subsection, we present insights into the aforementioned ED and DMRG results using perturbation theory based on different effective descriptions via the dynamics of TC and of the parton representation of spin. 
For $h=0$, the exactly solvable $Z_2$ QSL phase of the TC is gapped and hence stable against small magnetic fields. While the excitations of this gapped QSL are usually described in terms of the bosonic Ising electric charge $e$ on the vertices, and the magnetic flux $m$ on the plaquettes, of the square lattice in Fig. \ref{fig:fig1}(c) where  $A_s(B_p)=-1$~\cite{kitaev2003fault}, we present an alternative, but equivalent, description of this Ising gauge theory in terms of the fermion, $\epsilon=e\times m$, and the bosonic electric or magnetic charge, that allows for a transparent understanding of the effect of the magnetic field on the TC phase. The description in terms of the $\epsilon$ fermion and boson can also be successfully extended to the near isotropic limit.
\begin{figure*}[t]
    \centering
    \includegraphics[width=\textwidth]{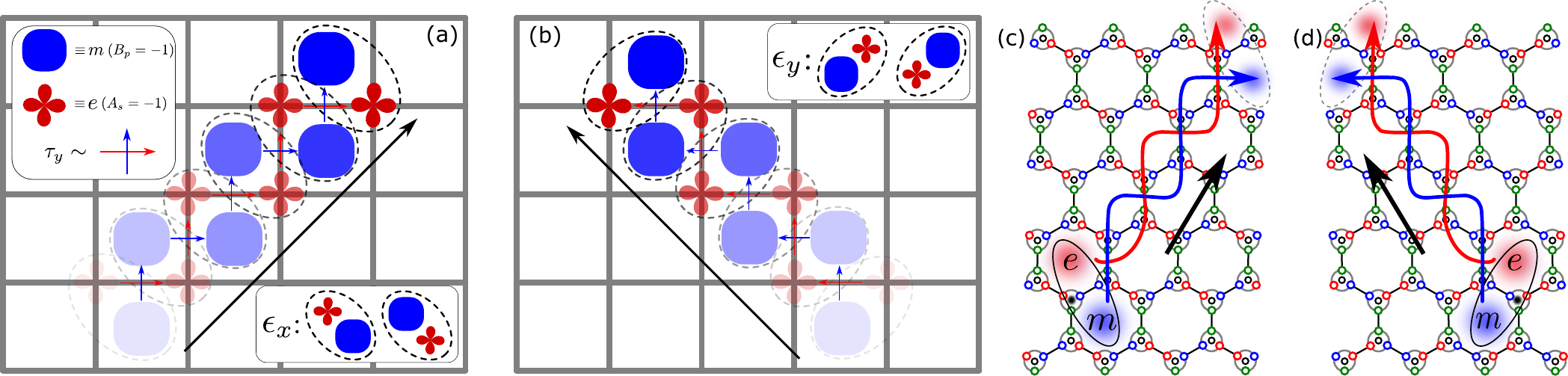}
    \caption{1D dispersion of e-m composite fermions at large $K_z/K$ (TC) limit and small magnetic field along [111] (Zeeman field $\tau_y$ in TC). (a,b) In the square lattice representation of TC, the $\tau_y$ perturbation reduces the gap along the diagonal directions which leads to $\epsilon_x$ and $\epsilon_y$ fermions dispersing along these fixed one-dimensional directions.
    (c) The corresponding directions of $\epsilon_x$ and (d) $\epsilon_y$ fermions in the honeycomb lattice; the $\epsilon$ particle can be excited by local $\sigma_x$ or $\sigma_y$, which respectively give $\epsilon_x$ and $\epsilon_y$ at the energy scale $O(J_{\rm TC})$, with majorana excited simultaneously at high energy sector $O(K_z/K)$.  }
    \label{fig:tcdyn}
\end{figure*}

While all $e,m$ and $\epsilon$ have a gapped flat band for $h=0$, the second term in Eq. \ref{TChpert}, provides, to leading order in $h$, dispersion to $\epsilon$, rather than individual $e$ and $m$ charges, along the diagonal directions $d_1$ or $d_2$ as shown in Fig. \ref{fig:tcdyn}. We label $\epsilon$ fermions that disperse along these two directions $\epsilon_x$ and $\epsilon_y$ respectively. To leading order, the one dimensional dispersions of $\epsilon_x$ and $\epsilon_y$ are given by
\begin{align}
    \varepsilon_x(\mathbf{k}) &= 4J_{\rm TC} - \frac{4h^2}{K_z} \cos(\frac{1}{2} k_x + \frac{\sqrt{3}}{2} k_y) \label{eq:dispersion1}\\
    \varepsilon_y(\mathbf{k}) &= 4J_{\rm TC}  - \frac{4h^2}{K_z} \cos(\frac{1}{2} k_x - \frac{\sqrt{3}}{2} k_y) \label{eq:dispersion2}
\end{align}
where $4J_{\rm TC}$ is the gap to an e-m excitation in absence of the magnetic field. These e-m composite fermions with extremely anisotropic, e.g. the dispersion in Eq. \ref{eq:dispersion1},  have a group velocity $\mathbf{v}(\mathbf{k}) = \partial_\mathbf{k} \varepsilon_x(\mathbf{k}) = \frac{4h^2}{K_z} \mathbf{d}_1 \sin(\mathbf{d}_1\cdot \mathbf{k})$ that propagate only in $\mathbf{d}_1$ direction, as demonstrated in Fig. \ref{fig:tcdyn}(c) and are in agreement with numerical results discussed in Fig. \ref{fig:dyn}(f). They develop zero modes at $ k_x \pm \sqrt{3} k_y = 4 n \pi $ for $4J_{\rm TC} = \frac{4h^2}{K_z}$ if the perturbation theory remains valid. 
Interestingly it is known that given the duality in the system \cite{Vidal_PRB_2009, Dusuel_PRL_2011}, a $y$ perturbation to the TC leads to a first order transition at $\frac{K^4}{16 K^3_z} = \frac{2 h^2}{K_z}$ i.e at $h_c = \frac{K^2}{ \sqrt{32}K_z}$ between a topologically ordered state and a $y$ polarized state. 
The critical value of the first-order transition is of the same order where the gap of the $\epsilon$ fermion reduces, suggesting that these fermions are energetically low lying as the system undergoes a phase transition to the VBS phase. Note that this scale is still within the perturbative regime in $h$ and is far from the $h \sim K_z$ scale where the microscopic spin operators polarize in the [111] direction.

The dispersing modes defined by Eq.~\ref{eq:dispersion1} and Eq.~\ref{eq:dispersion2} can be readily understood in the TC lattice or Wen's plaquette model \cite{You2013}. Indeed, a local $\tau_y$ can be decomposed into $\tau_x \tau_z$ up to a phase factor. It is then clear that a $\tau_z$, when acting on the shared link of a $e$-$m$ composite fermion, annihilates the $e$ charge on one end of the link and creates another on the other end of the link; by the same token $\tau_x$ causes the $m$ charge to hop to the nearest plaquette with which the existing $m$ charge shares a common link.  As shown in Fig. \ref{fig:tcdyn}(a,b), to leading order, a $\tau_y$ induces a fracton-like hopping of an $\epsilon$ particle in a fixed direction, with $e$ and $m$ exchanging their relative positions after each hopping; and depending on whether the initial local excitation is $\epsilon_x$ or $\epsilon_y$ as shown in insets of Fig. \ref{fig:tcdyn}(a,b), the transport directions are rotated by $90$ degrees on the square lattice.  

Such fracton-like hopping of $\epsilon$ anyons as described in Eq. \ref{eq:dispersion1} or Eq. \ref{eq:dispersion2}, depending on the initial local excitation, offers a definitively sharp signature of anyonic fractionalized spins in the abelian QSL with high anisotropy, whereby majoranas are gapped out at the high energy scale $O(K_z/K)$ while anyons dominate the low energy physics at $J_{\rm TC} \sim O(K^4/K_z^3)$. These types of excitations can be created, for example, by the local operator $\sigma_i^x \sigma_{i+z}^x$, i.e. two spin flips on a z bond, which creates a pair of $\epsilon$ anyons with a phase factor from the two majoranas. Upon introducing a [111] magnetic field, the $\epsilon$ excitation begins to hop in a fixed direction defined by the lowest lying soft mode, while majoranas disperse at much higher energy compared to that of the $\epsilon$ excitation, leaving the signature of the anyons unaffected at low energy. We will discuss this dynamics in greater detail in the following sections.   

\subsection{The gauge-matter hybridization in PF regime}
In order to discuss the dynamics of majoranas and anyons, and potential field-induced hybridization at lower anisotropy, here we briefly review the canonical fermion representation of the Kitaev model. This picture is an alternative to the majorana-flux picture up to a gauge transformation, but with better lucidity in formalizing the interplay between the two sectors. 
The integrable limit of the Kitaev QSL (KSL) is most naturally described in terms of fractionalized degrees of freedom ($c_i, b^\alpha_i$) where $\sigma^\alpha_i = i c_i b^\alpha_i$ \cite{kitaev2006anyons} and $c_i, b^\alpha_i$ are majorana fermions that are responsible for the matter majorana and gauge sector respectively. The sublattice character of the honeycomb lattice allows these  majoranas to be combined into canonical fermions
\bea
c_{i} \equiv  f_i + f^\dagger_i,~
c_{i +\hat{z}} \equiv i (f_i-f^\dagger_i) 
\label{majmap}
\eea
where $c_{i}$ denotes the majorana operator on the $A$ sublattice of the $i$-th site of the Bravais lattice; and $c_{i+\hat{z}}$ denotes that of $B$ sublattice thereof. Similarly for the bond fermions $\chi$ \cite{Baskaran2007}
\bea
b^\alpha_{i} = \chi_{i\alpha} + \chi^\dagger_{i\alpha},~
b^\alpha_{i + \hat{\alpha}} = i (\chi_{i\alpha} - \chi^\dagger_{i\alpha}) 
\label{bondmap}
\eea
This maps the Ising exchange on the $z$ bonds as
\beq
K_z \sigma^z_{i} \sigma^z_{i+\hat{z}} = -K_z (2n^f_i-1) (2n^{z}_i-1) 
\label{zbondscale}
\eeq
where $n_i^f$ and $n_i^z$ are the number operators of $f$ and $\chi_{iz}$ fermion. 
Hence it maps the spin states on the sites $i$ and $i+\hat{z}$ (labelled by $|\sigma^{A}_z \sigma^{B}_z \rangle $) to a occupation number basis $|n^f_i, n^z_i \rangle$ on every $z$-bond. The spin configurations  $\ket{\uparrow \downarrow} , \ket{\downarrow \uparrow }$ map to $|00\rangle , |11\rangle$ and  $\ket{\uparrow \uparrow } , \ket{\uparrow \uparrow}$ map to $|10\rangle , |01\rangle$ states. Similarly the other Kitaev exchanges $K_\alpha \sigma^\alpha_{i} \sigma^\alpha_{i+\alpha}$ are transformed to $-K_\alpha c_i c_{i+\alpha} (2n^{\alpha}_i-1)$. The static flux sector, whereby a uniform $n^\alpha_i=1$ is a valid choice of gauge, then leads to a free fermionic description for the $f$ fermions that at $K_z=1$ the model is effectively a $p+ip$ superconductor with Dirac cones at the $\pm \bK$ point \cite{Read2000,kitaev2006anyons}. Increasing $K_z$ displaces the Dirac cones, merging them via a semi-Dirac dispersion at $K_z=2$ (see Appendix. I). Increasing $K_z$ further opens up a gapped phase which is smoothly connected to the TC phase as detailed above. The operator correspondence between various ways of describing the excitations of the Kitaev model is illuminating. A two-spin flip operator $\sim \sigma^x_{i}\sigma^x_{i+\hat{z}}$  on $|\sigma \bar{\sigma} \rangle$ state corresponds to a two fermion excitation 
$\sim f_i^\dagger \chi_{iz}^\dagger + f_i \chi_{iz}$, 
in terms of $f$ and $\chi_z$ fermions, which in turn correspond to four flux excitations (two e-m pairs or a pair of $\epsilon$ particles) in terms of the $\tau^y$ operator acting on the ground state of the TC (see Eq.~\ref{TChpert} and Fig.~\ref{fig:dyn}(a)).


In the strong $K_z$ limit, given the mapping to the TC, it is natural to describe the excitations in terms of just $e$, and $m$  charges or the flux excitations, or their bound state  $\epsilon=e\times m$ which is a fermion. 
However, as $K_z$ is reduced,  we must include the effect of the itinerant majoranas ($f$) which becomes gapless at $K_z/K=2$ and remain gapless for $1< K_z/K <2$. Given that the $\epsilon$ particles and the $f$ fermions belong to the same super-selection sector they can, in general, hybridize via local spin-operators. Hence the strength of the hybridization can be tuned via local interactions in the spin Hamiltonian such as the magnetic field at a relatively low anisotropy. 
It is useful to interpret these in light of the emergent degrees discussed in Eq.~\ref{majmap} and Eq.~\ref{bondmap}. A magnetic field of the form $=\sum_i h^\alpha \sigma^\alpha_i $ leads to hybridization between $\chi_\alpha$ and $f$ fermions, though their behavior at low energies is significantly different. 
The magnetic field introduces a term of the form 
\begin{equation}
    h^\alpha \sigma^\alpha_i \sim
h^\alpha (i b^\alpha_{i} c_{i}) \sim i h^\alpha (\chi_{i\alpha}+ \chi^\dagger_{i\alpha})(f_i + f^\dagger_i)
\end{equation}
where $i$ belongs to A sublattice. This indicates that $h^\alpha$ mediates hybridization between $f$ fermions and $\chi_\alpha$. Including such contributions from both sublattices leads to hybridization,
\beq
ih [ (e^{ik.d_1}-1) \chi_{kx} f_{-k}  - (e^{ik.d_1}+1) \chi_{kx} f^\dagger_{k}  ]+{\rm H.c.}
\label{eq:hybrid}
\eeq
and similarly for $y$ ($x \rightarrow y, d_1 \rightarrow d_2$). For $1<K_z/K\leq 2$, the $f$ fermions are gapless and easily hybridize with the low energy $\chi_\alpha$ fermions via the above mechanism in the presence of the magnetic field. 

Such hybridization is negligibl in the TC limit since, contrary to the case in $1<K_z/K\leq 2$, the gap of the $f$ fermion in the limit $K_z \gg 2$ is large compared to that of the bond fermions. However, 
upon reducing the bond anisotropy from the TC limit, the gap of the $f$ fermions closes at the M$'$ point in the BZ and reopens at finite magnetic field through a change in the Chern number of the band. Under the static flux approximation \cite{kitaev2006anyons}, even at a finite field, the transition between the CSL and the abelian QSL occurs via a Dirac closing at the M$'$ point. In particular near M$'=\{0,\frac{2\pi}{3}\}$ where $f$ fermions are low-lying near $K_z/K \sim 2$, the hybridization leads to 
\beq \label{eq:hyb}
-2h i(\chi_{kx} f_{-k} - f^\dagger_{-k} \chi^\dagger_{kx}) + (x\leftrightarrow y)
\eeq
which implies instead of considering $\chi_{x/y}$ and $f$ fermions separately (as near $h\sim 0$), we should instead consider 
\beq
\psi \sim \frac{1}{\sqrt{2}} \big( (\chi^\dagger_{kx} + \chi^\dagger_{ky}) + i f_{-k} \big)
\label{hybridi2}
\eeq
as the low energy excitations of the $Z_2$ liquid for intermediate fields and $K_z \sim 2$. It is this regime of the phase diagram we dub the PF regime. Thus the PF regime we find is a generic $Z_2$ liquid, the primordial fractionalized $Z_2$ QSL, from which specific cases of $Z_2$ spin liquids as realized in the Kitaev model arise.
This qualitatively explains the fan-like shape of the PF regime which emanates from $K_z/K\sim2$ shown in Fig. \ref{fig:fig1}(b) and Fig. \ref{fig:static_all}(g). 

For large $K_z$, the single particle excitations for both $\chi_z$ and $f$ fermions are gapped at the $\sim K_z$ scale, which can be integrated out, consistent with Eq. \ref{zbondscale}. The low energy excitations are the $\chi_x$ and $\chi_y$ fermions that gain an independent dispersion due to the quadratic perturbation in $(h^x)^2$ and $(h^y)^2$. In fact a quadratic perturbation in $(h^x)$ gives $(h^x)^2 \sigma^x_i \sigma^x_{i+d_1}$
\beq
 \sim (h^x)^2 (2 n^f_i-1) (\chi^x_i \chi^x_{i+d_1} + \chi^{x\dagger}_i \chi^x_{i+d_1}  + \chi^{x\dagger}_{i+d_1}\chi^{x}_i +\chi^{x\dagger}_{i+d_1} \chi^{x\dagger}_i   )
 \label{xxpart}
\eeq
When acting on a single-fermion sector, it gives rise to the dispersion for the bond fermions albeit normalized by the occupancy of $f$ fermions within mean-field. The dispersion of $\chi_x$ fermions is of the form
\begin{equation} \label{eq:chix}
    \varepsilon_{\chi_x}(\mathbf{k}) \sim 2 (h^x)^2 \cos(\mathbf{k}\cdot \mathbf{d}_1)
\end{equation}
which is indeed the dispersion shown in Eq. \ref{eq:dispersion1}. An equivalent picture emerges when $x \rightarrow y$, $d_1 \rightarrow d_2$. Therefore the composite (e-m) pairs distilled in the large $K_z$ limit and the TC mapping in the previous subsection are in fact describable by the $\chi_x$ and $\chi_y$ fermions as discussed in Eq.~\ref{bondmap}. These excitations are energetically low lying close to the VBS phase at large $K_z$ when perturbed by $h$. Note that in this regime, $f$ fermions behave as spectators and do not mix with $\chi_\alpha$ fermions, and therefore do not affect the low energy physics except when confinement leads to the VBS phase.


\begin{table*}[t]
\caption{The summary of phases, the nature of low energy excitations, and their associated energy scales. The $\sim h^3/K^2$ scaling of $f$ fermions in $Z_2$ CSL is valid only in the perturbative regime where fluxes remain conserved \cite{kitaev2006anyons,Jiang_arXiv_2018}; this picture breaks down at larger $h$ whereby $f$ and $\chi$ hybridize. The central PF phase marked with an asterisk is connected to the TC QSL by a crossover. }
\begin{tabular}{|c|c|} 
\hline
Phase  &   Hierarchy of low energy excitations \\ [0.5ex] 
\hline
Isotropic KQSL & Gapless $c$ majoranas (or $f$ fermions), gapped flux excitations  \\ \hline
$Z_2$ CSL & $f$ fermions at energy $\sim h^3/K^2$; hybridized particle of $f$ and $\chi$ \\ \hline
$Z_2$ abelian QSL & Highly gapped $f$ fermions; gapped fluxes [1,e,m,$\epsilon$],  only $\epsilon$ disperses under [111] field \\ \hline 
VBS & one gapped ($\ket{\uparrow\downarrow} - \ket{\downarrow\uparrow}$) at energy scale $\sim h^2/K_z$; two gapped ($\ket{\uparrow\uparrow}, \ket{\downarrow\downarrow}$) at $\sim K_z/K$ \\ \hline
Gapless QSL & Neutral Fermi surface of $f$ fermions \\ \hline
Polarized & spin waves at $\sim h/K$\\ \hline
PF$^*$ & damped $\epsilon$ fermions, gapped $\psi$ fermions, and magnons at energy $\sim K^4/K_z^3$ \\ \hline
\end{tabular}
\label{table_phases}
\end{table*}
\subsection{Two-spin-flip dynamics in terms of partons}
Equipped with insights from perturbation theory in terms of partons, we can now understand the numerical results shown in Figs. \ref{fig:dyn} and \ref{fig:dyndmrg}. 
The excitation created by the operator $\hat{\mathcal{O}}_i^x = \sigma_i^x \sigma_{i+z}^x$ near $h=0$ is equivalent to the composite particle consisting of both occupancy of canonical $f$ fermion and a creation of a pair of $\epsilon$ particles (or equivalently $\chi_x$ of Eq. \ref{xxpart}). Further, the excitation created by $\hat{\mathcal{O}}_i^z = \sigma_i^z \sigma_{i+z}^z$ is equivalent to a product of number operators for both $f$ fermion and $\chi_z$ fermion (see Eq.~\ref{zbondscale}).
It is important to note that while the purity of these operators is true in the exactly-solvable limit ($h=0$), in intermediate fields these particles hybridize as discussed near Eq.~\ref{eq:hybrid}.

The large spectral peak on the order of $J_{\text{TC}}$ previously shown in Fig. \ref{fig:dyn}(b,c) 
at small $h$ can now be understood as arising from the composite object consisting of the $f$ fermion and two $\chi_x$ particles that disperse at low energy, even though a single $f$ excitation is gapped out on the order of $O(K_z/K)$.  Besides this $f-\chi_x$ composite object there is another signal, which is lowest in energy in Fig.~\ref{fig:dyn}(b) at around $\omega = 1.6\times 10^{-2}$ under low field, whose intensity grows as field increases. This signal rises due to the dispersion of $\epsilon$ anyons described previously in Eq.~\ref{eq:dispersion1} (or equivalently that of $\chi_x$ in Eq.~\ref{eq:chix}), which obviously agrees with the quadratically decreasing energy scale with the increasing $h$, as marked out by an eye-guiding dashed line in Fig.~\ref{fig:dyn}(b). Indeed, the $k$-resolved dynamics at the energy cut reveals the 1D dispersion of $\epsilon$ (see schematic in \Fig{fig:tcdyn}).
However, as is shown also in Fig.~\ref{fig:dyn}(b), for higher $h$ as the PF regime is approached, the highest peak relevant for $f-\chi_x$ composite in $S^{xx}(\omega)$ of $\sigma_{j}^x \sigma_{j+z}^x$ splits into several smaller peaks, that is, the composite particles consisting of the bound state of $f$ fermions and two $\chi_x$ further fractionalizes into its constituent parts, separating out the dispersion of $f$ fermion from the amalgam, as indicated in the next lowest spectral mode that branches out linearly with the increment of $h$ shown in Fig.~\ref{fig:dyn}(b). Similarly we find at the spectral mode of $\epsilon$ also branches into two pieces upon entering PF: while the lowest spectral mode is still dominated by $\epsilon$, the outer branch is primarily due to the hybrid mode $\psi$, whose energy scales linearly with $h$ and is consistent with Eq.~\ref{eq:hyb}.  
The spectral weight of $S^{zz}(\omega)$ at $h \gtrsim  0.16$ in Fig. \ref{fig:dyn}(c) is indicative of the $\psi$ fermion formed from the hybridization of $f-\epsilon$ by the [111] magnetic field. 
Furthermore, the energy scale of hybridized modes shown in Fig.~\ref{fig:dyn}(c) changes linearly with respect to $h$ in agreement with Eq.~\ref{eq:hyb}, which is in sharp contrast to the quadratic scaling of $\epsilon$ particle.    


The relation between PF and the CSL phase 
is revealed by higher $\omega$ cuts of $S^{xx}(\mathbf{k}, \omega)$, as is indicated by the intensity peaks around M$'$ points shown in Fig. \ref{fig:dyndmrg}(d,h). These modes are precisely dominated by the $\psi$ particle defined in Eq. \ref{hybridi2}, which would become the lowest-lying excitation dominated by majoranas in CSL near transition.  To understand the essence of this mode, recall that itinerant majorana fermions (or $f$ fermion) reside near the M$'$ points near the non-abelian to abelian transition at $K_z/K = 2$, which is given by the static flux sector calculation shown in Fig. \ref{fig:dispersion} of Appendix. I. Hence at perturbative field the majorana particle is responsible for signals at M$'$ points. However, as magnetic field increases, majoranas and fluxes (or bond fermions $\chi$) begin to hybridize, giving rise to the $\psi$ mode near M$'$ points whereby the intensity of $f$ is modified by $\chi$ -- as is shown in Fig. \ref{fig:dyndmrg}(d,h) where the spots near M$'$ points are distorted and stretched along the $d_1$ direction due to the 1D dispersion of $\chi_{x/y}$. Interestingly, this mode is also the lowest excitation in the CSL phase at a non-zero field near TC-CSL transition, which is already present in the PF regime of the abelian phase. 
The nature of excitations in all the phases is summarized in Table. \ref{table_phases}.



Besides the sharp flux signature reviewed in the dynamical structure factors, the excitations in the PF regime also provides the key to understanding the rich phase diagram in Fig.~\ref{fig:fig1}: the Zeeman field induced hybridization between the $\epsilon$ and majorana fermions resulting in a $\psi$ fermion. All the other phases in the $(h,K_z)$ plane are naturally obtained from this primodial soup:
the gapped PF liquid is continuously connected to the TC limit by a cross-over where the later is governed by $\epsilon$ fluxes; through continuous phase transitions to the CSL via a change in the topological invariant of the band structure of the $\psi$ fermions, which reduces to the usual majoranas in the weak field limit (A Chern number transition always occurs at M$'$ point, as is consistent with Fig. \ref{fig:dyndmrg}(d,h));
 to a gapless U(1) QSL with a Fermi surface via loss of $\langle\psi\psi\rangle$ pairing; and to the VBS phase with a dimer order parameter via confinement.

\section{Discussion and conclusion} \label{sec:discussion}

The possibility of observing signatures of majoranas in Kitaev QSL has been discussed in the context of spin dynamics and thermal Hall effects. It has been argued that the dynamical spin susceptibility \cite{Knolle2014,Song2016,Knolle2018}, which exploits the orthogonality between flux excitations, would allow the extraction of the two-point correlation between itinerant majoranas. Also, the non-zero Chern number of majorana bands in presence of gap-opening perturbations suggests non-trivial thermal Hall conductivity. As potential QSL candidates, the most studied is $\alpha$-RuCl$_3$~\cite{Khaliullin2009,PhysRevB.93.134423,ziatdinov2016atomic,Banerjee2017,Ponomaryov2017,Little2017,Hentrich2018,Kasahara2018,banerjee2018excitations,Banerjee2016,Holleis2021,Yokoi2021} in which some recent experiments appear to see a half-quantized Hall plateau of thermal edge conduction indicative of chiral gapless edge modes of majoranas \cite{Kasahara2018}. While the experimental situation is currently unclear~\cite{czajka2021oscillations}, given various material growth issues and the presence of non-Kitaev spin interactions, it raises another theoretical aspect less studied in the field:  the physics of fluxes and their interplay with majoranas. 

In this work we have probed this rich physics with an experimentally relevant perturbation-- an external magnetic field-- for various parameter regimes of the Kitaev model to reveal the structure of the underlying fractionalized excitations in terms of their response to the magnetic field. We have focused on the resultant phases and the dynamics of the excitations generated from the interplay of bond anisotropy and the magnetic field.

The central result reported in this paper is that the effect of an external magnetic field in the $[111]$ direction is remarkably different on the gapless and the gapped $Z_2$ QSL. While the former shows transitions to a gapped non-abelian chiral QSL, followed by a further transition to a gapless U(1) QSL, before entering the polarized phase, the latter gives way to $\epsilon$ fluxes dispersing in fixed one dimensional directions before transitioning to a valence bond solid phase.
The significance of our results is that typically one expects in a QSL phase, the energy and momentum imparted to the system in an inelastic scattering experiment will be shared among the fractionalized components, leading to broad features in spectroscopy. However, we find quite remarkably, that there are indeed sharp signatures of the different anyon fractons in the linear response of appropriate local spin flip operators that are further tunable by a magnetic field. In addition to linear spectroscopy, time-resolved non-linear pump-probe experiments that can explore the dynamics of two anyons created at different times braiding around each other~\cite{Parameswaran2022a,Parameswaran2022} could be a powerful probe of the fractionalized excitations and ultimately provide the smoking-gun signatures of anyons in a QSL. 
In particular, it is shown in \cite{Parameswaran2022} that in a 2D system with non-trivial braiding statistics, the non-linear response is divergent in time according to $\sim t^{1/2}$. Given the unique 1D dispersion of $\epsilon$ quasiparticles induced by the [111] field, the scaling in time or frequency of non-linear response functions can be distinct from general 2D fermionic systems. We also expect that the non-linear response signal in the abelian QSL can be tuned by tilting the magnetic field.

We have also reported our discovery of a gapped primordial $Z_2$ fractionalized (PF) phase, the generalized abelian $Z_2$ phase with coupled matter and gauge degrees of freedom at intermediate bond anisotropy and magnetic field in the center of the phase diagram in the anisotropy-field plane. Key to this phase is the twin role of the magnetic field that-- (1) provides dispersion to the $Z_2$ fluxes which in turn selectively provides dispersion to the $\epsilon=e\times m$ fermions in the anisotropic limit, and, (2) provides hybridization between the $\epsilon$ and the majorana fermions-- to produce new hybridized fermions whose properties naturally explain the PF phase.
The significance of this finding is that all the phases surrounding this central region: the gapless Kitaev spin liquid, the gapped abelian QSL, the TC phase, the gapped chiral non-abelian QSL, the gapless U(1) QSL, the dimer or valence bond phase, and the polarized phase, emerge from the primordial fractionalized phase. We have therefore identified the essential coupled matter and $Z_2$ gauge degrees of freedom in the PF regime that produce the surrounding gapped phases with topological order, gapless phases with spinon Fermi surfaces, and first order transition driven by $\epsilon$ to the VBS order. The most direct information on the nature of the PF regime has come from the dynamics and their dispersion in the Brillouin zone of different combinations of spin operators that create particular fractionalized excitations. By observing the peaks of the structure factor corresponding to these spin operators as a function of the magnetic field and anisotropy, we have been able to track their evolution across phase transitions. 
Since the manipulation of an anisotropy in the exchange coupling was recently proposed in the realistic materials by means of the light irradiation \cite{Arakawa2021}, and that the toric code (TC) topological phase was recently realized in cold atom setup \cite{Semeghini2021}, we expect our results can inspire relevant experiments on Kitaev QSLs in both quantum materials and cold atom platforms.

\section*{Appendix I: Non-interacting majoranas at the Low Field Limit}
\begin{figure}[t]
    \centering
    \includegraphics[width=\linewidth]{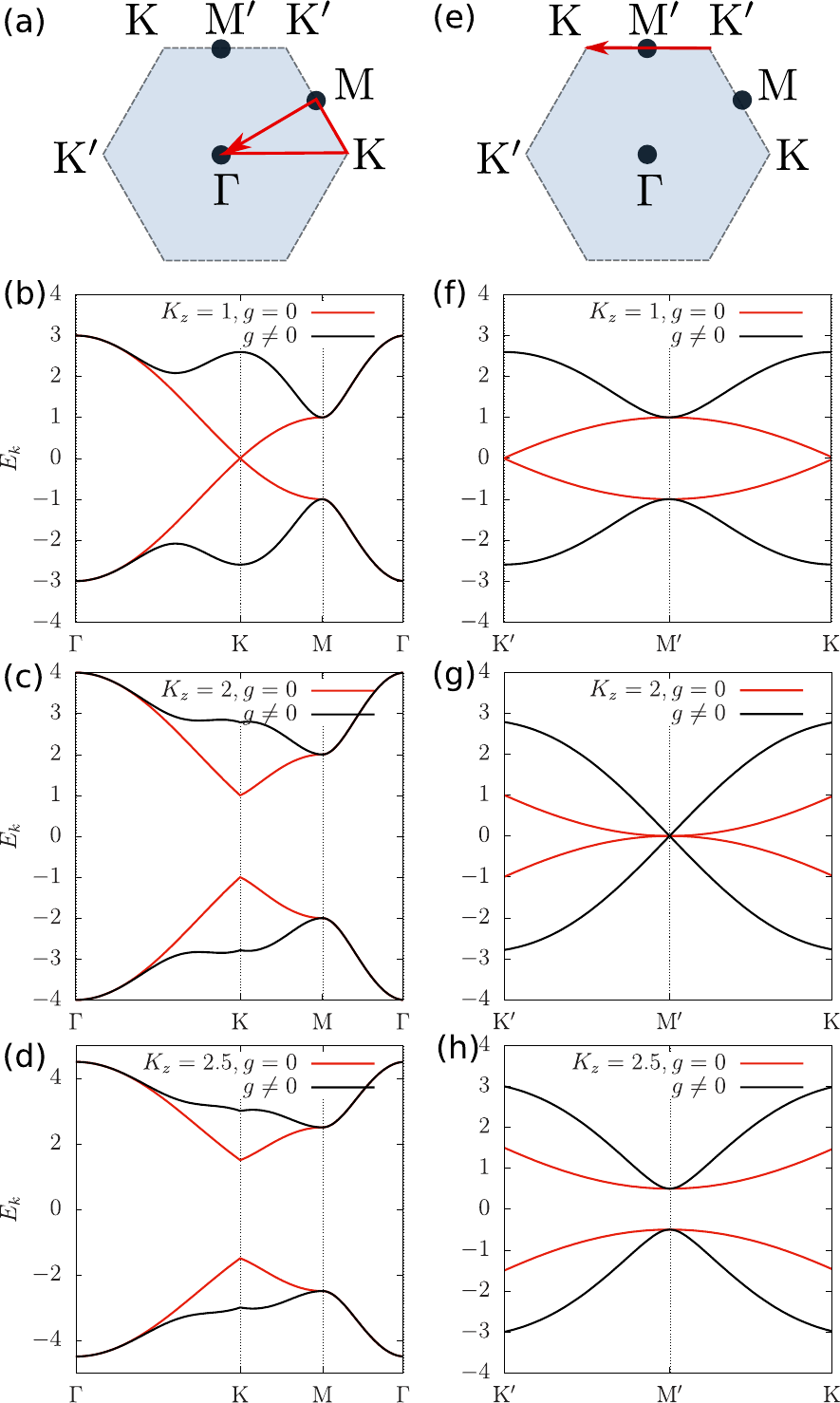}
    \caption{Dispersion for the itinerant majoranas in the static flux sector near $K_z/K=1, 2, 2.5$ in presence and absence of field dependent coupling $g \sim h^3/(K K_z) = 0.5$, showing the low energy itinerant fermions near the  K and M$'$ points. Panel (b-d) shows the dispersion through the Brillouin zone cut in (a); panel (f-h) shows that through the cut in (e).  }
    \label{fig:dispersion}
\end{figure}
Although the focus of this paper is the flux dynamics, it is necessary to explain the dynamics of majoranas in the low field limit which provide the anchor point for understanding the dynamical signatures of hybridized excitations of fluxes and majoranas at higher magnetic field.  
In this appendix we explain the low energy dispersion of majoranas (or $f$ fermions) at integrable limit and in the regime of low field perturbation theory where fluxes are static ($Z_2$ fluxes remain conserved), as is shown schematically in the inset bands of Fig.~\ref{fig:fig1}(b) in the main text. At this limit the majorana and gauge sector remain separated from each other.
The model is integrable in absence of magnetic field, where the low energy Hamiltonian is governed by the matter majoranas in the zero flux sector. The Kitaev Hamiltonian then becomes quadratic in majorana operators \cite{kitaev2006anyons} $\mathcal{H}_K = \sum_{i\in A,\delta} i c_{i,A} c_{i+ \delta,B}$, with $\delta = \hat{x}, \hat{y}$ or $\hat{z}$. In momentum space the Kitaev Hamiltonian reads as
\begin{equation}
    \mathcal{H}_K = \frac{1}{2} \sum_{\mathbf{k}} \left[ it(\mathbf{k}; K_x,K_y,K_z) c_{\mathbf{k},A} c_{-\mathbf{k},B} + {\rm H.c.} \right]
\end{equation}
where for convenience we set $K_x = K_y \equiv K = 1$, and the $t(\mathbf{k})$ above is given by
\begin{equation} \label{eq:tk}
    t(\mathbf{k};K_z) = K_z + 2\exp(i\frac{3}{2}k_y)\cos(\frac{\sqrt{3}}{2}k_x)
\end{equation}
At isotropic point ($K_x = K_y = K_z = 1$) and without magnetic field, Eq. \ref{eq:tk} gives two Dirac mode at K and K$'$ points as shown Fig. \ref{fig:dispersion}(b,f) and also illustrated in the Fig. \ref{fig:fig1}(b) of the main text. However, for a generic anisotropic $K_z/K$, Eq. \ref{eq:tk} has gapless modes at
\begin{equation}
    k_y = 0,~ k_x = \pm \arccos(-\frac{K_z}{2})\frac{2}{\sqrt{3}}
\end{equation}
hence the gapless modes defined by Eq. \ref{eq:tk} shift from K and K$'$ points and move towards each other until they meet and merge at the gapless M$'$ points when $K_z/K = 2$ (Fig. \ref{fig:dispersion}(c,g)), beyond which there is no gapless solution and majoranas become gapped; and the previously gapless momenta K, K$'$ become gapped as is shown in Fig. \ref{fig:dispersion}(d,h). At $h=0$ and $K_z/K = 2$, 
the soft mode expansion for $f$ fermions near M$'$ gives the effective Hamiltonian
\begin{equation}
    \mathcal{H}_{\text{M}'} = -\frac{1}{2}\sum_{\mathbf{k}} (f_\mathbf{k}^\dagger
    ~ f_{-\mathbf{k}})
    \begin{pmatrix}
        \frac{3}{2}k_x^2 & -3ik_y \\
        3ik_y & -\frac{3}{2}k_x^2
    \end{pmatrix}
    \begin{pmatrix}
        f_\mathbf{k} \\ f_{-\mathbf{k}}^\dagger
    \end{pmatrix}
\end{equation}
where all momenta are measured with respect to M$'$. 
Hence the majorana excitations form semi-Dirac instead of Dirac cones at M$'$ point shown in Fig. \ref{fig:dispersion}(b).
Interestingly, in presence of a weak [111] magnetic field whereby the third order perturbation breaks TR while keeps the integrability \cite{kitaev2006anyons}, 
the $\mathcal{H}_{\text{M}'}$ becomes
\begin{equation}
    \tilde{\mathcal{H}}_{\text{M}'} = \mathcal{H}_{\text{M}'}  + 
    4\sqrt{3}\sum_{\mathbf{k}} (f_\mathbf{k}^\dagger
    ~ f_{-\mathbf{k}})
    g k_x \sigma^x
    \begin{pmatrix}
        f_\mathbf{k} \\ f_{-\mathbf{k}}^\dagger
    \end{pmatrix}
\end{equation}
where $g = \frac{h^3}{KK_z}$; hence the semi-Dirac cone becomes a linear Dirac cone at low energy near the M$'$ point as presented by the black solid line in Fig. \ref{fig:dispersion}(g) and also schematically in Fig. \ref{fig:fig1}(b). Such low energy excitation at the M$'$ point near $K_z/K = 2$ transition qualitatively agrees with the low lying excitation in the strongly hybridized regime near transition, where the dynamical signals are centered at the M$'$ points while distorted and stretched along the $d_1$ direction due to the 1D dispersion of $\epsilon$ fermions.


\section*{Appendix II: Computational Details} \label{sec:dmrg}
\begin{figure}[t]
    \centering
    \includegraphics[width=0.48\textwidth]{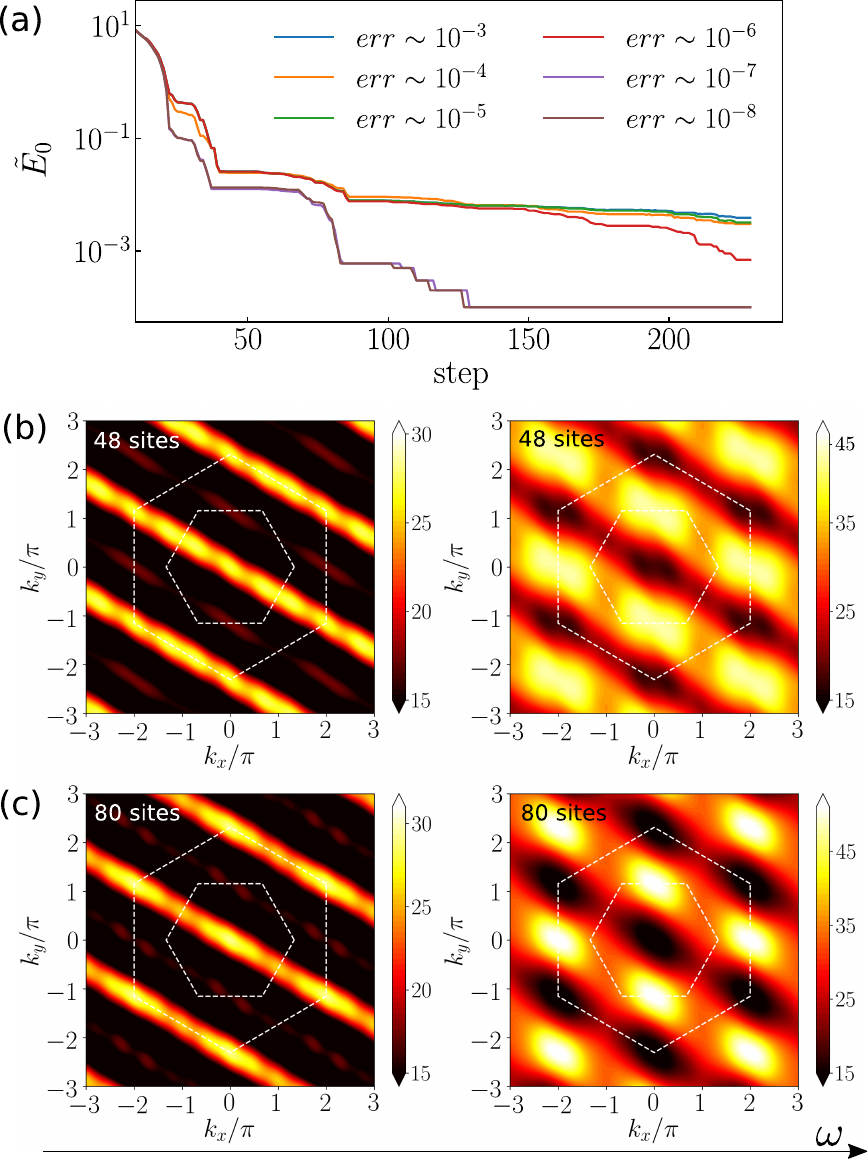}
    \caption{(a) Convergence of ground state energy $\tilde{E}_0$ (shifted for log scale) in 96-site DMRG with cylinder geometry. $\tilde{E}_0$ converges within 150 steps for truncation tolerance that is smaller than $err \sim 10^{-7}$. (b,c) Two energy cuts of $S^{xx}(\mathbf{k}, \omega)$ at $\omega = 0.0054$ and $\omega = 0.0072$ obtained by (b) 48-site DMRG and (c) that by 80-site DMRG. }
    \label{fig:Econverge}
\end{figure}
In this appendix we present details of DMRG and ED used to calculate the dynamics of anyons in the main text. 
\begin{figure}[h]
    \centering
    \includegraphics[width=0.45\textwidth]{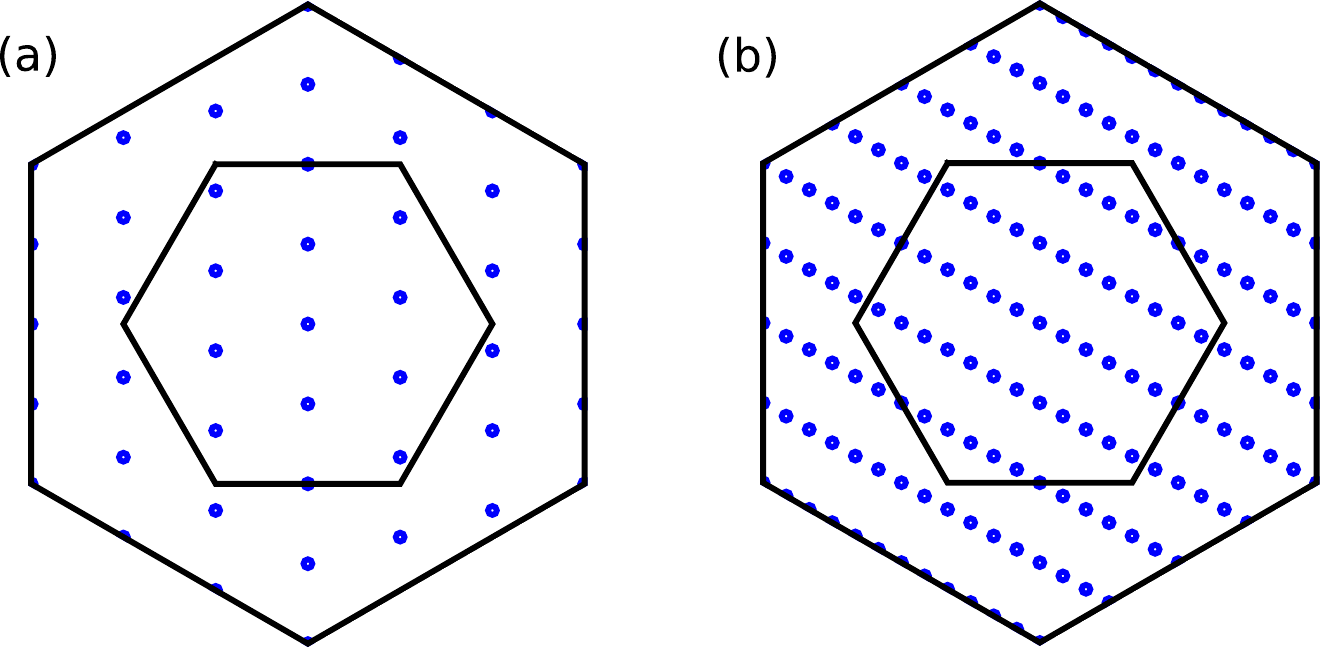}
    \caption{The momentum space resolution of (a) the  $24$-site lattice with $3\times 4$ unit cells used in ED; and (b) the $96$-site lattice with $12\times 4$ unit cells used in DMRG. Blue dots denote the available momenta in the first and second Brillouin zones of the corresponding clusters.}
    \label{fig:kres}
\end{figure}
The dynamical structure factor $S(\mathbf{k},\omega)$ as a function of 
frequency $\omega$ and momentum $q$ can be
measured with relevant inelastic scattering experiments. $S(\mathbf{k},\omega)$ is defined as usual
\begin{equation} \label{eq:sqw}
        S^{\alpha\beta}(\mathbf{k},\omega) = \sum_{r} e^{-i\mathbf{k}\cdot \mathbf{r}} \int_{-\infty}^{\infty} dt \langle \hat{\mathcal{O}}_c^\alpha (t) \hat{\mathcal{O}}_{c+\mathbf{r}}^{\beta}(0)\rangle e^{i\omega t}
\end{equation} 
where $\hat{\mathcal{O}}$ corresponds to two-spin flip operators defined on the Bravais lattice as described in the main text. The density of states can be derived accordingly by $S(\omega) = \int d\mathbf{k} ~S(\mathbf{k}, \omega)$. The numerical results are obtained by both ED and DMRG, and interpolated according to the momentum space resolution shown in Fig. \ref{fig:kres}. 
To evalute Eq. \ref{eq:sqw} under cylinderical geometry by DMRG, we take the central site of reference $c$, and compute the dynamical structure factor by its analytic continuation which is given by the real space function:
\begin{equation} \label{eq:Sdmrg}
S^{\alpha\beta}(r, c, \omega) \sim \langle{\rm g.s.}|\hat{\mathcal{O}}^\alpha_r \frac{1}{\omega + i\eta + H - E_0 }\hat{\mathcal{O}}^\beta_c|{\rm g.s.}\rangle ,
\end{equation}
with respect to all sites at $r$ and $c$, where $|{\rm g.s.}\rangle $ is the ground state of the
Hamiltonian $H$ in the abelian phase of the Kitaev model, with or without magnetic
field, $E_0$ the corresponding ground state energy, and  $\eta$ a small broadening factor to ensure the convergence of the Green's function.
From the Fourier transform we obtain $S(q, \omega )$ and by integrating over all momenta, the density of states $S(\omega) \sim S(c, c, \omega )$. Note that we are interested in the dynamical signatures of gapped excitations, thus the operator should be normal ordered to remove the ground state contribution. We therefore replace the operator $\hat{\mathcal{O}}$ in Eq. \ref{eq:Sdmrg} by the fluctuation operators $\delta \hat{\mathcal{O}} = \hat{\mathcal{O}} - \expval*{\hat{\mathcal{O}}}_0$ where $\expval*{\hat{\mathcal{O}}}_0$ denotes the ground state expectation. This is equivalent to the Lehmann representation of dynamical structure factor excluding the ground state contribution: $S(r,c, \omega) - \expval*{\hat{\mathcal{O}}_r}_0 \expval*{\hat{\mathcal{O}}_c}_0\delta(\omega)$, as is implemented in our computation, 
where we use the Lorentzian function with broadning factor $\eta=1.2\times 10^{-3}$ for the Dirac delta, and scan the frequencies in increments of $\Delta \omega = 4\times 10^{-4}$ in units of exchange energy $K$.
We use the Krylov-space approach of dynamical DMRG which is described and implemented in Ref. \cite{Alvarez2016,alvarez08} (See also Ref. \cite{Feng2022gapless} and supplemental materials thereof). We here also provides evidence of convergence with the number of states $m$ kept within DMRG, and shows when finite size effects in the dynamical structure factor can be neglected. 
As shown in Fig. \ref{fig:Econverge}(a), the largest truncation error $\sim 10^{-8}$ that requires a maximum number of $m \sim 700$ kept states; and in Fig. \ref{fig:Econverge}(b,c) where the dynamics of anyons remains robust with respect to the increasing size of clusters.

\section{Acknowledgment}
We thank Niravkumar Patel and Cullen Gantenberg for numerical data during an early stage of the project. We thank Animesh Nanda and Yuan-Ming Lu for discussions and comments. S.F. acknowledges support from NSF Materials Research Science and Engineering Center (MRSEC) Grant No. DMR-2011876, N.T. from NSF-DMR 2138905. A.A. acknowledges support from IIT Kanpur Initiation Grant (IITK/PHY/2022010) and thanks ICTS, Bangalore for hospitality. S. B. acknowledges funding from Max Planck Partner group Grant at ICTS, Swarna jayanti fellowship grant of SERB-DST (India) Grant No. SB/SJF/2021-22/12 and the Department of Atomic Energy, Government of India, under Project No. RTI4001.

\bibliography{biblio.bib}

\begin{thebibliography}{80}%
\makeatletter
\providecommand \@ifxundefined [1]{%
 \@ifx{#1\undefined}
}%
\providecommand \@ifnum [1]{%
 \ifnum #1\expandafter \@firstoftwo
 \else \expandafter \@secondoftwo
 \fi
}%
\providecommand \@ifx [1]{%
 \ifx #1\expandafter \@firstoftwo
 \else \expandafter \@secondoftwo
 \fi
}%
\providecommand \natexlab [1]{#1}%
\providecommand \enquote  [1]{``#1''}%
\providecommand \bibnamefont  [1]{#1}%
\providecommand \bibfnamefont [1]{#1}%
\providecommand \citenamefont [1]{#1}%
\providecommand \href@noop [0]{\@secondoftwo}%
\providecommand \href [0]{\begingroup \@sanitize@url \@href}%
\providecommand \@href[1]{\@@startlink{#1}\@@href}%
\providecommand \@@href[1]{\endgroup#1\@@endlink}%
\providecommand \@sanitize@url [0]{\catcode `\\12\catcode `\$12\catcode
  `\&12\catcode `\#12\catcode `\^12\catcode `\_12\catcode `\%12\relax}%
\providecommand \@@startlink[1]{}%
\providecommand \@@endlink[0]{}%
\providecommand \url  [0]{\begingroup\@sanitize@url \@url }%
\providecommand \@url [1]{\endgroup\@href {#1}{\urlprefix }}%
\providecommand \urlprefix  [0]{URL }%
\providecommand \Eprint [0]{\href }%
\providecommand \doibase [0]{http://dx.doi.org/}%
\providecommand \selectlanguage [0]{\@gobble}%
\providecommand \bibinfo  [0]{\@secondoftwo}%
\providecommand \bibfield  [0]{\@secondoftwo}%
\providecommand \translation [1]{[#1]}%
\providecommand \BibitemOpen [0]{}%
\providecommand \bibitemStop [0]{}%
\providecommand \bibitemNoStop [0]{.\EOS\space}%
\providecommand \EOS [0]{\spacefactor3000\relax}%
\providecommand \BibitemShut  [1]{\csname bibitem#1\endcsname}%
\let\auto@bib@innerbib\@empty
\bibitem [{\citenamefont {Mott}(1949)}]{Mott_1949}%
  \BibitemOpen
  \bibfield  {author} {\bibinfo {author} {\bibfnamefont {N~F}\ \bibnamefont
  {Mott}},\ }\bibfield  {title} {\enquote {\bibinfo {title} {The basis of the
  electron theory of metals, with special reference to the transition
  metals},}\ }\href {\doibase 10.1088/0370-1298/62/7/303} {\bibfield  {journal}
  {\bibinfo  {journal} {Proceedings of the Physical Society. Section A}\
  }\textbf {\bibinfo {volume} {62}},\ \bibinfo {pages} {416} (\bibinfo {year}
  {1949})}\BibitemShut {NoStop}%
\bibitem [{\citenamefont {Kimchi}\ \emph {et~al.}(2013)\citenamefont {Kimchi},
  \citenamefont {Parameswaran}, \citenamefont {Turner}, \citenamefont {Wang},\
  and\ \citenamefont {Vishwanath}}]{Vishwanath}%
  \BibitemOpen
  \bibfield  {author} {\bibinfo {author} {\bibfnamefont {Itamar}\ \bibnamefont
  {Kimchi}}, \bibinfo {author} {\bibfnamefont {S.~A.}\ \bibnamefont
  {Parameswaran}}, \bibinfo {author} {\bibfnamefont {Ari~M.}\ \bibnamefont
  {Turner}}, \bibinfo {author} {\bibfnamefont {Fa}~\bibnamefont {Wang}}, \ and\
  \bibinfo {author} {\bibfnamefont {Ashvin}\ \bibnamefont {Vishwanath}},\
  }\bibfield  {title} {\enquote {\bibinfo {title} {Featureless and
  nonfractionalized mott insulators on the honeycomb lattice at 1/2 site
  filling},}\ }\href {\doibase 10.1073/pnas.1307245110} {\bibfield  {journal}
  {\bibinfo  {journal} {Proceedings of the National Academy of Sciences}\
  }\textbf {\bibinfo {volume} {110}},\ \bibinfo {pages} {16378--16383}
  (\bibinfo {year} {2013})}\BibitemShut {NoStop}%
\bibitem [{\citenamefont {Wen}\ and\ \citenamefont {Niu}(1990)}]{Wen1990}%
  \BibitemOpen
  \bibfield  {author} {\bibinfo {author} {\bibfnamefont {X.~G.}\ \bibnamefont
  {Wen}}\ and\ \bibinfo {author} {\bibfnamefont {Q.}~\bibnamefont {Niu}},\
  }\bibfield  {title} {\enquote {\bibinfo {title} {Ground-state degeneracy of
  the fractional quantum hall states in the presence of a random potential and
  on high-genus riemann surfaces},}\ }\href {\doibase 10.1103/PhysRevB.41.9377}
  {\bibfield  {journal} {\bibinfo  {journal} {Phys. Rev. B}\ }\textbf {\bibinfo
  {volume} {41}},\ \bibinfo {pages} {9377--9396} (\bibinfo {year}
  {1990})}\BibitemShut {NoStop}%
\bibitem [{\citenamefont {Wen}(2002)}]{wen2002quantum}%
  \BibitemOpen
  \bibfield  {author} {\bibinfo {author} {\bibfnamefont {Xiao-Gang}\
  \bibnamefont {Wen}},\ }\bibfield  {title} {\enquote {\bibinfo {title}
  {Quantum orders and symmetric spin liquids},}\ }\href {\doibase
  10.1103/PhysRevB.65.165113} {\bibfield  {journal} {\bibinfo  {journal} {Phys.
  Rev. B}\ }\textbf {\bibinfo {volume} {65}},\ \bibinfo {pages} {165113}
  (\bibinfo {year} {2002})}\BibitemShut {NoStop}%
\bibitem [{\citenamefont {Chen}\ \emph {et~al.}(2010)\citenamefont {Chen},
  \citenamefont {Gu},\ and\ \citenamefont {Wen}}]{Wen2010}%
  \BibitemOpen
  \bibfield  {author} {\bibinfo {author} {\bibfnamefont {Xie}\ \bibnamefont
  {Chen}}, \bibinfo {author} {\bibfnamefont {Zheng-Cheng}\ \bibnamefont {Gu}},
  \ and\ \bibinfo {author} {\bibfnamefont {Xiao-Gang}\ \bibnamefont {Wen}},\
  }\bibfield  {title} {\enquote {\bibinfo {title} {Local unitary
  transformation, long-range quantum entanglement, wave function
  renormalization, and topological order},}\ }\href {\doibase
  10.1103/PhysRevB.82.155138} {\bibfield  {journal} {\bibinfo  {journal} {Phys.
  Rev. B}\ }\textbf {\bibinfo {volume} {82}},\ \bibinfo {pages} {155138}
  (\bibinfo {year} {2010})}\BibitemShut {NoStop}%
\bibitem [{\citenamefont {Savary}\ and\ \citenamefont
  {Balents}(2016)}]{savary2016quantum}%
  \BibitemOpen
  \bibfield  {author} {\bibinfo {author} {\bibfnamefont {Lucile}\ \bibnamefont
  {Savary}}\ and\ \bibinfo {author} {\bibfnamefont {Leon}\ \bibnamefont
  {Balents}},\ }\bibfield  {title} {\enquote {\bibinfo {title} {Quantum spin
  liquids: a review},}\ }\href {\doibase
  https://doi.org/10.1088/0034-4885/80/1/016502} {\bibfield  {journal}
  {\bibinfo  {journal} {Reports on Progress in Physics}\ }\textbf {\bibinfo
  {volume} {80}},\ \bibinfo {pages} {016502} (\bibinfo {year}
  {2016})}\BibitemShut {NoStop}%
\bibitem [{\citenamefont {Zhou}\ \emph {et~al.}(2017)\citenamefont {Zhou},
  \citenamefont {Kanoda},\ and\ \citenamefont {Ng}}]{ZhouRMP}%
  \BibitemOpen
  \bibfield  {author} {\bibinfo {author} {\bibfnamefont {Yi}~\bibnamefont
  {Zhou}}, \bibinfo {author} {\bibfnamefont {Kazushi}\ \bibnamefont {Kanoda}},
  \ and\ \bibinfo {author} {\bibfnamefont {Tai-Kai}\ \bibnamefont {Ng}},\
  }\bibfield  {title} {\enquote {\bibinfo {title} {Quantum spin liquid
  states},}\ }\href {\doibase 10.1103/RevModPhys.89.025003} {\bibfield
  {journal} {\bibinfo  {journal} {Rev. Mod. Phys.}\ }\textbf {\bibinfo {volume}
  {89}},\ \bibinfo {pages} {025003} (\bibinfo {year} {2017})},\ \bibinfo {note}
  {see also references therein.}\BibitemShut {Stop}%
\bibitem [{\citenamefont {Knolle}\ and\ \citenamefont
  {Moessner}(2019)}]{KnolleReview}%
  \BibitemOpen
  \bibfield  {author} {\bibinfo {author} {\bibfnamefont {J.}~\bibnamefont
  {Knolle}}\ and\ \bibinfo {author} {\bibfnamefont {R.}~\bibnamefont
  {Moessner}},\ }\bibfield  {title} {\enquote {\bibinfo {title} {A field guide
  to spin liquids},}\ }\href {\doibase
  10.1146/annurev-conmatphys-031218-013401} {\bibfield  {journal} {\bibinfo
  {journal} {Annual Review of Condensed Matter Physics}\ }\textbf {\bibinfo
  {volume} {10}},\ \bibinfo {pages} {451--472} (\bibinfo {year} {2019})},\
  \Eprint
  {http://arxiv.org/abs/https://doi.org/10.1146/annurev-conmatphys-031218-013401}
  {https://doi.org/10.1146/annurev-conmatphys-031218-013401} \BibitemShut
  {NoStop}%
\bibitem [{\citenamefont {Kitaev}(2006)}]{kitaev2006anyons}%
  \BibitemOpen
  \bibfield  {author} {\bibinfo {author} {\bibfnamefont {Alexei}\ \bibnamefont
  {Kitaev}},\ }\bibfield  {title} {\enquote {\bibinfo {title} {Anyons in an
  exactly solved model and beyond},}\ }\href {\doibase
  https://doi.org/10.1016/j.aop.2005.10.005} {\bibfield  {journal} {\bibinfo
  {journal} {Annals of Physics}\ }\textbf {\bibinfo {volume} {321}},\ \bibinfo
  {pages} {2--111} (\bibinfo {year} {2006})}\BibitemShut {NoStop}%
\bibitem [{\citenamefont {Kells}\ \emph {et~al.}(2008)\citenamefont {Kells},
  \citenamefont {Bolukbasi}, \citenamefont {Lahtinen}, \citenamefont
  {Slingerland}, \citenamefont {Pachos},\ and\ \citenamefont
  {Vala}}]{Vala2008}%
  \BibitemOpen
  \bibfield  {author} {\bibinfo {author} {\bibfnamefont {G.}~\bibnamefont
  {Kells}}, \bibinfo {author} {\bibfnamefont {A.~T.}\ \bibnamefont
  {Bolukbasi}}, \bibinfo {author} {\bibfnamefont {V.}~\bibnamefont {Lahtinen}},
  \bibinfo {author} {\bibfnamefont {J.~K.}\ \bibnamefont {Slingerland}},
  \bibinfo {author} {\bibfnamefont {J.~K.}\ \bibnamefont {Pachos}}, \ and\
  \bibinfo {author} {\bibfnamefont {J.}~\bibnamefont {Vala}},\ }\bibfield
  {title} {\enquote {\bibinfo {title} {Topological degeneracy and vortex
  manipulation in kitaev's honeycomb model},}\ }\href {\doibase
  10.1103/PhysRevLett.101.240404} {\bibfield  {journal} {\bibinfo  {journal}
  {Phys. Rev. Lett.}\ }\textbf {\bibinfo {volume} {101}},\ \bibinfo {pages}
  {240404} (\bibinfo {year} {2008})}\BibitemShut {NoStop}%
\bibitem [{\citenamefont {Bolukbasi}\ and\ \citenamefont
  {Vala}(2012)}]{Bolukbasi_2012}%
  \BibitemOpen
  \bibfield  {author} {\bibinfo {author} {\bibfnamefont {Ahmet~Tuna}\
  \bibnamefont {Bolukbasi}}\ and\ \bibinfo {author} {\bibfnamefont {Jiri}\
  \bibnamefont {Vala}},\ }\bibfield  {title} {\enquote {\bibinfo {title}
  {Rigorous calculations of non-abelian statistics in the kitaev honeycomb
  model},}\ }\href {\doibase 10.1088/1367-2630/14/4/045007} {\bibfield
  {journal} {\bibinfo  {journal} {New Journal of Physics}\ }\textbf {\bibinfo
  {volume} {14}},\ \bibinfo {pages} {045007} (\bibinfo {year}
  {2012})}\BibitemShut {NoStop}%
\bibitem [{\citenamefont {Nayak}\ \emph {et~al.}(2008)\citenamefont {Nayak},
  \citenamefont {Simon}, \citenamefont {Stern}, \citenamefont {Freedman},\ and\
  \citenamefont {Das~Sarma}}]{Sankar2008}%
  \BibitemOpen
  \bibfield  {author} {\bibinfo {author} {\bibfnamefont {Chetan}\ \bibnamefont
  {Nayak}}, \bibinfo {author} {\bibfnamefont {Steven~H.}\ \bibnamefont
  {Simon}}, \bibinfo {author} {\bibfnamefont {Ady}\ \bibnamefont {Stern}},
  \bibinfo {author} {\bibfnamefont {Michael}\ \bibnamefont {Freedman}}, \ and\
  \bibinfo {author} {\bibfnamefont {Sankar}\ \bibnamefont {Das~Sarma}},\
  }\bibfield  {title} {\enquote {\bibinfo {title} {Non-abelian anyons and
  topological quantum computation},}\ }\href {\doibase
  10.1103/RevModPhys.80.1083} {\bibfield  {journal} {\bibinfo  {journal} {Rev.
  Mod. Phys.}\ }\textbf {\bibinfo {volume} {80}},\ \bibinfo {pages}
  {1083--1159} (\bibinfo {year} {2008})}\BibitemShut {NoStop}%
\bibitem [{\citenamefont {Semeghini}\ \emph {et~al.}(2021)\citenamefont
  {Semeghini}, \citenamefont {Levine}, \citenamefont {Keesling}, \citenamefont
  {Ebadi}, \citenamefont {Wang}, \citenamefont {Bluvstein}, \citenamefont
  {Verresen}, \citenamefont {Pichler}, \citenamefont {Kalinowski},
  \citenamefont {Samajdar}, \citenamefont {Omran}, \citenamefont {Sachdev},
  \citenamefont {Vishwanath}, \citenamefont {Greiner}, \citenamefont
  {Vuletić},\ and\ \citenamefont {Lukin}}]{Semeghini2021}%
  \BibitemOpen
  \bibfield  {author} {\bibinfo {author} {\bibfnamefont {G.}~\bibnamefont
  {Semeghini}}, \bibinfo {author} {\bibfnamefont {H.}~\bibnamefont {Levine}},
  \bibinfo {author} {\bibfnamefont {A.}~\bibnamefont {Keesling}}, \bibinfo
  {author} {\bibfnamefont {S.}~\bibnamefont {Ebadi}}, \bibinfo {author}
  {\bibfnamefont {T.~T.}\ \bibnamefont {Wang}}, \bibinfo {author}
  {\bibfnamefont {D.}~\bibnamefont {Bluvstein}}, \bibinfo {author}
  {\bibfnamefont {R.}~\bibnamefont {Verresen}}, \bibinfo {author}
  {\bibfnamefont {H.}~\bibnamefont {Pichler}}, \bibinfo {author} {\bibfnamefont
  {M.}~\bibnamefont {Kalinowski}}, \bibinfo {author} {\bibfnamefont
  {R.}~\bibnamefont {Samajdar}}, \bibinfo {author} {\bibfnamefont
  {A.}~\bibnamefont {Omran}}, \bibinfo {author} {\bibfnamefont
  {S.}~\bibnamefont {Sachdev}}, \bibinfo {author} {\bibfnamefont
  {A.}~\bibnamefont {Vishwanath}}, \bibinfo {author} {\bibfnamefont
  {M.}~\bibnamefont {Greiner}}, \bibinfo {author} {\bibfnamefont
  {V.}~\bibnamefont {Vuletić}}, \ and\ \bibinfo {author} {\bibfnamefont
  {M.~D.}\ \bibnamefont {Lukin}},\ }\bibfield  {title} {\enquote {\bibinfo
  {title} {Probing topological spin liquids on a programmable quantum
  simulator},}\ }\href {\doibase 10.1126/science.abi8794} {\bibfield  {journal}
  {\bibinfo  {journal} {Science}\ }\textbf {\bibinfo {volume} {374}},\ \bibinfo
  {pages} {1242--1247} (\bibinfo {year} {2021})}\BibitemShut {NoStop}%
\bibitem [{\citenamefont {Jang}\ \emph {et~al.}(2021)\citenamefont {Jang},
  \citenamefont {Kato},\ and\ \citenamefont {Motome}}]{Kato2021}%
  \BibitemOpen
  \bibfield  {author} {\bibinfo {author} {\bibfnamefont {Seong-Hoon}\
  \bibnamefont {Jang}}, \bibinfo {author} {\bibfnamefont {Yasuyuki}\
  \bibnamefont {Kato}}, \ and\ \bibinfo {author} {\bibfnamefont {Yukitoshi}\
  \bibnamefont {Motome}},\ }\bibfield  {title} {\enquote {\bibinfo {title}
  {Vortex creation and control in the kitaev spin liquid by local bond
  modulations},}\ }\href {\doibase 10.1103/PhysRevB.104.085142} {\bibfield
  {journal} {\bibinfo  {journal} {Phys. Rev. B}\ }\textbf {\bibinfo {volume}
  {104}},\ \bibinfo {pages} {085142} (\bibinfo {year} {2021})}\BibitemShut
  {NoStop}%
\bibitem [{\citenamefont {Liu}\ \emph {et~al.}(2022)\citenamefont {Liu},
  \citenamefont {Slagle}, \citenamefont {Burch},\ and\ \citenamefont
  {Alicea}}]{Jason2022}%
  \BibitemOpen
  \bibfield  {author} {\bibinfo {author} {\bibfnamefont {Yue}\ \bibnamefont
  {Liu}}, \bibinfo {author} {\bibfnamefont {Kevin}\ \bibnamefont {Slagle}},
  \bibinfo {author} {\bibfnamefont {Kenneth~S.}\ \bibnamefont {Burch}}, \ and\
  \bibinfo {author} {\bibfnamefont {Jason}\ \bibnamefont {Alicea}},\ }\bibfield
   {title} {\enquote {\bibinfo {title} {Dynamical anyon generation in kitaev
  honeycomb non-abelian spin liquids},}\ }\href {\doibase
  10.1103/PhysRevLett.129.037201} {\bibfield  {journal} {\bibinfo  {journal}
  {Phys. Rev. Lett.}\ }\textbf {\bibinfo {volume} {129}},\ \bibinfo {pages}
  {037201} (\bibinfo {year} {2022})}\BibitemShut {NoStop}%
\bibitem [{\citenamefont {Lieb}(1994)}]{Lieb1994}%
  \BibitemOpen
  \bibfield  {author} {\bibinfo {author} {\bibfnamefont {Elliott}\ \bibnamefont
  {Lieb}},\ }\bibfield  {title} {\enquote {\bibinfo {title} {Flux phase of the
  half-filled band},}\ }\href {\doibase 10.1103/PhysRevLett.73.2158} {\bibfield
   {journal} {\bibinfo  {journal} {Phys. Rev. Lett.}\ }\textbf {\bibinfo
  {volume} {73}},\ \bibinfo {pages} {2158--2161} (\bibinfo {year}
  {1994})}\BibitemShut {NoStop}%
\bibitem [{\citenamefont {Kitaev}(2003)}]{kitaev2003fault}%
  \BibitemOpen
  \bibfield  {author} {\bibinfo {author} {\bibfnamefont {A~Yu}\ \bibnamefont
  {Kitaev}},\ }\bibfield  {title} {\enquote {\bibinfo {title} {Fault-tolerant
  quantum computation by anyons},}\ }\href {\doibase
  https://doi.org/10.1016/S0003-4916(02)00018-0} {\bibfield  {journal}
  {\bibinfo  {journal} {Annals of Physics}\ }\textbf {\bibinfo {volume}
  {303}},\ \bibinfo {pages} {2--30} (\bibinfo {year} {2003})}\BibitemShut
  {NoStop}%
\bibitem [{\citenamefont {Nanda}\ \emph {et~al.}(2020)\citenamefont {Nanda},
  \citenamefont {Dhochak},\ and\ \citenamefont
  {Bhattacharjee}}]{nanda2020phases}%
  \BibitemOpen
  \bibfield  {author} {\bibinfo {author} {\bibfnamefont {Animesh}\ \bibnamefont
  {Nanda}}, \bibinfo {author} {\bibfnamefont {Kusum}\ \bibnamefont {Dhochak}},
  \ and\ \bibinfo {author} {\bibfnamefont {Subhro}\ \bibnamefont
  {Bhattacharjee}},\ }\bibfield  {title} {\enquote {\bibinfo {title} {Phases
  and quantum phase transitions in an anisotropic ferromagnetic
  kitaev-heisenberg-$\mathrm{\ensuremath{\Gamma}}$ magnet},}\ }\href {\doibase
  10.1103/PhysRevB.102.235124} {\bibfield  {journal} {\bibinfo  {journal}
  {Phys. Rev. B}\ }\textbf {\bibinfo {volume} {102}},\ \bibinfo {pages}
  {235124} (\bibinfo {year} {2020})}\BibitemShut {NoStop}%
\bibitem [{\citenamefont {Joy}\ and\ \citenamefont {Rosch}(2022)}]{Joy2022}%
  \BibitemOpen
  \bibfield  {author} {\bibinfo {author} {\bibfnamefont {Aprem~P.}\
  \bibnamefont {Joy}}\ and\ \bibinfo {author} {\bibfnamefont {Achim}\
  \bibnamefont {Rosch}},\ }\bibfield  {title} {\enquote {\bibinfo {title}
  {Dynamics of visons and thermal hall effect in perturbed {K}itaev models},}\
  }\href {\doibase 10.1103/PhysRevX.12.041004} {\bibfield  {journal} {\bibinfo
  {journal} {Phys. Rev. X}\ }\textbf {\bibinfo {volume} {12}},\ \bibinfo
  {pages} {041004} (\bibinfo {year} {2022})}\BibitemShut {NoStop}%
\bibitem [{\citenamefont {Song}\ and\ \citenamefont
  {Senthil}(2022)}]{Song2022}%
  \BibitemOpen
  \bibfield  {author} {\bibinfo {author} {\bibfnamefont {Xue-Yang}\
  \bibnamefont {Song}}\ and\ \bibinfo {author} {\bibfnamefont {T.}~\bibnamefont
  {Senthil}},\ }\bibfield  {title} {\enquote {\bibinfo {title}
  {Translation-enriched ${Z}_2$ spin liquids and topological vison bands:
  Possible application to $\alpha$-{R}u{C}l$_3$},}\ }\href {\doibase
  10.48550/ARXIV.2206.14197} {\  (\bibinfo {year} {2022}),\
  10.48550/ARXIV.2206.14197}\BibitemShut {NoStop}%
\bibitem [{\citenamefont {Nanda}\ \emph {et~al.}(2021)\citenamefont {Nanda},
  \citenamefont {Agarwala},\ and\ \citenamefont
  {Bhattacharjee}}]{Nanda_PRB_2021}%
  \BibitemOpen
  \bibfield  {author} {\bibinfo {author} {\bibfnamefont {Animesh}\ \bibnamefont
  {Nanda}}, \bibinfo {author} {\bibfnamefont {Adhip}\ \bibnamefont {Agarwala}},
  \ and\ \bibinfo {author} {\bibfnamefont {Subhro}\ \bibnamefont
  {Bhattacharjee}},\ }\bibfield  {title} {\enquote {\bibinfo {title} {Phases
  and quantum phase transitions in the anisotropic antiferromagnetic
  kitaev-heisenberg-$\mathrm{\ensuremath{\Gamma}}$ magnet},}\ }\href {\doibase
  10.1103/PhysRevB.104.195115} {\bibfield  {journal} {\bibinfo  {journal}
  {Phys. Rev. B}\ }\textbf {\bibinfo {volume} {104}},\ \bibinfo {pages}
  {195115} (\bibinfo {year} {2021})}\BibitemShut {NoStop}%
\bibitem [{\citenamefont {Patel}\ and\ \citenamefont
  {Trivedi}(2019)}]{Patel12199}%
  \BibitemOpen
  \bibfield  {author} {\bibinfo {author} {\bibfnamefont {Niravkumar~D.}\
  \bibnamefont {Patel}}\ and\ \bibinfo {author} {\bibfnamefont {Nandini}\
  \bibnamefont {Trivedi}},\ }\bibfield  {title} {\enquote {\bibinfo {title}
  {Magnetic field-induced intermediate quantum spin liquid with a spinon fermi
  surface},}\ }\href {\doibase 10.1073/pnas.1821406116} {\bibfield  {journal}
  {\bibinfo  {journal} {Proceedings of the National Academy of Sciences}\
  }\textbf {\bibinfo {volume} {116}},\ \bibinfo {pages} {12199--12203}
  (\bibinfo {year} {2019})}\BibitemShut {NoStop}%
\bibitem [{\citenamefont {Wan}\ and\ \citenamefont
  {Armitage}(2019)}]{Armitage2019}%
  \BibitemOpen
  \bibfield  {author} {\bibinfo {author} {\bibfnamefont {Yuan}\ \bibnamefont
  {Wan}}\ and\ \bibinfo {author} {\bibfnamefont {N.~P.}\ \bibnamefont
  {Armitage}},\ }\bibfield  {title} {\enquote {\bibinfo {title} {Resolving
  continua of fractional excitations by spinon echo in thz 2d coherent
  spectroscopy},}\ }\href {\doibase 10.1103/PhysRevLett.122.257401} {\bibfield
  {journal} {\bibinfo  {journal} {Phys. Rev. Lett.}\ }\textbf {\bibinfo
  {volume} {122}},\ \bibinfo {pages} {257401} (\bibinfo {year}
  {2019})}\BibitemShut {NoStop}%
\bibitem [{\citenamefont {Choi}\ \emph {et~al.}(2020)\citenamefont {Choi},
  \citenamefont {Lee},\ and\ \citenamefont {Kim}}]{Choi2020}%
  \BibitemOpen
  \bibfield  {author} {\bibinfo {author} {\bibfnamefont {Wonjune}\ \bibnamefont
  {Choi}}, \bibinfo {author} {\bibfnamefont {Ki~Hoon}\ \bibnamefont {Lee}}, \
  and\ \bibinfo {author} {\bibfnamefont {Yong~Baek}\ \bibnamefont {Kim}},\
  }\bibfield  {title} {\enquote {\bibinfo {title} {Theory of two-dimensional
  nonlinear spectroscopy for the kitaev spin liquid},}\ }\href {\doibase
  10.1103/PhysRevLett.124.117205} {\bibfield  {journal} {\bibinfo  {journal}
  {Phys. Rev. Lett.}\ }\textbf {\bibinfo {volume} {124}},\ \bibinfo {pages}
  {117205} (\bibinfo {year} {2020})}\BibitemShut {NoStop}%
\bibitem [{\citenamefont {Burnell}\ and\ \citenamefont
  {Nayak}(2011)}]{Burnell2011}%
  \BibitemOpen
  \bibfield  {author} {\bibinfo {author} {\bibfnamefont {F.~J.}\ \bibnamefont
  {Burnell}}\ and\ \bibinfo {author} {\bibfnamefont {Chetan}\ \bibnamefont
  {Nayak}},\ }\bibfield  {title} {\enquote {\bibinfo {title} {Su(2) slave
  fermion solution of the kitaev honeycomb lattice model},}\ }\href {\doibase
  10.1103/PhysRevB.84.125125} {\bibfield  {journal} {\bibinfo  {journal} {Phys.
  Rev. B}\ }\textbf {\bibinfo {volume} {84}},\ \bibinfo {pages} {125125}
  (\bibinfo {year} {2011})}\BibitemShut {NoStop}%
\bibitem [{\citenamefont {Rao}\ and\ \citenamefont
  {Sodemann}(2021)}]{PhysRevResearch.3.023120}%
  \BibitemOpen
  \bibfield  {author} {\bibinfo {author} {\bibfnamefont {Peng}\ \bibnamefont
  {Rao}}\ and\ \bibinfo {author} {\bibfnamefont {Inti}\ \bibnamefont
  {Sodemann}},\ }\bibfield  {title} {\enquote {\bibinfo {title} {Theory of weak
  symmetry breaking of translations in ${\mathbb{z}}_{2}$ topologically ordered
  states and its relation to topological superconductivity from an exact
  lattice ${\mathbb{z}}_{2}$ charge-flux attachment},}\ }\href {\doibase
  10.1103/PhysRevResearch.3.023120} {\bibfield  {journal} {\bibinfo  {journal}
  {Phys. Rev. Res.}\ }\textbf {\bibinfo {volume} {3}},\ \bibinfo {pages}
  {023120} (\bibinfo {year} {2021})}\BibitemShut {NoStop}%
\bibitem [{\citenamefont {{Jiang}}\ \emph {et~al.}(2018)\citenamefont
  {{Jiang}}, \citenamefont {{Wang}}, \citenamefont {{Huang}},\ and\
  \citenamefont {{Lu}}}]{Jiang_arXiv_2018}%
  \BibitemOpen
  \bibfield  {author} {\bibinfo {author} {\bibfnamefont {Hong-Chen}\
  \bibnamefont {{Jiang}}}, \bibinfo {author} {\bibfnamefont {Chang-Yan}\
  \bibnamefont {{Wang}}}, \bibinfo {author} {\bibfnamefont {Biao}\ \bibnamefont
  {{Huang}}}, \ and\ \bibinfo {author} {\bibfnamefont {Yuan-Ming}\ \bibnamefont
  {{Lu}}},\ }\bibfield  {title} {\enquote {\bibinfo {title} {{Field induced
  quantum spin liquid with spinon Fermi surfaces in the Kitaev model}},}\
  }\href@noop {} {\bibfield  {journal} {\bibinfo  {journal} {arXiv e-prints}\
  ,\ \bibinfo {eid} {arXiv:1809.08247}} (\bibinfo {year} {2018})},\ \Eprint
  {http://arxiv.org/abs/1809.08247} {arXiv:1809.08247 [cond-mat.str-el]}
  \BibitemShut {NoStop}%
\bibitem [{\citenamefont {Hickey}\ and\ \citenamefont
  {Trebst}(2019)}]{hickey2019emergence}%
  \BibitemOpen
  \bibfield  {author} {\bibinfo {author} {\bibfnamefont {Ciar{\'a}n}\
  \bibnamefont {Hickey}}\ and\ \bibinfo {author} {\bibfnamefont {Simon}\
  \bibnamefont {Trebst}},\ }\bibfield  {title} {\enquote {\bibinfo {title}
  {Emergence of a field-driven u (1) spin liquid in the kitaev honeycomb
  model},}\ }\href@noop {} {\bibfield  {journal} {\bibinfo  {journal} {Nature
  communications}\ }\textbf {\bibinfo {volume} {10}},\ \bibinfo {pages} {1--10}
  (\bibinfo {year} {2019})}\BibitemShut {NoStop}%
\bibitem [{\citenamefont {Zhang}\ \emph {et~al.}(2023)\citenamefont {Zhang},
  \citenamefont {Feng}, \citenamefont {Lensky}, \citenamefont {Trivedi},\ and\
  \citenamefont {Kim}}]{zhang2023machine}%
  \BibitemOpen
  \bibfield  {author} {\bibinfo {author} {\bibfnamefont {Kevin}\ \bibnamefont
  {Zhang}}, \bibinfo {author} {\bibfnamefont {Shi}\ \bibnamefont {Feng}},
  \bibinfo {author} {\bibfnamefont {Yuri~D.}\ \bibnamefont {Lensky}}, \bibinfo
  {author} {\bibfnamefont {Nandini}\ \bibnamefont {Trivedi}}, \ and\ \bibinfo
  {author} {\bibfnamefont {Eun-Ah}\ \bibnamefont {Kim}},\ }\href@noop {}
  {\enquote {\bibinfo {title} {Machine learning feature discovery of spinon
  fermi surface},}\ } (\bibinfo {year} {2023}),\ \Eprint
  {http://arxiv.org/abs/2306.03143} {arXiv:2306.03143 [cond-mat.str-el]}
  \BibitemShut {NoStop}%
\bibitem [{\citenamefont {Jiang}\ \emph {et~al.}(2020)\citenamefont {Jiang},
  \citenamefont {Liang}, \citenamefont {Chen}, \citenamefont {Qi},
  \citenamefont {Li},\ and\ \citenamefont {Wang}}]{Jiang2020Phys.Rev.Lett.}%
  \BibitemOpen
  \bibfield  {author} {\bibinfo {author} {\bibfnamefont {Ming-Hong}\
  \bibnamefont {Jiang}}, \bibinfo {author} {\bibfnamefont {Shuang}\
  \bibnamefont {Liang}}, \bibinfo {author} {\bibfnamefont {Wei}\ \bibnamefont
  {Chen}}, \bibinfo {author} {\bibfnamefont {Yang}\ \bibnamefont {Qi}},
  \bibinfo {author} {\bibfnamefont {Jian-Xin}\ \bibnamefont {Li}}, \ and\
  \bibinfo {author} {\bibfnamefont {Qiang-Hua}\ \bibnamefont {Wang}},\
  }\bibfield  {title} {\enquote {\bibinfo {title} {Tuning {{Topological
  Orders}} by a {{Conical Magnetic Field}} in the {{Kitaev Model}}},}\ }\href
  {\doibase 10.1103/PhysRevLett.125.177203} {\bibfield  {journal} {\bibinfo
  {journal} {Phys. Rev. Lett.}\ }\textbf {\bibinfo {volume} {125}},\ \bibinfo
  {pages} {177203} (\bibinfo {year} {2020})}\BibitemShut {NoStop}%
\bibitem [{\citenamefont {Zhang}\ \emph {et~al.}(2022)\citenamefont {Zhang},
  \citenamefont {Hal{\'a}sz},\ and\ \citenamefont
  {Batista}}]{Zhang2022NatCommun}%
  \BibitemOpen
  \bibfield  {author} {\bibinfo {author} {\bibfnamefont {Shang-Shun}\
  \bibnamefont {Zhang}}, \bibinfo {author} {\bibfnamefont {G{\'a}bor~B.}\
  \bibnamefont {Hal{\'a}sz}}, \ and\ \bibinfo {author} {\bibfnamefont
  {Cristian~D.}\ \bibnamefont {Batista}},\ }\bibfield  {title} {\enquote
  {\bibinfo {title} {Theory of the {{Kitaev}} model in a [111] magnetic
  field},}\ }\href {\doibase 10.1038/s41467-022-28014-3} {\bibfield  {journal}
  {\bibinfo  {journal} {Nat Commun}\ }\textbf {\bibinfo {volume} {13}},\
  \bibinfo {pages} {399} (\bibinfo {year} {2022})}\BibitemShut {NoStop}%
\bibitem [{\citenamefont {Ronquillo}\ \emph {et~al.}(2019)\citenamefont
  {Ronquillo}, \citenamefont {Vengal},\ and\ \citenamefont
  {Trivedi}}]{David2019}%
  \BibitemOpen
  \bibfield  {author} {\bibinfo {author} {\bibfnamefont {David~C.}\
  \bibnamefont {Ronquillo}}, \bibinfo {author} {\bibfnamefont {Adu}\
  \bibnamefont {Vengal}}, \ and\ \bibinfo {author} {\bibfnamefont {Nandini}\
  \bibnamefont {Trivedi}},\ }\bibfield  {title} {\enquote {\bibinfo {title}
  {Signatures of magnetic-field-driven quantum phase transitions in the
  entanglement entropy and spin dynamics of the kitaev honeycomb model},}\
  }\href {\doibase 10.1103/PhysRevB.99.140413} {\bibfield  {journal} {\bibinfo
  {journal} {Phys. Rev. B}\ }\textbf {\bibinfo {volume} {99}},\ \bibinfo
  {pages} {140413} (\bibinfo {year} {2019})}\BibitemShut {NoStop}%
\bibitem [{\citenamefont {Pradhan}\ \emph {et~al.}(2020)\citenamefont
  {Pradhan}, \citenamefont {Patel},\ and\ \citenamefont
  {Trivedi}}]{Pradhan_PRB_2020}%
  \BibitemOpen
  \bibfield  {author} {\bibinfo {author} {\bibfnamefont {Subhasree}\
  \bibnamefont {Pradhan}}, \bibinfo {author} {\bibfnamefont {Niravkumar~D.}\
  \bibnamefont {Patel}}, \ and\ \bibinfo {author} {\bibfnamefont {Nandini}\
  \bibnamefont {Trivedi}},\ }\bibfield  {title} {\enquote {\bibinfo {title}
  {Two-magnon bound states in the kitaev model in a [111] field},}\ }\href
  {\doibase 10.1103/PhysRevB.101.180401} {\bibfield  {journal} {\bibinfo
  {journal} {Phys. Rev. B}\ }\textbf {\bibinfo {volume} {101}},\ \bibinfo
  {pages} {180401} (\bibinfo {year} {2020})}\BibitemShut {NoStop}%
\bibitem [{\citenamefont {Nasu}\ \emph {et~al.}(2018)\citenamefont {Nasu},
  \citenamefont {Kato}, \citenamefont {Kamiya},\ and\ \citenamefont
  {Motome}}]{Nasu_PRB_2018}%
  \BibitemOpen
  \bibfield  {author} {\bibinfo {author} {\bibfnamefont {Joji}\ \bibnamefont
  {Nasu}}, \bibinfo {author} {\bibfnamefont {Yasuyuki}\ \bibnamefont {Kato}},
  \bibinfo {author} {\bibfnamefont {Yoshitomo}\ \bibnamefont {Kamiya}}, \ and\
  \bibinfo {author} {\bibfnamefont {Yukitoshi}\ \bibnamefont {Motome}},\
  }\bibfield  {title} {\enquote {\bibinfo {title} {Successive majorana
  topological transitions driven by a magnetic field in the kitaev model},}\
  }\href {\doibase 10.1103/PhysRevB.98.060416} {\bibfield  {journal} {\bibinfo
  {journal} {Phys. Rev. B}\ }\textbf {\bibinfo {volume} {98}},\ \bibinfo
  {pages} {060416} (\bibinfo {year} {2018})}\BibitemShut {NoStop}%
\bibitem [{\citenamefont {Liang}\ \emph {et~al.}(2018)\citenamefont {Liang},
  \citenamefont {Jiang}, \citenamefont {Chen}, \citenamefont {Li},\ and\
  \citenamefont {Wang}}]{Liang_PRB_2018}%
  \BibitemOpen
  \bibfield  {author} {\bibinfo {author} {\bibfnamefont {Shuang}\ \bibnamefont
  {Liang}}, \bibinfo {author} {\bibfnamefont {Ming-Hong}\ \bibnamefont
  {Jiang}}, \bibinfo {author} {\bibfnamefont {Wei}\ \bibnamefont {Chen}},
  \bibinfo {author} {\bibfnamefont {Jian-Xin}\ \bibnamefont {Li}}, \ and\
  \bibinfo {author} {\bibfnamefont {Qiang-Hua}\ \bibnamefont {Wang}},\
  }\bibfield  {title} {\enquote {\bibinfo {title} {Intermediate gapless phase
  and topological phase transition of the kitaev model in a uniform magnetic
  field},}\ }\href {\doibase 10.1103/PhysRevB.98.054433} {\bibfield  {journal}
  {\bibinfo  {journal} {Phys. Rev. B}\ }\textbf {\bibinfo {volume} {98}},\
  \bibinfo {pages} {054433} (\bibinfo {year} {2018})}\BibitemShut {NoStop}%
\bibitem [{\citenamefont {Gohlke}\ \emph {et~al.}(2018)\citenamefont {Gohlke},
  \citenamefont {Wachtel}, \citenamefont {Yamaji}, \citenamefont {Pollmann},\
  and\ \citenamefont {Kim}}]{Gohlke_PRB_2018}%
  \BibitemOpen
  \bibfield  {author} {\bibinfo {author} {\bibfnamefont {Matthias}\
  \bibnamefont {Gohlke}}, \bibinfo {author} {\bibfnamefont {Gideon}\
  \bibnamefont {Wachtel}}, \bibinfo {author} {\bibfnamefont {Youhei}\
  \bibnamefont {Yamaji}}, \bibinfo {author} {\bibfnamefont {Frank}\
  \bibnamefont {Pollmann}}, \ and\ \bibinfo {author} {\bibfnamefont
  {Yong~Baek}\ \bibnamefont {Kim}},\ }\bibfield  {title} {\enquote {\bibinfo
  {title} {Quantum spin liquid signatures in kitaev-like frustrated magnets},}\
  }\href {\doibase 10.1103/PhysRevB.97.075126} {\bibfield  {journal} {\bibinfo
  {journal} {Phys. Rev. B}\ }\textbf {\bibinfo {volume} {97}},\ \bibinfo
  {pages} {075126} (\bibinfo {year} {2018})}\BibitemShut {NoStop}%
\bibitem [{\citenamefont {Berke}\ \emph {et~al.}(2020)\citenamefont {Berke},
  \citenamefont {Trebst},\ and\ \citenamefont {Hickey}}]{Berke_PRB_2020}%
  \BibitemOpen
  \bibfield  {author} {\bibinfo {author} {\bibfnamefont {Christoph}\
  \bibnamefont {Berke}}, \bibinfo {author} {\bibfnamefont {Simon}\ \bibnamefont
  {Trebst}}, \ and\ \bibinfo {author} {\bibfnamefont {Ciar\'an}\ \bibnamefont
  {Hickey}},\ }\bibfield  {title} {\enquote {\bibinfo {title} {Field stability
  of majorana spin liquids in antiferromagnetic kitaev models},}\ }\href
  {\doibase 10.1103/PhysRevB.101.214442} {\bibfield  {journal} {\bibinfo
  {journal} {Phys. Rev. B}\ }\textbf {\bibinfo {volume} {101}},\ \bibinfo
  {pages} {214442} (\bibinfo {year} {2020})}\BibitemShut {NoStop}%
\bibitem [{\citenamefont {Zhu}\ \emph {et~al.}(2018)\citenamefont {Zhu},
  \citenamefont {Kimchi}, \citenamefont {Sheng},\ and\ \citenamefont
  {Fu}}]{Zhu_PRB_2018}%
  \BibitemOpen
  \bibfield  {author} {\bibinfo {author} {\bibfnamefont {Zheng}\ \bibnamefont
  {Zhu}}, \bibinfo {author} {\bibfnamefont {Itamar}\ \bibnamefont {Kimchi}},
  \bibinfo {author} {\bibfnamefont {D.~N.}\ \bibnamefont {Sheng}}, \ and\
  \bibinfo {author} {\bibfnamefont {Liang}\ \bibnamefont {Fu}},\ }\bibfield
  {title} {\enquote {\bibinfo {title} {Robust non-abelian spin liquid and a
  possible intermediate phase in the antiferromagnetic kitaev model with
  magnetic field},}\ }\href {\doibase 10.1103/PhysRevB.97.241110} {\bibfield
  {journal} {\bibinfo  {journal} {Phys. Rev. B}\ }\textbf {\bibinfo {volume}
  {97}},\ \bibinfo {pages} {241110} (\bibinfo {year} {2018})}\BibitemShut
  {NoStop}%
\bibitem [{\citenamefont {Kitaev}\ and\ \citenamefont
  {Preskill}(2006)}]{kitaev2006topological}%
  \BibitemOpen
  \bibfield  {author} {\bibinfo {author} {\bibfnamefont {Alexei}\ \bibnamefont
  {Kitaev}}\ and\ \bibinfo {author} {\bibfnamefont {John}\ \bibnamefont
  {Preskill}},\ }\bibfield  {title} {\enquote {\bibinfo {title} {Topological
  entanglement entropy},}\ }\href {\doibase 10.1103/PhysRevLett.96.110404}
  {\bibfield  {journal} {\bibinfo  {journal} {Physical review letters}\
  }\textbf {\bibinfo {volume} {96}},\ \bibinfo {pages} {110404} (\bibinfo
  {year} {2006})}\BibitemShut {NoStop}%
\bibitem [{\citenamefont {Levin}\ and\ \citenamefont
  {Wen}(2006)}]{levin2006detecting}%
  \BibitemOpen
  \bibfield  {author} {\bibinfo {author} {\bibfnamefont {Michael}\ \bibnamefont
  {Levin}}\ and\ \bibinfo {author} {\bibfnamefont {Xiao-Gang}\ \bibnamefont
  {Wen}},\ }\bibfield  {title} {\enquote {\bibinfo {title} {Detecting
  topological order in a ground state wave function},}\ }\href {\doibase
  10.1103/PhysRevLett.96.110405} {\bibfield  {journal} {\bibinfo  {journal}
  {Physical review letters}\ }\textbf {\bibinfo {volume} {96}},\ \bibinfo
  {pages} {110405} (\bibinfo {year} {2006})}\BibitemShut {NoStop}%
\bibitem [{\citenamefont {Feng}\ \emph {et~al.}(2023)\citenamefont {Feng},
  \citenamefont {Kong},\ and\ \citenamefont {Trivedi}}]{feng2023statistical}%
  \BibitemOpen
  \bibfield  {author} {\bibinfo {author} {\bibfnamefont {Shi}\ \bibnamefont
  {Feng}}, \bibinfo {author} {\bibfnamefont {Deqian}\ \bibnamefont {Kong}}, \
  and\ \bibinfo {author} {\bibfnamefont {Nandini}\ \bibnamefont {Trivedi}},\
  }\href@noop {} {\enquote {\bibinfo {title} {A statistical approach to
  topological entanglement: Boltzmann machine representation of higher-order
  irreducible correlation},}\ } (\bibinfo {year} {2023}),\ \Eprint
  {http://arxiv.org/abs/2302.03212} {arXiv:2302.03212 [quant-ph]} \BibitemShut
  {NoStop}%
\bibitem [{\citenamefont {Furukawa}\ and\ \citenamefont
  {Misguich}(2007)}]{Furukawa2007}%
  \BibitemOpen
  \bibfield  {author} {\bibinfo {author} {\bibfnamefont {Shunsuke}\
  \bibnamefont {Furukawa}}\ and\ \bibinfo {author} {\bibfnamefont {Gr\'egoire}\
  \bibnamefont {Misguich}},\ }\bibfield  {title} {\enquote {\bibinfo {title}
  {Topological entanglement entropy in the quantum dimer model on the
  triangular lattice},}\ }\href {\doibase 10.1103/PhysRevB.75.214407}
  {\bibfield  {journal} {\bibinfo  {journal} {Phys. Rev. B}\ }\textbf {\bibinfo
  {volume} {75}},\ \bibinfo {pages} {214407} (\bibinfo {year}
  {2007})}\BibitemShut {NoStop}%
\bibitem [{\citenamefont {Vidal}\ \emph {et~al.}(2009)\citenamefont {Vidal},
  \citenamefont {Thomale}, \citenamefont {Schmidt},\ and\ \citenamefont
  {Dusuel}}]{Vidal_PRB_2009}%
  \BibitemOpen
  \bibfield  {author} {\bibinfo {author} {\bibfnamefont {Julien}\ \bibnamefont
  {Vidal}}, \bibinfo {author} {\bibfnamefont {Ronny}\ \bibnamefont {Thomale}},
  \bibinfo {author} {\bibfnamefont {Kai~Phillip}\ \bibnamefont {Schmidt}}, \
  and\ \bibinfo {author} {\bibfnamefont {S\'ebastien}\ \bibnamefont {Dusuel}},\
  }\bibfield  {title} {\enquote {\bibinfo {title} {Self-duality and bound
  states of the toric code model in a transverse field},}\ }\href {\doibase
  10.1103/PhysRevB.80.081104} {\bibfield  {journal} {\bibinfo  {journal} {Phys.
  Rev. B}\ }\textbf {\bibinfo {volume} {80}},\ \bibinfo {pages} {081104}
  (\bibinfo {year} {2009})}\BibitemShut {NoStop}%
\bibitem [{\citenamefont {Xu}\ and\ \citenamefont
  {Moore}(2004)}]{PhysRevLett.93.047003}%
  \BibitemOpen
  \bibfield  {author} {\bibinfo {author} {\bibfnamefont {Cenke}\ \bibnamefont
  {Xu}}\ and\ \bibinfo {author} {\bibfnamefont {J.~E.}\ \bibnamefont {Moore}},\
  }\bibfield  {title} {\enquote {\bibinfo {title} {Strong-weak coupling
  self-duality in the two-dimensional quantum phase transition of $p+ip$
  superconducting arrays},}\ }\href {\doibase 10.1103/PhysRevLett.93.047003}
  {\bibfield  {journal} {\bibinfo  {journal} {Phys. Rev. Lett.}\ }\textbf
  {\bibinfo {volume} {93}},\ \bibinfo {pages} {047003} (\bibinfo {year}
  {2004})}\BibitemShut {NoStop}%
\bibitem [{\citenamefont {Nussinov}\ and\ \citenamefont
  {Fradkin}(2005)}]{PhysRevB.71.195120}%
  \BibitemOpen
  \bibfield  {author} {\bibinfo {author} {\bibfnamefont {Zohar}\ \bibnamefont
  {Nussinov}}\ and\ \bibinfo {author} {\bibfnamefont {Eduardo}\ \bibnamefont
  {Fradkin}},\ }\bibfield  {title} {\enquote {\bibinfo {title} {Discrete
  sliding symmetries, dualities, and self-dualities of quantum orbital compass
  models and $p+ip$ superconducting arrays},}\ }\href {\doibase
  10.1103/PhysRevB.71.195120} {\bibfield  {journal} {\bibinfo  {journal} {Phys.
  Rev. B}\ }\textbf {\bibinfo {volume} {71}},\ \bibinfo {pages} {195120}
  (\bibinfo {year} {2005})}\BibitemShut {NoStop}%
\bibitem [{\citenamefont {Feng}\ \emph
  {et~al.}(2022{\natexlab{a}})\citenamefont {Feng}, \citenamefont {He},\ and\
  \citenamefont {Trivedi}}]{Feng2022}%
  \BibitemOpen
  \bibfield  {author} {\bibinfo {author} {\bibfnamefont {Shi}\ \bibnamefont
  {Feng}}, \bibinfo {author} {\bibfnamefont {Yanjun}\ \bibnamefont {He}}, \
  and\ \bibinfo {author} {\bibfnamefont {Nandini}\ \bibnamefont {Trivedi}},\
  }\bibfield  {title} {\enquote {\bibinfo {title} {Detection of long-range
  entanglement in gapped quantum spin liquids by local measurements},}\ }\href
  {\doibase 10.1103/PhysRevA.106.042417} {\bibfield  {journal} {\bibinfo
  {journal} {Phys. Rev. A}\ }\textbf {\bibinfo {volume} {106}},\ \bibinfo
  {pages} {042417} (\bibinfo {year} {2022}{\natexlab{a}})}\BibitemShut
  {NoStop}%
\bibitem [{\citenamefont {Hosur}\ \emph {et~al.}(2016)\citenamefont {Hosur},
  \citenamefont {Qi}, \citenamefont {Roberts},\ and\ \citenamefont
  {Yoshida}}]{Hosur2016}%
  \BibitemOpen
  \bibfield  {author} {\bibinfo {author} {\bibfnamefont {Pavan}\ \bibnamefont
  {Hosur}}, \bibinfo {author} {\bibfnamefont {Xiao-Liang}\ \bibnamefont {Qi}},
  \bibinfo {author} {\bibfnamefont {Daniel~A.}\ \bibnamefont {Roberts}}, \ and\
  \bibinfo {author} {\bibfnamefont {Beni}\ \bibnamefont {Yoshida}},\ }\bibfield
   {title} {\enquote {\bibinfo {title} {Chaos in quantum channels},}\ }\href
  {\doibase 10.1007/JHEP02(2016)004} {\bibfield  {journal} {\bibinfo  {journal}
  {Journal of High Energy Physics}\ }\textbf {\bibinfo {volume} {2016}},\
  \bibinfo {pages} {4} (\bibinfo {year} {2016})}\BibitemShut {NoStop}%
\bibitem [{\citenamefont {Kudler-Flam}\ \emph {et~al.}(2020)\citenamefont
  {Kudler-Flam}, \citenamefont {Nozaki}, \citenamefont {Ryu},\ and\
  \citenamefont {Tan}}]{Ryu2020}%
  \BibitemOpen
  \bibfield  {author} {\bibinfo {author} {\bibfnamefont {Jonah}\ \bibnamefont
  {Kudler-Flam}}, \bibinfo {author} {\bibfnamefont {Masahiro}\ \bibnamefont
  {Nozaki}}, \bibinfo {author} {\bibfnamefont {Shinsei}\ \bibnamefont {Ryu}}, \
  and\ \bibinfo {author} {\bibfnamefont {Mao~Tian}\ \bibnamefont {Tan}},\
  }\bibfield  {title} {\enquote {\bibinfo {title} {Quantum vs. classical
  information: operator negativity as a probe of scrambling},}\ }\href
  {\doibase 10.1007/JHEP01(2020)031} {\bibfield  {journal} {\bibinfo  {journal}
  {Journal of High Energy Physics}\ }\textbf {\bibinfo {volume} {2020}},\
  \bibinfo {pages} {31} (\bibinfo {year} {2020})}\BibitemShut {NoStop}%
\bibitem [{\citenamefont {Balents}(2010)}]{balents2010spin}%
  \BibitemOpen
  \bibfield  {author} {\bibinfo {author} {\bibfnamefont {Leon}\ \bibnamefont
  {Balents}},\ }\bibfield  {title} {\enquote {\bibinfo {title} {Spin liquids in
  frustrated magnets},}\ }\href {\doibase https://doi.org/10.1038/nature08917}
  {\bibfield  {journal} {\bibinfo  {journal} {Nature}\ }\textbf {\bibinfo
  {volume} {464}},\ \bibinfo {pages} {199--208} (\bibinfo {year}
  {2010})}\BibitemShut {NoStop}%
\bibitem [{\citenamefont {Knolle}\ \emph {et~al.}(2014)\citenamefont {Knolle},
  \citenamefont {Kovrizhin}, \citenamefont {Chalker},\ and\ \citenamefont
  {Moessner}}]{Knolle2014}%
  \BibitemOpen
  \bibfield  {author} {\bibinfo {author} {\bibfnamefont {J.}~\bibnamefont
  {Knolle}}, \bibinfo {author} {\bibfnamefont {D.~L.}\ \bibnamefont
  {Kovrizhin}}, \bibinfo {author} {\bibfnamefont {J.~T.}\ \bibnamefont
  {Chalker}}, \ and\ \bibinfo {author} {\bibfnamefont {R.}~\bibnamefont
  {Moessner}},\ }\bibfield  {title} {\enquote {\bibinfo {title} {Dynamics of a
  two-dimensional quantum spin liquid: Signatures of emergent majorana fermions
  and fluxes},}\ }\href {\doibase 10.1103/PhysRevLett.112.207203} {\bibfield
  {journal} {\bibinfo  {journal} {Phys. Rev. Lett.}\ }\textbf {\bibinfo
  {volume} {112}},\ \bibinfo {pages} {207203} (\bibinfo {year}
  {2014})}\BibitemShut {NoStop}%
\bibitem [{\citenamefont {Kanega}\ \emph {et~al.}(2021)\citenamefont {Kanega},
  \citenamefont {Ikeda},\ and\ \citenamefont {Sato}}]{Sato2021}%
  \BibitemOpen
  \bibfield  {author} {\bibinfo {author} {\bibfnamefont {Minoru}\ \bibnamefont
  {Kanega}}, \bibinfo {author} {\bibfnamefont {Tatsuhiko~N.}\ \bibnamefont
  {Ikeda}}, \ and\ \bibinfo {author} {\bibfnamefont {Masahiro}\ \bibnamefont
  {Sato}},\ }\bibfield  {title} {\enquote {\bibinfo {title} {Linear and
  nonlinear optical responses in kitaev spin liquids},}\ }\href {\doibase
  10.1103/PhysRevResearch.3.L032024} {\bibfield  {journal} {\bibinfo  {journal}
  {Phys. Rev. Res.}\ }\textbf {\bibinfo {volume} {3}},\ \bibinfo {pages}
  {L032024} (\bibinfo {year} {2021})}\BibitemShut {NoStop}%
\bibitem [{\citenamefont {White}(1992)}]{re:white92}%
  \BibitemOpen
  \bibfield  {author} {\bibinfo {author} {\bibfnamefont {Steven~R.}\
  \bibnamefont {White}},\ }\bibfield  {title} {\enquote {\bibinfo {title}
  {Density matrix formulation for quantum renormalization groups},}\ }\href
  {\doibase 10.1103/PhysRevLett.69.2863} {\bibfield  {journal} {\bibinfo
  {journal} {Phys. Rev. Lett.}\ }\textbf {\bibinfo {volume} {69}},\ \bibinfo
  {pages} {2863--2866} (\bibinfo {year} {1992})}\BibitemShut {NoStop}%
\bibitem [{\citenamefont {White}(1993)}]{re:white93}%
  \BibitemOpen
  \bibfield  {author} {\bibinfo {author} {\bibfnamefont {Steven~R.}\
  \bibnamefont {White}},\ }\bibfield  {title} {\enquote {\bibinfo {title}
  {Density-matrix algorithms for quantum renormalization groups},}\ }\href
  {\doibase 10.1103/PhysRevB.48.10345} {\bibfield  {journal} {\bibinfo
  {journal} {Phys. Rev. B}\ }\textbf {\bibinfo {volume} {48}},\ \bibinfo
  {pages} {10345--10356} (\bibinfo {year} {1993})}\BibitemShut {NoStop}%
\bibitem [{\citenamefont {Alvarez}(2009)}]{alvarez09}%
  \BibitemOpen
  \bibfield  {author} {\bibinfo {author} {\bibfnamefont {G.}~\bibnamefont
  {Alvarez}},\ }\bibfield  {title} {\enquote {\bibinfo {title} {The density
  matrix renormalization group for strongly correlated electron systems: A
  generic implementation},}\ }\href@noop {} {\bibfield  {journal} {\bibinfo
  {journal} {Computer Physics Communications}\ }\textbf {\bibinfo {volume}
  {180}},\ \bibinfo {pages} {1572} (\bibinfo {year} {2009})}\BibitemShut
  {NoStop}%
\bibitem [{\citenamefont {Alvarez}\ \emph {et~al.}(2011)\citenamefont
  {Alvarez}, \citenamefont {da~Silva}, \citenamefont {Ponce},\ and\
  \citenamefont {Dagotto}}]{alvarez0311}%
  \BibitemOpen
  \bibfield  {author} {\bibinfo {author} {\bibfnamefont {G.}~\bibnamefont
  {Alvarez}}, \bibinfo {author} {\bibfnamefont {L.~G. G. V.~Dias}\ \bibnamefont
  {da~Silva}}, \bibinfo {author} {\bibfnamefont {E.}~\bibnamefont {Ponce}}, \
  and\ \bibinfo {author} {\bibfnamefont {E.}~\bibnamefont {Dagotto}},\
  }\bibfield  {title} {\enquote {\bibinfo {title} {Time evolution with the dmrg
  algorithm: A generic implementation for strongly correlated electronic
  systems},}\ }\href@noop {} {\bibfield  {journal} {\bibinfo  {journal} {Phys.
  Rev. E}\ }\textbf {\bibinfo {volume} {84}},\ \bibinfo {pages} {056706}
  (\bibinfo {year} {2011})}\BibitemShut {NoStop}%
\bibitem [{\citenamefont {Nocera}\ and\ \citenamefont
  {Alvarez}(2016)}]{Alvarez2016}%
  \BibitemOpen
  \bibfield  {author} {\bibinfo {author} {\bibfnamefont {A.}~\bibnamefont
  {Nocera}}\ and\ \bibinfo {author} {\bibfnamefont {G.}~\bibnamefont
  {Alvarez}},\ }\bibfield  {title} {\enquote {\bibinfo {title} {Spectral
  functions with the density matrix renormalization group: Krylov-space
  approach for correction vectors},}\ }\href {\doibase
  10.1103/PhysRevE.94.053308} {\bibfield  {journal} {\bibinfo  {journal} {Phys.
  Rev. E}\ }\textbf {\bibinfo {volume} {94}},\ \bibinfo {pages} {053308}
  (\bibinfo {year} {2016})}\BibitemShut {NoStop}%
\bibitem [{\citenamefont {Dusuel}\ \emph {et~al.}(2011)\citenamefont {Dusuel},
  \citenamefont {Kamfor}, \citenamefont {Or\'us}, \citenamefont {Schmidt},\
  and\ \citenamefont {Vidal}}]{Dusuel_PRL_2011}%
  \BibitemOpen
  \bibfield  {author} {\bibinfo {author} {\bibfnamefont {S\'ebastien}\
  \bibnamefont {Dusuel}}, \bibinfo {author} {\bibfnamefont {Michael}\
  \bibnamefont {Kamfor}}, \bibinfo {author} {\bibfnamefont {Rom\'an}\
  \bibnamefont {Or\'us}}, \bibinfo {author} {\bibfnamefont {Kai~Phillip}\
  \bibnamefont {Schmidt}}, \ and\ \bibinfo {author} {\bibfnamefont {Julien}\
  \bibnamefont {Vidal}},\ }\bibfield  {title} {\enquote {\bibinfo {title}
  {Robustness of a perturbed topological phase},}\ }\href {\doibase
  10.1103/PhysRevLett.106.107203} {\bibfield  {journal} {\bibinfo  {journal}
  {Phys. Rev. Lett.}\ }\textbf {\bibinfo {volume} {106}},\ \bibinfo {pages}
  {107203} (\bibinfo {year} {2011})}\BibitemShut {NoStop}%
\bibitem [{\citenamefont {You}\ \emph {et~al.}(2013)\citenamefont {You},
  \citenamefont {Jian},\ and\ \citenamefont {Wen}}]{You2013}%
  \BibitemOpen
  \bibfield  {author} {\bibinfo {author} {\bibfnamefont {Yi-Zhuang}\
  \bibnamefont {You}}, \bibinfo {author} {\bibfnamefont {Chao-Ming}\
  \bibnamefont {Jian}}, \ and\ \bibinfo {author} {\bibfnamefont {Xiao-Gang}\
  \bibnamefont {Wen}},\ }\bibfield  {title} {\enquote {\bibinfo {title}
  {Synthetic non-abelian statistics by abelian anyon condensation},}\ }\href
  {\doibase 10.1103/PhysRevB.87.045106} {\bibfield  {journal} {\bibinfo
  {journal} {Phys. Rev. B}\ }\textbf {\bibinfo {volume} {87}},\ \bibinfo
  {pages} {045106} (\bibinfo {year} {2013})}\BibitemShut {NoStop}%
\bibitem [{\citenamefont {Baskaran}\ \emph {et~al.}(2007)\citenamefont
  {Baskaran}, \citenamefont {Mandal},\ and\ \citenamefont
  {Shankar}}]{Baskaran2007}%
  \BibitemOpen
  \bibfield  {author} {\bibinfo {author} {\bibfnamefont {G.}~\bibnamefont
  {Baskaran}}, \bibinfo {author} {\bibfnamefont {Saptarshi}\ \bibnamefont
  {Mandal}}, \ and\ \bibinfo {author} {\bibfnamefont {R.}~\bibnamefont
  {Shankar}},\ }\bibfield  {title} {\enquote {\bibinfo {title} {Exact results
  for spin dynamics and fractionalization in the kitaev model},}\ }\href
  {\doibase 10.1103/PhysRevLett.98.247201} {\bibfield  {journal} {\bibinfo
  {journal} {Phys. Rev. Lett.}\ }\textbf {\bibinfo {volume} {98}},\ \bibinfo
  {pages} {247201} (\bibinfo {year} {2007})}\BibitemShut {NoStop}%
\bibitem [{\citenamefont {Read}\ and\ \citenamefont {Green}(2000)}]{Read2000}%
  \BibitemOpen
  \bibfield  {author} {\bibinfo {author} {\bibfnamefont {N.}~\bibnamefont
  {Read}}\ and\ \bibinfo {author} {\bibfnamefont {Dmitry}\ \bibnamefont
  {Green}},\ }\bibfield  {title} {\enquote {\bibinfo {title} {Paired states of
  fermions in two dimensions with breaking of parity and time-reversal
  symmetries and the fractional quantum hall effect},}\ }\href {\doibase
  10.1103/PhysRevB.61.10267} {\bibfield  {journal} {\bibinfo  {journal} {Phys.
  Rev. B}\ }\textbf {\bibinfo {volume} {61}},\ \bibinfo {pages} {10267--10297}
  (\bibinfo {year} {2000})}\BibitemShut {NoStop}%
\bibitem [{\citenamefont {Song}\ \emph {et~al.}(2016)\citenamefont {Song},
  \citenamefont {You},\ and\ \citenamefont {Balents}}]{Song2016}%
  \BibitemOpen
  \bibfield  {author} {\bibinfo {author} {\bibfnamefont {Xue-Yang}\
  \bibnamefont {Song}}, \bibinfo {author} {\bibfnamefont {Yi-Zhuang}\
  \bibnamefont {You}}, \ and\ \bibinfo {author} {\bibfnamefont {Leon}\
  \bibnamefont {Balents}},\ }\bibfield  {title} {\enquote {\bibinfo {title}
  {Low-energy spin dynamics of the honeycomb spin liquid beyond the kitaev
  limit},}\ }\href {\doibase 10.1103/PhysRevLett.117.037209} {\bibfield
  {journal} {\bibinfo  {journal} {Phys. Rev. Lett.}\ }\textbf {\bibinfo
  {volume} {117}},\ \bibinfo {pages} {037209} (\bibinfo {year}
  {2016})}\BibitemShut {NoStop}%
\bibitem [{\citenamefont {Hermanns}\ \emph {et~al.}(2018)\citenamefont
  {Hermanns}, \citenamefont {Kimchi},\ and\ \citenamefont
  {Knolle}}]{Knolle2018}%
  \BibitemOpen
  \bibfield  {author} {\bibinfo {author} {\bibfnamefont {M.}~\bibnamefont
  {Hermanns}}, \bibinfo {author} {\bibfnamefont {I.}~\bibnamefont {Kimchi}}, \
  and\ \bibinfo {author} {\bibfnamefont {J.}~\bibnamefont {Knolle}},\
  }\bibfield  {title} {\enquote {\bibinfo {title} {Physics of the {K}itaev
  model: Fractionalization, dynamic correlations, and material connections},}\
  }\href {\doibase 10.1146/annurev-conmatphys-033117-053934} {\bibfield
  {journal} {\bibinfo  {journal} {Annual Review of Condensed Matter Physics}\
  }\textbf {\bibinfo {volume} {9}},\ \bibinfo {pages} {17--33} (\bibinfo {year}
  {2018})}\BibitemShut {NoStop}%
\bibitem [{\citenamefont {Jackeli}\ and\ \citenamefont
  {Khaliullin}(2009)}]{Khaliullin2009}%
  \BibitemOpen
  \bibfield  {author} {\bibinfo {author} {\bibfnamefont {G.}~\bibnamefont
  {Jackeli}}\ and\ \bibinfo {author} {\bibfnamefont {G.}~\bibnamefont
  {Khaliullin}},\ }\bibfield  {title} {\enquote {\bibinfo {title} {Mott
  insulators in the strong spin-orbit coupling limit: From heisenberg to a
  quantum compass and {K}itaev models},}\ }\href {\doibase
  10.1103/PhysRevLett.102.017205} {\bibfield  {journal} {\bibinfo  {journal}
  {Phys. Rev. Lett.}\ }\textbf {\bibinfo {volume} {102}},\ \bibinfo {pages}
  {017205} (\bibinfo {year} {2009})}\BibitemShut {NoStop}%
\bibitem [{\citenamefont {Cao}\ \emph {et~al.}(2016)\citenamefont {Cao},
  \citenamefont {Banerjee}, \citenamefont {Yan}, \citenamefont {Bridges},
  \citenamefont {Lumsden}, \citenamefont {Mandrus}, \citenamefont {Tennant},
  \citenamefont {Chakoumakos},\ and\ \citenamefont
  {Nagler}}]{PhysRevB.93.134423}%
  \BibitemOpen
  \bibfield  {author} {\bibinfo {author} {\bibfnamefont {H.~B.}\ \bibnamefont
  {Cao}}, \bibinfo {author} {\bibfnamefont {A.}~\bibnamefont {Banerjee}},
  \bibinfo {author} {\bibfnamefont {J.-Q.}\ \bibnamefont {Yan}}, \bibinfo
  {author} {\bibfnamefont {C.~A.}\ \bibnamefont {Bridges}}, \bibinfo {author}
  {\bibfnamefont {M.~D.}\ \bibnamefont {Lumsden}}, \bibinfo {author}
  {\bibfnamefont {D.~G.}\ \bibnamefont {Mandrus}}, \bibinfo {author}
  {\bibfnamefont {D.~A.}\ \bibnamefont {Tennant}}, \bibinfo {author}
  {\bibfnamefont {B.~C.}\ \bibnamefont {Chakoumakos}}, \ and\ \bibinfo {author}
  {\bibfnamefont {S.~E.}\ \bibnamefont {Nagler}},\ }\bibfield  {title}
  {\enquote {\bibinfo {title} {Low-temperature crystal and magnetic structure
  of $\alpha$-{R}u{C}l$_3$},}\ }\href {\doibase 10.1103/PhysRevB.93.134423}
  {\bibfield  {journal} {\bibinfo  {journal} {Phys. Rev. B}\ }\textbf {\bibinfo
  {volume} {93}},\ \bibinfo {pages} {134423} (\bibinfo {year}
  {2016})}\BibitemShut {NoStop}%
\bibitem [{\citenamefont {Ziatdinov}\ \emph {et~al.}(2016)\citenamefont
  {Ziatdinov}, \citenamefont {Banerjee}, \citenamefont {Maksov}, \citenamefont
  {Berlijn}, \citenamefont {Zhou}, \citenamefont {Cao}, \citenamefont {Yan},
  \citenamefont {Bridges}, \citenamefont {Mandrus}, \citenamefont {Nagler}
  \emph {et~al.}}]{ziatdinov2016atomic}%
  \BibitemOpen
  \bibfield  {author} {\bibinfo {author} {\bibfnamefont {M}~\bibnamefont
  {Ziatdinov}}, \bibinfo {author} {\bibfnamefont {Arnab}\ \bibnamefont
  {Banerjee}}, \bibinfo {author} {\bibfnamefont {A}~\bibnamefont {Maksov}},
  \bibinfo {author} {\bibfnamefont {Tom}\ \bibnamefont {Berlijn}}, \bibinfo
  {author} {\bibfnamefont {Wu}~\bibnamefont {Zhou}}, \bibinfo {author}
  {\bibfnamefont {HB}~\bibnamefont {Cao}}, \bibinfo {author} {\bibfnamefont
  {J-Q}\ \bibnamefont {Yan}}, \bibinfo {author} {\bibfnamefont {Craig~A}\
  \bibnamefont {Bridges}}, \bibinfo {author} {\bibfnamefont {DG}~\bibnamefont
  {Mandrus}}, \bibinfo {author} {\bibfnamefont {Stephen~E}\ \bibnamefont
  {Nagler}},  \emph {et~al.},\ }\bibfield  {title} {\enquote {\bibinfo {title}
  {Atomic-scale observation of structural and electronic orders in the layered
  compound $\alpha$-{R}u{C}l 3},}\ }\href@noop {} {\bibfield  {journal}
  {\bibinfo  {journal} {Nature communications}\ }\textbf {\bibinfo {volume}
  {7}},\ \bibinfo {pages} {13774} (\bibinfo {year} {2016})}\BibitemShut
  {NoStop}%
\bibitem [{\citenamefont {Banerjee}\ \emph {et~al.}(2017)\citenamefont
  {Banerjee}, \citenamefont {Yan}, \citenamefont {Knolle}, \citenamefont
  {Bridges}, \citenamefont {Stone}, \citenamefont {Lumsden}, \citenamefont
  {Mandrus}, \citenamefont {Tennant}, \citenamefont {Moessner},\ and\
  \citenamefont {Nagler}}]{Banerjee2017}%
  \BibitemOpen
  \bibfield  {author} {\bibinfo {author} {\bibfnamefont {Arnab}\ \bibnamefont
  {Banerjee}}, \bibinfo {author} {\bibfnamefont {Jiaqiang}\ \bibnamefont
  {Yan}}, \bibinfo {author} {\bibfnamefont {Johannes}\ \bibnamefont {Knolle}},
  \bibinfo {author} {\bibfnamefont {Craig~A}\ \bibnamefont {Bridges}}, \bibinfo
  {author} {\bibfnamefont {Matthew~B}\ \bibnamefont {Stone}}, \bibinfo {author}
  {\bibfnamefont {Mark~D}\ \bibnamefont {Lumsden}}, \bibinfo {author}
  {\bibfnamefont {David~G}\ \bibnamefont {Mandrus}}, \bibinfo {author}
  {\bibfnamefont {David~A}\ \bibnamefont {Tennant}}, \bibinfo {author}
  {\bibfnamefont {Roderich}\ \bibnamefont {Moessner}}, \ and\ \bibinfo {author}
  {\bibfnamefont {Stephen~E}\ \bibnamefont {Nagler}},\ }\bibfield  {title}
  {\enquote {\bibinfo {title} {Neutron scattering in the proximate quantum spin
  liquid $\alpha$-$\text{{R}u{C}l}_3$},}\ }\href {\doibase
  10.1126/science.aah6015} {\bibfield  {journal} {\bibinfo  {journal}
  {Science}\ }\textbf {\bibinfo {volume} {356}},\ \bibinfo {pages} {1055--1059}
  (\bibinfo {year} {2017})}\BibitemShut {NoStop}%
\bibitem [{\citenamefont {Ponomaryov}\ \emph {et~al.}(2017)\citenamefont
  {Ponomaryov}, \citenamefont {Schulze}, \citenamefont {Wosnitza},
  \citenamefont {Lampen-Kelley}, \citenamefont {Banerjee}, \citenamefont {Yan},
  \citenamefont {Bridges}, \citenamefont {Mandrus}, \citenamefont {Nagler},
  \citenamefont {Kolezhuk},\ and\ \citenamefont {Zvyagin}}]{Ponomaryov2017}%
  \BibitemOpen
  \bibfield  {author} {\bibinfo {author} {\bibfnamefont {A.~N.}\ \bibnamefont
  {Ponomaryov}}, \bibinfo {author} {\bibfnamefont {E.}~\bibnamefont {Schulze}},
  \bibinfo {author} {\bibfnamefont {J.}~\bibnamefont {Wosnitza}}, \bibinfo
  {author} {\bibfnamefont {P.}~\bibnamefont {Lampen-Kelley}}, \bibinfo {author}
  {\bibfnamefont {A.}~\bibnamefont {Banerjee}}, \bibinfo {author}
  {\bibfnamefont {J.-Q.}\ \bibnamefont {Yan}}, \bibinfo {author} {\bibfnamefont
  {C.~A.}\ \bibnamefont {Bridges}}, \bibinfo {author} {\bibfnamefont {D.~G.}\
  \bibnamefont {Mandrus}}, \bibinfo {author} {\bibfnamefont {S.~E.}\
  \bibnamefont {Nagler}}, \bibinfo {author} {\bibfnamefont {A.~K.}\
  \bibnamefont {Kolezhuk}}, \ and\ \bibinfo {author} {\bibfnamefont {S.~A.}\
  \bibnamefont {Zvyagin}},\ }\bibfield  {title} {\enquote {\bibinfo {title}
  {Unconventional spin dynamics in the honeycomb-lattice material
  $\ensuremath{\alpha}\text{\ensuremath{-}}{\mathrm{{r}u{c}l}}_{3}$: High-field
  electron spin resonance studies},}\ }\href {\doibase
  10.1103/PhysRevB.96.241107} {\bibfield  {journal} {\bibinfo  {journal} {Phys.
  Rev. B}\ }\textbf {\bibinfo {volume} {96}},\ \bibinfo {pages} {241107}
  (\bibinfo {year} {2017})}\BibitemShut {NoStop}%
\bibitem [{\citenamefont {Little}\ \emph {et~al.}(2017)\citenamefont {Little},
  \citenamefont {Wu}, \citenamefont {Lampen-Kelley}, \citenamefont {Banerjee},
  \citenamefont {Patankar}, \citenamefont {Rees}, \citenamefont {Bridges},
  \citenamefont {Yan}, \citenamefont {Mandrus}, \citenamefont {Nagler},\ and\
  \citenamefont {Orenstein}}]{Little2017}%
  \BibitemOpen
  \bibfield  {author} {\bibinfo {author} {\bibfnamefont {A.}~\bibnamefont
  {Little}}, \bibinfo {author} {\bibfnamefont {Liang}\ \bibnamefont {Wu}},
  \bibinfo {author} {\bibfnamefont {P.}~\bibnamefont {Lampen-Kelley}}, \bibinfo
  {author} {\bibfnamefont {A.}~\bibnamefont {Banerjee}}, \bibinfo {author}
  {\bibfnamefont {S.}~\bibnamefont {Patankar}}, \bibinfo {author}
  {\bibfnamefont {D.}~\bibnamefont {Rees}}, \bibinfo {author} {\bibfnamefont
  {C.~A.}\ \bibnamefont {Bridges}}, \bibinfo {author} {\bibfnamefont {J.-Q.}\
  \bibnamefont {Yan}}, \bibinfo {author} {\bibfnamefont {D.}~\bibnamefont
  {Mandrus}}, \bibinfo {author} {\bibfnamefont {S.~E.}\ \bibnamefont {Nagler}},
  \ and\ \bibinfo {author} {\bibfnamefont {J.}~\bibnamefont {Orenstein}},\
  }\bibfield  {title} {\enquote {\bibinfo {title} {Antiferromagnetic resonance
  and terahertz continuum in
  $\ensuremath{\alpha}\text{\ensuremath{-}}{\mathrm{{r}u{c}l}}_{3}$},}\ }\href
  {\doibase 10.1103/PhysRevLett.119.227201} {\bibfield  {journal} {\bibinfo
  {journal} {Phys. Rev. Lett.}\ }\textbf {\bibinfo {volume} {119}},\ \bibinfo
  {pages} {227201} (\bibinfo {year} {2017})}\BibitemShut {NoStop}%
\bibitem [{\citenamefont {Hentrich}\ \emph {et~al.}(2018)\citenamefont
  {Hentrich}, \citenamefont {Wolter}, \citenamefont {Zotos}, \citenamefont
  {Brenig}, \citenamefont {Nowak}, \citenamefont {Isaeva}, \citenamefont
  {Doert}, \citenamefont {Banerjee}, \citenamefont {Lampen-Kelley},
  \citenamefont {Mandrus}, \citenamefont {Nagler}, \citenamefont {Sears},
  \citenamefont {Kim}, \citenamefont {B\"uchner},\ and\ \citenamefont
  {Hess}}]{Hentrich2018}%
  \BibitemOpen
  \bibfield  {author} {\bibinfo {author} {\bibfnamefont {Richard}\ \bibnamefont
  {Hentrich}}, \bibinfo {author} {\bibfnamefont {Anja U.~B.}\ \bibnamefont
  {Wolter}}, \bibinfo {author} {\bibfnamefont {Xenophon}\ \bibnamefont
  {Zotos}}, \bibinfo {author} {\bibfnamefont {Wolfram}\ \bibnamefont {Brenig}},
  \bibinfo {author} {\bibfnamefont {Domenic}\ \bibnamefont {Nowak}}, \bibinfo
  {author} {\bibfnamefont {Anna}\ \bibnamefont {Isaeva}}, \bibinfo {author}
  {\bibfnamefont {Thomas}\ \bibnamefont {Doert}}, \bibinfo {author}
  {\bibfnamefont {Arnab}\ \bibnamefont {Banerjee}}, \bibinfo {author}
  {\bibfnamefont {Paula}\ \bibnamefont {Lampen-Kelley}}, \bibinfo {author}
  {\bibfnamefont {David~G.}\ \bibnamefont {Mandrus}}, \bibinfo {author}
  {\bibfnamefont {Stephen~E.}\ \bibnamefont {Nagler}}, \bibinfo {author}
  {\bibfnamefont {Jennifer}\ \bibnamefont {Sears}}, \bibinfo {author}
  {\bibfnamefont {Young-June}\ \bibnamefont {Kim}}, \bibinfo {author}
  {\bibfnamefont {Bernd}\ \bibnamefont {B\"uchner}}, \ and\ \bibinfo {author}
  {\bibfnamefont {Christian}\ \bibnamefont {Hess}},\ }\bibfield  {title}
  {\enquote {\bibinfo {title} {Unusual phonon heat transport in
  $\ensuremath{\alpha}\text{\ensuremath{-}}{\mathrm{{r}u{c}l}}_{3}$: Strong
  spin-phonon scattering and field-induced spin gap},}\ }\href {\doibase
  10.1103/PhysRevLett.120.117204} {\bibfield  {journal} {\bibinfo  {journal}
  {Phys. Rev. Lett.}\ }\textbf {\bibinfo {volume} {120}},\ \bibinfo {pages}
  {117204} (\bibinfo {year} {2018})}\BibitemShut {NoStop}%
\bibitem [{\citenamefont {Kasahara}\ \emph {et~al.}(2018)\citenamefont
  {Kasahara}, \citenamefont {Ohnishi}, \citenamefont {Mizukami}, \citenamefont
  {Tanaka}, \citenamefont {Ma}, \citenamefont {Sugii}, \citenamefont {Kurita},
  \citenamefont {Tanaka}, \citenamefont {Nasu}, \citenamefont {Motome},
  \citenamefont {Shibauchi},\ and\ \citenamefont {Matsuda}}]{Kasahara2018}%
  \BibitemOpen
  \bibfield  {author} {\bibinfo {author} {\bibfnamefont {Y.}~\bibnamefont
  {Kasahara}}, \bibinfo {author} {\bibfnamefont {T.}~\bibnamefont {Ohnishi}},
  \bibinfo {author} {\bibfnamefont {Y.}~\bibnamefont {Mizukami}}, \bibinfo
  {author} {\bibfnamefont {O.}~\bibnamefont {Tanaka}}, \bibinfo {author}
  {\bibfnamefont {Sixiao}\ \bibnamefont {Ma}}, \bibinfo {author} {\bibfnamefont
  {K.}~\bibnamefont {Sugii}}, \bibinfo {author} {\bibfnamefont
  {N.}~\bibnamefont {Kurita}}, \bibinfo {author} {\bibfnamefont
  {H.}~\bibnamefont {Tanaka}}, \bibinfo {author} {\bibfnamefont
  {J.}~\bibnamefont {Nasu}}, \bibinfo {author} {\bibfnamefont {Y.}~\bibnamefont
  {Motome}}, \bibinfo {author} {\bibfnamefont {T.}~\bibnamefont {Shibauchi}}, \
  and\ \bibinfo {author} {\bibfnamefont {Y.}~\bibnamefont {Matsuda}},\
  }\bibfield  {title} {\enquote {\bibinfo {title} {Majorana quantization and
  half-integer thermal quantum hall effect in a kitaev spin liquid},}\ }\href
  {\doibase 10.1038/s41586-018-0274-0} {\bibfield  {journal} {\bibinfo
  {journal} {Nature}\ }\textbf {\bibinfo {volume} {559}},\ \bibinfo {pages}
  {227--231} (\bibinfo {year} {2018})}\BibitemShut {NoStop}%
\bibitem [{\citenamefont {Banerjee}\ \emph {et~al.}(2018)\citenamefont
  {Banerjee}, \citenamefont {Lampen-Kelley}, \citenamefont {Knolle},
  \citenamefont {Balz}, \citenamefont {Aczel}, \citenamefont {Winn},
  \citenamefont {Liu}, \citenamefont {Pajerowski}, \citenamefont {Yan},
  \citenamefont {Bridges} \emph {et~al.}}]{banerjee2018excitations}%
  \BibitemOpen
  \bibfield  {author} {\bibinfo {author} {\bibfnamefont {Arnab}\ \bibnamefont
  {Banerjee}}, \bibinfo {author} {\bibfnamefont {Paula}\ \bibnamefont
  {Lampen-Kelley}}, \bibinfo {author} {\bibfnamefont {Johannes}\ \bibnamefont
  {Knolle}}, \bibinfo {author} {\bibfnamefont {Christian}\ \bibnamefont
  {Balz}}, \bibinfo {author} {\bibfnamefont {Adam~Anthony}\ \bibnamefont
  {Aczel}}, \bibinfo {author} {\bibfnamefont {Barry}\ \bibnamefont {Winn}},
  \bibinfo {author} {\bibfnamefont {Yaohua}\ \bibnamefont {Liu}}, \bibinfo
  {author} {\bibfnamefont {Daniel}\ \bibnamefont {Pajerowski}}, \bibinfo
  {author} {\bibfnamefont {Jiaqiang}\ \bibnamefont {Yan}}, \bibinfo {author}
  {\bibfnamefont {Craig~A}\ \bibnamefont {Bridges}},  \emph {et~al.},\
  }\bibfield  {title} {\enquote {\bibinfo {title} {Excitations in the
  field-induced quantum spin liquid state of $\alpha$-{R}u{C}l 3},}\ }\href
  {\doibase https://doi.org/10.1038/s41535-018-0079-2} {\bibfield  {journal}
  {\bibinfo  {journal} {npj Quantum Materials}\ }\textbf {\bibinfo {volume}
  {3}},\ \bibinfo {pages} {8} (\bibinfo {year} {2018})}\BibitemShut {NoStop}%
\bibitem [{\citenamefont {Banerjee}\ \emph {et~al.}(2016)\citenamefont
  {Banerjee}, \citenamefont {Bridges}, \citenamefont {Yan}, \citenamefont
  {Aczel}, \citenamefont {Li}, \citenamefont {Stone}, \citenamefont {Granroth},
  \citenamefont {Lumsden}, \citenamefont {Yiu}, \citenamefont {Knolle} \emph
  {et~al.}}]{Banerjee2016}%
  \BibitemOpen
  \bibfield  {author} {\bibinfo {author} {\bibfnamefont {A}~\bibnamefont
  {Banerjee}}, \bibinfo {author} {\bibfnamefont {CA}~\bibnamefont {Bridges}},
  \bibinfo {author} {\bibfnamefont {J-Q}\ \bibnamefont {Yan}}, \bibinfo
  {author} {\bibfnamefont {AA}~\bibnamefont {Aczel}}, \bibinfo {author}
  {\bibfnamefont {L}~\bibnamefont {Li}}, \bibinfo {author} {\bibfnamefont
  {MB}~\bibnamefont {Stone}}, \bibinfo {author} {\bibfnamefont
  {GE}~\bibnamefont {Granroth}}, \bibinfo {author} {\bibfnamefont
  {MD}~\bibnamefont {Lumsden}}, \bibinfo {author} {\bibfnamefont
  {Y}~\bibnamefont {Yiu}}, \bibinfo {author} {\bibfnamefont {J}~\bibnamefont
  {Knolle}},  \emph {et~al.},\ }\bibfield  {title} {\enquote {\bibinfo {title}
  {Proximate kitaev quantum spin liquid behaviour in a honeycomb magnet},}\
  }\href {\doibase https://doi.org/10.1038/nmat4604} {\bibfield  {journal}
  {\bibinfo  {journal} {Nature materials}\ }\textbf {\bibinfo {volume} {15}},\
  \bibinfo {pages} {733--740} (\bibinfo {year} {2016})}\BibitemShut {NoStop}%
\bibitem [{\citenamefont {Holleis}\ \emph {et~al.}(2021)\citenamefont
  {Holleis}, \citenamefont {Prestigiacomo}, \citenamefont {Fan}, \citenamefont
  {Nishimoto}, \citenamefont {Osofsky}, \citenamefont {Chern}, \citenamefont
  {van~den Brink},\ and\ \citenamefont {Shivaram}}]{Holleis2021}%
  \BibitemOpen
  \bibfield  {author} {\bibinfo {author} {\bibfnamefont {Ludwig}\ \bibnamefont
  {Holleis}}, \bibinfo {author} {\bibfnamefont {Joseph~C.}\ \bibnamefont
  {Prestigiacomo}}, \bibinfo {author} {\bibfnamefont {Zhijie}\ \bibnamefont
  {Fan}}, \bibinfo {author} {\bibfnamefont {Satoshi}\ \bibnamefont
  {Nishimoto}}, \bibinfo {author} {\bibfnamefont {Michael}\ \bibnamefont
  {Osofsky}}, \bibinfo {author} {\bibfnamefont {Gia-Wei}\ \bibnamefont
  {Chern}}, \bibinfo {author} {\bibfnamefont {Jeroen}\ \bibnamefont {van~den
  Brink}}, \ and\ \bibinfo {author} {\bibfnamefont {B.~S.}\ \bibnamefont
  {Shivaram}},\ }\bibfield  {title} {\enquote {\bibinfo {title} {Anomalous and
  anisotropic nonlinear susceptibility in the proximate kitaev magnet
  $\alpha$-{R}u{C}l3},}\ }\href {\doibase 10.1038/s41535-021-00364-z}
  {\bibfield  {journal} {\bibinfo  {journal} {npj Quantum Materials}\ }\textbf
  {\bibinfo {volume} {6}},\ \bibinfo {pages} {66} (\bibinfo {year}
  {2021})}\BibitemShut {NoStop}%
\bibitem [{\citenamefont {Yokoi}\ \emph {et~al.}(2021)\citenamefont {Yokoi},
  \citenamefont {Ma}, \citenamefont {Kasahara}, \citenamefont {Kasahara},
  \citenamefont {Shibauchi}, \citenamefont {Kurita}, \citenamefont {Tanaka},
  \citenamefont {Nasu}, \citenamefont {Motome}, \citenamefont {Hickey},
  \citenamefont {Trebst},\ and\ \citenamefont {Matsuda}}]{Yokoi2021}%
  \BibitemOpen
  \bibfield  {author} {\bibinfo {author} {\bibfnamefont {T.}~\bibnamefont
  {Yokoi}}, \bibinfo {author} {\bibfnamefont {S.}~\bibnamefont {Ma}}, \bibinfo
  {author} {\bibfnamefont {Y.}~\bibnamefont {Kasahara}}, \bibinfo {author}
  {\bibfnamefont {S.}~\bibnamefont {Kasahara}}, \bibinfo {author}
  {\bibfnamefont {T.}~\bibnamefont {Shibauchi}}, \bibinfo {author}
  {\bibfnamefont {N.}~\bibnamefont {Kurita}}, \bibinfo {author} {\bibfnamefont
  {H.}~\bibnamefont {Tanaka}}, \bibinfo {author} {\bibfnamefont
  {J.}~\bibnamefont {Nasu}}, \bibinfo {author} {\bibfnamefont {Y.}~\bibnamefont
  {Motome}}, \bibinfo {author} {\bibfnamefont {C.}~\bibnamefont {Hickey}},
  \bibinfo {author} {\bibfnamefont {S.}~\bibnamefont {Trebst}}, \ and\ \bibinfo
  {author} {\bibfnamefont {Y.}~\bibnamefont {Matsuda}},\ }\bibfield  {title}
  {\enquote {\bibinfo {title} {Half-integer quantized anomalous thermal hall
  effect in the {K}itaev material candidate rucl$_3$},}\ }\href {\doibase
  10.1126/science.aay5551} {\bibfield  {journal} {\bibinfo  {journal}
  {Science}\ }\textbf {\bibinfo {volume} {373}},\ \bibinfo {pages} {568--572}
  (\bibinfo {year} {2021})},\ \Eprint
  {http://arxiv.org/abs/https://www.science.org/doi/pdf/10.1126/science.aay5551}
  {https://www.science.org/doi/pdf/10.1126/science.aay5551} \BibitemShut
  {NoStop}%
\bibitem [{\citenamefont {Czajka}\ \emph {et~al.}(2021)\citenamefont {Czajka},
  \citenamefont {Gao}, \citenamefont {Hirschberger}, \citenamefont
  {Lampen-Kelley}, \citenamefont {Banerjee}, \citenamefont {Yan}, \citenamefont
  {Mandrus}, \citenamefont {Nagler},\ and\ \citenamefont
  {Ong}}]{czajka2021oscillations}%
  \BibitemOpen
  \bibfield  {author} {\bibinfo {author} {\bibfnamefont {Peter}\ \bibnamefont
  {Czajka}}, \bibinfo {author} {\bibfnamefont {Tong}\ \bibnamefont {Gao}},
  \bibinfo {author} {\bibfnamefont {Max}\ \bibnamefont {Hirschberger}},
  \bibinfo {author} {\bibfnamefont {Paula}\ \bibnamefont {Lampen-Kelley}},
  \bibinfo {author} {\bibfnamefont {Arnab}\ \bibnamefont {Banerjee}}, \bibinfo
  {author} {\bibfnamefont {Jiaqiang}\ \bibnamefont {Yan}}, \bibinfo {author}
  {\bibfnamefont {David~G}\ \bibnamefont {Mandrus}}, \bibinfo {author}
  {\bibfnamefont {Stephen~E}\ \bibnamefont {Nagler}}, \ and\ \bibinfo {author}
  {\bibfnamefont {NP}~\bibnamefont {Ong}},\ }\bibfield  {title} {\enquote
  {\bibinfo {title} {Oscillations of the thermal conductivity in the
  spin-liquid state of $\alpha$-{R}u{C}l3},}\ }\href@noop {} {\bibfield
  {journal} {\bibinfo  {journal} {Nature Physics}\ }\textbf {\bibinfo {volume}
  {17}},\ \bibinfo {pages} {915--919} (\bibinfo {year} {2021})}\BibitemShut
  {NoStop}%
\bibitem [{\citenamefont {Fava}\ \emph {et~al.}(2022)\citenamefont {Fava},
  \citenamefont {Gopalakrishnan}, \citenamefont {Vasseur}, \citenamefont
  {Essler},\ and\ \citenamefont {Parameswaran}}]{Parameswaran2022a}%
  \BibitemOpen
  \bibfield  {author} {\bibinfo {author} {\bibfnamefont {Michele}\ \bibnamefont
  {Fava}}, \bibinfo {author} {\bibfnamefont {Sarang}\ \bibnamefont
  {Gopalakrishnan}}, \bibinfo {author} {\bibfnamefont {Romain}\ \bibnamefont
  {Vasseur}}, \bibinfo {author} {\bibfnamefont {Fabian H.~L.}\ \bibnamefont
  {Essler}}, \ and\ \bibinfo {author} {\bibfnamefont {S.~A.}\ \bibnamefont
  {Parameswaran}},\ }\bibfield  {title} {\enquote {\bibinfo {title} {Divergent
  nonlinear response from quasiparticle interactions},}\ }\href {\doibase
  10.48550/ARXIV.2208.09490} {\  (\bibinfo {year} {2022}),\
  10.48550/ARXIV.2208.09490}\BibitemShut {NoStop}%
\bibitem [{\citenamefont {McGinley}\ \emph {et~al.}(2022)\citenamefont
  {McGinley}, \citenamefont {Fava},\ and\ \citenamefont
  {Parameswaran}}]{Parameswaran2022}%
  \BibitemOpen
  \bibfield  {author} {\bibinfo {author} {\bibfnamefont {Max}\ \bibnamefont
  {McGinley}}, \bibinfo {author} {\bibfnamefont {Michele}\ \bibnamefont
  {Fava}}, \ and\ \bibinfo {author} {\bibfnamefont {S.~A.}\ \bibnamefont
  {Parameswaran}},\ }\bibfield  {title} {\enquote {\bibinfo {title} {Signatures
  of fractional statistics in nonlinear pump-probe spectroscopy},}\ }\href
  {\doibase 10.48550/ARXIV.2210.16249} {\  (\bibinfo {year} {2022}),\
  10.48550/ARXIV.2210.16249}\BibitemShut {NoStop}%
\bibitem [{\citenamefont {Arakawa}\ and\ \citenamefont
  {Yonemitsu}(2021)}]{Arakawa2021}%
  \BibitemOpen
  \bibfield  {author} {\bibinfo {author} {\bibfnamefont {Naoya}\ \bibnamefont
  {Arakawa}}\ and\ \bibinfo {author} {\bibfnamefont {Kenji}\ \bibnamefont
  {Yonemitsu}},\ }\bibfield  {title} {\enquote {\bibinfo {title} {Floquet
  engineering of mott insulators with strong spin-orbit coupling},}\ }\href
  {\doibase 10.1103/PhysRevB.103.L100408} {\bibfield  {journal} {\bibinfo
  {journal} {Phys. Rev. B}\ }\textbf {\bibinfo {volume} {103}},\ \bibinfo
  {pages} {L100408} (\bibinfo {year} {2021})}\BibitemShut {NoStop}%
\bibitem [{\citenamefont {Alvarez}()}]{alvarez08}%
  \BibitemOpen
  \bibfield  {author} {\bibinfo {author} {\bibfnamefont {G.}~\bibnamefont
  {Alvarez}},\ }\bibfield  {title} {\enquote {\bibinfo {title} {Dmrg++
  website},}\ }\href {https://g1257.github.com/dmrgPlusPlus} {\ }\BibitemShut
  {NoStop}%
\bibitem [{\citenamefont {Feng}\ \emph
  {et~al.}(2022{\natexlab{b}})\citenamefont {Feng}, \citenamefont {Alvarez},\
  and\ \citenamefont {Trivedi}}]{Feng2022gapless}%
  \BibitemOpen
  \bibfield  {author} {\bibinfo {author} {\bibfnamefont {Shi}\ \bibnamefont
  {Feng}}, \bibinfo {author} {\bibfnamefont {Gonzalo}\ \bibnamefont {Alvarez}},
  \ and\ \bibinfo {author} {\bibfnamefont {Nandini}\ \bibnamefont {Trivedi}},\
  }\bibfield  {title} {\enquote {\bibinfo {title} {Gapless to gapless phase
  transitions in quantum spin chains},}\ }\href {\doibase
  10.1103/PhysRevB.105.014435} {\bibfield  {journal} {\bibinfo  {journal}
  {Phys. Rev. B}\ }\textbf {\bibinfo {volume} {105}},\ \bibinfo {pages}
  {014435} (\bibinfo {year} {2022}{\natexlab{b}})}\BibitemShut {NoStop}%
\end{thebibliography}%

\end{document}